%% file: Tesis_v_Arxiv.tex
\documentclass[A4paper,twoside,openright,11pt]{report}

\usepackage[latin1]{inputenc}
\usepackage{amssymb}
\usepackage{latexsym}
\usepackage{mathrsfs}
\usepackage{euscript}
\usepackage{amsmath}
\usepackage{graphicx}
\usepackage{psfrag}
\usepackage{epsfig}
\usepackage{slashed}
\usepackage[dvips]{color}
\usepackage{fancyhdr}
\usepackage{hyperref}

\renewcommand{\chaptermark}[1]{\markboth{#1}{}}
\renewcommand{\sectionmark}[1]{\markright{\thesection\ #1}}
\renewcommand{\headrulewidth}{0.5pt}
\renewcommand{\footrulewidth}{0pt}
\fancypagestyle{plain}{\fancyhead{}\renewcommand{\headrulewidth}{0pt}}

\newcommand{\clearemptydoublepage}{}
\newcommand{\be}{\begin{equation}}
 \newcommand{\ee}{\end{equation}}
 \def\bea{\begin{eqnarray}}
 \def\eea{\end{eqnarray}}
 
 \newcommand{\f}[2]{\frac{#1}{#2}}
 \newcommand{\go}{\rightarrow}
 \newcommand{\p}{\phi}
 \newcommand{\g}{\gamma}
 \newcommand{\ep}{\epsilon}
 
 \newcommand{\E}[1]{10^{#1}\,}
 \newcommand{\e}[1]{10^{- #1}\,}
 \renewcommand{\sin}{\mathrm{sin}}
 \newcommand{\pa}[1]{\left( #1 \right)}
 \newcommand{\br}[1]{\left[ #1 \right]}

\begin{document}

\thispagestyle{plain}

\ \vspace{-1cm}

\begin{center}

\begin{center}
\includegraphics[height=5cm,width=7cm]{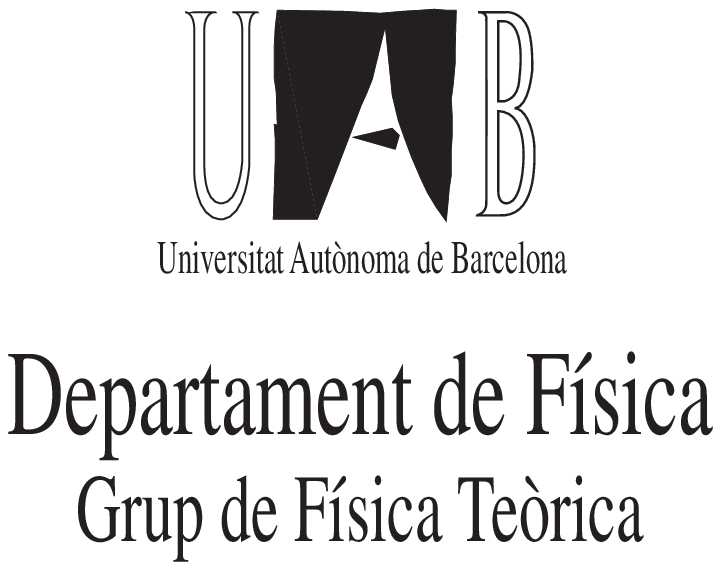} 
\includegraphics{ifae.ps}
\end{center}

\vspace{2cm}
{\LARGE \textbf{Ph.D. Thesis:  }}
\vspace{2cm}

{\huge \textbf{Can the PVLAS particle be compatible \\ with the astrophysical bounds? }}

\vspace{2cm}
{\large \textbf{by  }}
\vspace{2cm}

{\LARGE \textbf{Javier Redondo}}

\vspace{3cm}

{\Large \textbf{Bellaterra, 2007}}
\end{center}
\newpage
\thispagestyle{empty}

\tableofcontents

\input{chapter0.tex}

\newpage

\renewcommand{\chaptermark}[1]{\markboth{#1}{}}
\renewcommand{\sectionmark}[1]{\markright{\thesection\ #1}}

\fancyhf{}
\fancyhead[LE,RO]{\it \thepage}
\fancyhead[LO]{\it \rightmark}
\fancyhead[RE]{\it \leftmark}
\renewcommand{\headrulewidth}{0.5pt}
\renewcommand{\footrulewidth}{0pt}
\addtolength{\headheight}{0.5pt}
\fancypagestyle{plain}{\fancyhead{}\renewcommand{\headrulewidth}{0pt}}


\pagenumbering{arabic}

\input{chapter1.tex}

\input{chapter3.tex}

\input{chapter2.tex}
\input{chapter4.tex}
\newpage\thispagestyle{empty}
\input{JCAP.tex}
\input{PRL.tex}
\newpage\thispagestyle{empty}
\input{PRD-II.tex}
\newpage\thispagestyle{empty}
\input{PRL2.tex}

\input{chapter5.tex}
\newpage\thispagestyle{empty}
\appendix
\input{squizons}

\end{document}

%% file: chapter0.tex
\chapter*{Preface}
\addcontentsline{toc}{chapter}{Preface}
\thispagestyle{plain}
\hspace{1cm}

Very recently, the PVLAS collaboration has reported the observation of two unexpected effects.
Studying the propagation of linearly polarized laser light through a strong transverse magnetic
field in vacuum they find an anomalous rotation of the polarization plane as well as an induced
ellipticity of the outgoing beam. None of these two effects has a standard explanation within
conventional physics at this moment, but they converge into a frequent prediction of physics beyond
the standard model of particle physics: the existence of neutral and nearly massless bosons coupled
to light.

The key word in the above paragraph is ``unexpected". Indeed, the small collaboration was actively
looking for these so called axionlike particles (ALPs) leading to the two mentioned effects. The
important point is that these particles seem to have couplings to the electromagnetic field which
are far to be allowed by other experiments and, in particular by astrophysical arguments.

A conservative look would then discard immediately the ALP interpretation of the PVLAS signal but I
shall show in these few pages that this decision would be a bit premature. In these years under the
supervision of Dr. Mass\'o we have showed that there are particle physics models in which the
discrepancies coming from astrophysics and other laboratory experiments can be circumvented.
Notably, these models involve new physics at low energy scales. In this PhD thesis I address the
issue of presenting all our conclusions. \thispagestyle{plain}

The realization of this work has been a great pleasure for me, in part because of the confluence of
the many fields of physics involved, optics, astrophysics, cosmology and, of course, particle
physics, but mostly because it has been carried out in friendly touch with many experts on all
these fields. It has been then twice an enriching experience also, but moreover, I think that has
given our work a soundness which otherwise would be lacking. The multidisciplinarity of this thesis
has a natural drawback, however, one would be tempted to review all the fields involved and produce
a too long writing. I was not really seduced with this idea and I have tried only to touch the
points I really consider important to understand my original work. Wherever I feel there is more to
be said I give the needed references.

Since the publication of the PVLAS result we have been witnesses of a real ``boom" in the field
with both, theoretical efforts and the advent of many experimental proposals to test the PVLAS ALP
interpretation. This reflects, among other things, the great importance of having new tests of
physics beyond the standard model. In the years before the Large Hadron Collider, which will scan
TeV energies with an unprecedent budget in the history of physics, we could be finding a new
frontier of knowledge in a much more modest experiment, and at very low energies.

In Chapter \ref{general concerns} I briefly motivate low mass particles coupled to light and
introduce the required ingredients that will show up in my models. I am concerned with scalar
particles such as pseudoGoldstone bosons, candidates for the PVLAS ALP, paraphotons and
millicharged particles. Chapter \ref{PVLAS} is devoted to introduce the PVLAS experiment as well as
its ALP interpretation in terms of a new scalar particle coupled to two photons. Further comments
will be made about other possible interpretations. Finally, in Chapter \ref{constraints} I review
the experimental knowledge about the low mass particles coupled to light that could be involved in
the PVLAS signal and state the inconsistency of the bare PVLAS ALP interpretation with astrophysics
and two very sensitive experiments: ALP helioscopes and 5$^\mathrm{th}$ force searches.

Chapter 4 constitutes an introduction for the original work presented in this thesis. My
contributions are organized as a compendium of articles published between the years 2005 and 2007,
each in a separate Chapter. The chronological order in which they appear reflects somehow the
evolution of the theoretical efforts required during these last three years. Chapter 5 (reference
\cite{Masso:2005ym}) gives two general ideas for reconciling the ALP interpretation of PVLAS with
the physics of stars introducing new particles and interactions at low, accessible, energy scales
$\sim$ eV. The concrete models proposed seem to be now a bit obsolete but the message imprinted
remains valid, there is need for additional low energy physics in order to evade astrophysical
bounds on the PVLAS particle. \thispagestyle{plain}

A further refined model is presented in Chapter 6 (reference \cite{Masso:2006gc}). There we include
two paraphotons and a millicharged particle, together with the PVLAS ALP, in order to circumvent
the astrophysical constraints. The model is simple and, although it has some fine-tuned quantities,
turns out to be one of the few models that could to the job if the PVLAS ALP is confirmed.

Even before the time this paper was published, the community was starting to be active proposing
other models and we had the feeling that some model-independent study of evading astrophysical
bounds could be of interest. Therefore we prepared \cite{Jaeckel:2006xm}, which is presented in
Chapter 7, as a tool of evaluating more precisely the models in \cite{Masso:2005ym,Masso:2006gc}
(which lack precise calculations) as well as any other of similar
characteristics\footnote{Interestingly enough, we noticed a new way of evading the astrophysical
bounds if the PVLAS particle is a Chameleon field.}. The general conclusion is again that the new
physics should appear at scales accessible for Earth-ground laboratory experiments (even lower than
naively expected) and therefore experimental efforts were strongly encouraged.

Last summer, the PVLAS collaboration released new data suggesting the parity of the ALP.
Measurements performed in gas indicated that the ALP should be a parity-even particle and therefore
it would mediate long range forces between macroscopic bodies. In the last article presented in
this thesis (Chapter 8 from reference \cite{Dupays:2006dp}) we analyzed such a possibility. The
very precise measurements of the Newton's law and Casimir effect rule out contributions of the
PVLAS particle and the impact in the parameter space of ALPs is really impressive. Moreover, we
realized than in the model presented in Chapter 6, the force is suppressed with respect to the
naive expectations and the ALP interpretation of PVLAS is possible. \thispagestyle{plain}

Finally, In Chapter 9 a short summary and a final discussion are presented. \thispagestyle{plain}

%% file: chapter1.tex
\chapter{General concerns about light weakly coupled particles \label{general concerns}}


\section{Introduction}
Despite the considerable success of explaining almost every observation within particle physics,
the standard model (SM), based on the gauge symmetry $SU(3)_C\otimes SU(2)_I\otimes U(1)_Y$, is
generally thought not to be the final theory explaining the interactions between elementary
particles. There are several reasons for that, and we can arrange them into three big sets of
problems and objections.

The first one is concerned with very fundamental objections to the quantum theory: the postulate of
measurement, the lack of a direction of time, etc. The second is less fundamental and simply
concerns the elusion of the gravitational interactions. This problem is addressed at present within
quantum gravity and string theory, being the last the most promising but still far from providing
mature answers.

Finally, the third set are problems regarding the concrete form of the standard model as a quantum
field theory. Most of them can be regarded as purely ``aesthetical" features. As examples we find
the large number of parameters (coming from our ignorance of a possible flavor structure at higher
energies?), the flavor problem (why three generations?), the hierarchy problem (why such a
difference between the electroweak (EW) and the Planck scales?), the missing Higgs boson (where is
it?), the strong CP problem (why does QCD respect CP?), etc.

While the first two sets of ``fundamental problems" of the SM seem difficult to solve in a quantum
field theory framework, those from the third are the subject of many speculative but ingenious
proposals beyond the SM. For instance, the existence of a symmetry between bosons and fermions, the
so called Supersymmetry, could help to stabilize the EW-Planck hierarchy and theories of dynamical
breaking of the EW symmetry do not need a Higgs boson. These models generally introduce new fields
and gauge symmetries beyond the SM. Most of the new particles and gauge bosons are extremely
massive and we only have chances to discover them by building huge accelerators or by precision
measurements of carefully chosen observables.

However there are several of such theories in which also low mass particles are predicted. If these
particles exist they must necessarily be very weakly interacting, otherwise they would have been
already discovered\footnote{This last assertion acts as a definition of ``very weakly interacting"
and ``low mass" in a precise sense.}. On the other hand they can be active and stable at the low
energies of our universe and give rise to a completely different phenomenology from their massive
companions.

This Chapter is devoted to review some relevant scenarios where those particles arise while I leave
the study of their peculiar phenomenology to Chapter \ref{constraints}. Particular emphasis will be
made on their interactions with the electromagnetic field because these turn out to be specially
interesting from the experimental point of view. Indeed many of the experimental efforts looking
for these particles have been focused on these couplings, and one of those experiments, performed
by the PVLAS collaboration, has recently reported a positive signal.

From a theoretical viewpoint, the existence of low mass particles is closely related to symmetries.
Regarding the renormalization group approach, masses are dimensionful parameters of the theory that
tend to develop values near the ultraviolet cut-off unless some symmetry forbids them. Put it
another way: if we intend to force some dimensionful parameter to be much smaller than the higher
energy scale in the theory $\Lambda$, radiative corrections involving particles related to
$\Lambda$ will induce contributions proportional to powers of $\Lambda$ which will have to be
canceled at a very precise level with counterterms of the same order. Such a tuned cancelation is
generally though to be unnatural and unaesthetical, and it engenders the so called
\textit{hierarchy problem}. The only way to control these radiative corrections is to protect the
small scales with symmetries. Indeed, supersymmetric theories were motivated in part to avoid the
electroweak-planck hierarchy. As notorious examples we find also the masses of fermions in chiral
theories and Goldstone and gauge bosons. Let me say a few words about them.

The lowest-dimension (non trivial) irreducible representations of the Lorentz group in 3+1
space-time dimensions are spinors of left and right handed chirality. Parity transformations
convert one set into the other but, although parity is approximately conserved, it seems not to be
a good quantum number at the most fundamental level. It is allowed then for LH and RH fermions to
be in different representations of the gauge group, whatever they are. Dirac mass terms involve
just one LH and other RH spinor so they have to combine forming a gauge invariant term. From this
viewpoint, a pair of LH and RH fermion fields living in representations that allow a scalar of the
gauge group will acquire a mass of the order of the cut-off, $\Lambda$, but LH and RH fermion
fields which cannot be paired for any reason\footnote{For instance, there could be more LH than RH
fields in nature, they could belong to representations of different range,etc.} would lead to
massless modes. On the other hand, Majorana mass terms are allowed only for fields completely
neutral under the gauge group. This is the case, for instance, of the RH neutrinos of the standard
model which for the above reasoning are supposed to be very massive. Interestingly, a huge Majorana
mass for the RH neutrino field produces a suppression of the Dirac neutrino masses by means of the
see-saw mechanism.

This picture is valid as long as the gauge symmetry is unbroken. In the standard model, the Higgs field acquires a
vacuum expectation value that breaks the $SU(2)_I\otimes U(1)_Y$ sector to $U(1)_{QED}$. In this way, interactions of
Yukawa type of the Higgs with a pair (LH-RH) of fermion fields lead to \textit{effective mass terms} that are naturally
much smaller than the cut-off scale of the theory as long as the vacuum expectation value of the Higgs, or the Yukawa
couplings, are small. Explaining why a scalar field like the Higgs could develop a small expectation value $\sim 200$
GeV (small with respect to the Planck scale $m_\mathrm{Pl}\sim 10^{19}$ GeV) is again the hierarchy problem.

Furthermore, the Nambu-Goldstone theorem tells us that theories with a \textit{global} symmetry
spontaneously broken in the vacuum contain massless particles, known as Goldstone bosons. This
result goes beyond perturbation theory so radiative corrections do not spoil this conclusion.
Nevertheless most of the known examples of Goldstone bosons come from slightly explicitly broken
symmetries which allow for small masses that can be again sensitive to the hierarchy.

Finally, the gauge bosons of a local unbroken symmetry are naturally massless at all orders being their masses
protected by the gauge invariance of the dynamics. There is a nice exception in pure $U(1)$ symmetries because a mass
term is also allowed \cite{Stuckelberg:1937,Stuckelberg:1938a,Stuckelberg:1938b,Stuckelberg:1938c}. From this point of
view it seems \textit{natural} to think that $U(1)_{QED}$ or $U(1)_Y$ are not fundamental $U(1)$ symmetries but the low
energy residual of a higher embedding non-abelian group.

In the remaining of this Chapter I will go deeper into the important theoretical frameworks that
provide small mass particles that are crucial for this thesis. I begin with an exposition of the
Nambu-Goldstone theorem followed by the related history of two of the most important examples,
\textit{pions} coming from the breaking of the axial part of the isospin symmetry of QCD and
\textit{axions}, Goldstone bosons of a hypothetical broken axial $U(1)$ symmetry proposed to solve
the strong CP problem. The discussion is focused on this last case, since it has become a kind of
paradigm in low mass bosons searches.

Next I consider the phenomenology of new $U(1)$ symmetries in a hidden sector, giving rise to
paraphotons and millicharged particles. These provide the framework for our solution to the
PVLAS-Astrophysics puzzle so I describe them with greater detail.

\section{The Nambu-Goldstone theorem}

The Nambu-Goldstone theorem states that:

\textit{whenever a global continuous symmetry of the action is not respected by the ground state,
there appear massless particles, one for each broken generator of the symmetry group.}

The demonstration of this result inhabits central pages of many text books, so I believe it is not
worth to be reviewed here. However, it will prove convenient to express it in mathematical form. To
do so, consider the global infinitesimal transformation
\be \phi_i(x) \rightarrow \phi_i(x) + \alpha_a \mathsf{t}^a_{ij}\phi_j(x) \ee
where $\phi_i(x)$ are fields (of any type, although only scalar fields are allowed to have nonzero
vacuum expectation values), $\alpha_a$ is an array of arbitrary infinitesimals, and
$\mathsf{t}^a_{ij}$ are the generators of the transformations with the index $a$ running over such
transformations. If these transformations are symmetries of the lagrangian there are associated
currents $J^{a\mu}$ which are conserved in the sense that $\partial_\mu J^{a\mu}=0$. We can write
the Goldstone theorem\footnote{I display the result for hermitian $J_a^\mu$ and $\phi_i$s only for
simplicity.} as
\bea && \langle [J^{a\mu}(x),\phi_i(0)]\rangle_0 = \int \f{d^4p}{(2\pi)^3} \left[
p^\mu\rho^a_{i}(p^2)e^{ip\cdot x}+h.c.\right] \label{Nambu-Gold-theo:current-field-correlator} \\
 && \rho_i^a(\mu^2)=i\delta(\mu^2)\sum_j \mathsf{t}^a_{ij}\langle\phi(0)_j\rangle_0 \eea
We see that the K\"all\'en-Lehmann spectral density $\rho^a_i(\mu^2)$ contains a pole in $\mu^2=0$
 only if the associated symmetry generator is broken in the vacuum, $\sum_i
\mathsf{t}^a_{ij}\langle\phi(0)_j\rangle_0\neq 0$. The corresponding massless particles, known as
Goldstone bosons (GB), are created by combinations of the fields that will develop a v.e.v.
\be \pi^a_l(x)\propto \mathsf{t}^a_{lj} \phi_j(x)  \ .\ee
It is clear that there are as many GBs as broken generators. The contribution of the massless modes
to the two-point function \eqref{Nambu-Gold-theo:current-field-correlator} is proportional to
$\langle 0 | J^{a\mu}(x)|\pi^a_l\rangle \langle \pi^a_l | \phi_i(0)|0\rangle$. A nonzero value
requires that $\pi^a_l$ must be spin-zero fields (because $\phi_i(0)|0\rangle$ is rotationally
invariant) and have the same conserved quantum numbers\footnote{The quantum numbers that are
conserved also by the vacuum.} than $J^{a0}$ (otherwise $\langle 0 | J^{a0}(x)|\pi^a_l\rangle$
vanishes\footnote{Provided $\pi^a_l$ has spin-zero, $\mu\neq 0$ can not contribute because it
transforms as a spin-1 field.}).

\subsection{Approximate symmetries and pseudo Goldstone Bosons.}

It is usually the case that global continuous symmetries present in our theories are just
approximate. The potential $V(\phi)$ is then divided into a ``symmetric" part $V_0(\phi)$ and a
breaking term $V_1(\phi)$. If the breaking term $V_1$ is small we also find GB but the new terms
can include non-zero (but small) masses. We name such particles \textit{pseudo}-Goldstone bosons
(pGBs). Also, typically, a reduction of the degeneracy of the vacuum configurations is found (the
so called \textit{vacuum alignment}).

\subsection{Axions and the strong CP problem}

Consider quantum chromodynamics with only two flavors of massless quarks, $u$ and $d$. The
lagrangian of this theory has a global symmetry $U(2)_L\otimes U(2)_R$ that allows to perform
rotations of LH and RH fields independently, namely
\be \left( \begin{array}{c} u_{L} \\ d_L \end{array}\right) \rightarrow e^{i \lambda^{\small L}_a
T^a} \left(
\begin{array}{c} u_L \\ d_L \end{array}\right) \ \ ; \ \
\left( \begin{array}{c} u_{R} \\ d_R \end{array}\right) \rightarrow e^{i \lambda^{\small R}_a T^a}
\left(
\begin{array}{c} u_R \\ d_R \end{array}\right) \label{rotationsU2xU2}\ee
where $T_a=\{1,\overrightarrow{\tau}\}$ ($a=0,1,2,3$) with $\overrightarrow{\tau}$ the Pauli
matrices and $\lambda^{\small L,R}_a$ free parameters. This symmetry can be also decomposed into
vector and axial parts $SU(2)_V\otimes SU(2)_A\otimes U(1)_V \otimes U(1)_A$. We can easily
associate the $SU(2)_V$ part with the isospin symmetry\footnote{In the standard model this symmetry
is only approximate because $u$ and $d$ quarks have different electric charge and a ``small" mass
difference compared with $\Lambda_\mathrm{QCD}$, both due to the electroweak interactions.} of the
strong interactions. Moreover, $U(1)_V$ leads to conservation of the baryon number, which is an
observed property of the strong and electroweak interactions\footnote{This is valid up to the
electroweak scale where instantons alter the picture.}.

At first sight there is no symmetry (not even approximate) in nature related to the axial
subgroups. At low energies, a quark condensate forms making
$\langle\overline{u}u\rangle_0=\langle\overline{d}d\rangle_0\neq 0$ and thus breaking spontaneously
this part of the symmetry. The axial symmetries must then show up in the appearance of the
corresponding Goldstone bosons. This seems to be the case for the $SU(2)_A$ part because we observe
in nature three light bosons, the pions, one for each broken generator of $SU(2)_A$. But there is
no such Goldstone boson for the $U(1)_A$ symmetry\footnote{The lightest candidate was the $\eta$,
but it turns out to be too big. Weinberg showed \cite{Weinberg:1975ui} that the squared mass of the
$U(1)_A$ Goldstone should be smaller than three times the square of the pion mass.}. Weinberg
stated this puzzle as the $U(1)$ problem and suggested that there could be no $U(1)_A$ symmetry in
the strong interactions.

\begin{figure}   \centering \vspace{-1cm}
  \includegraphics[width=6cm]{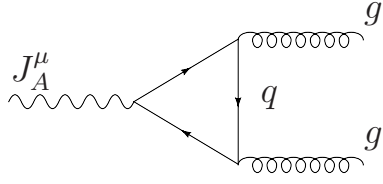}\vspace{-.5cm}
  \caption{\small \label{anomalia} Feynman graph for the color anomaly. The bosonic line labeled
  $J_A^\mu$ stands for a \textit{fictitious} gauge field coupled to $J_A^\mu$. The gluon field is labeled $g$ and it
  is assumed a sum over all fermions $q$ carrying color charge. There is an
  additional graph with the fermions circulating in opposite sense.}
\end{figure}

The solution of this problem was found by 't Hooft who realized that the $U(1)_A$ subgroup was
indeed badly broken in QCD due to its non-trivial vacuum structure \cite{tHooft:1986nc}. To
understand the main points, first realize that $U(1)_A$ suffers an anomaly due to a triangle loop
diagram (Fig. \ref{anomalia}) with quarks circulating on it. This anomaly contributes to the
divergence of its related current $J_A^\mu$ with
\be \partial_\mu J^\mu_A = \f{\alpha_s }{8\pi}{\cal N}G_a^{\mu\nu}\widetilde G_{a\mu\nu}
\label{anomaly}\ee
where  $G_a^{\mu\nu}$ is the gluon field strength, and $\widetilde
G_a^{\mu\nu}=\f{1}{2}\epsilon_{\mu\nu\alpha\beta}G_a^{\alpha\beta}$ is its dual with
$\epsilon_{\mu\nu\alpha\beta}$ the total antisymmetric tensor in 4-dimensions. ($\cal N$ is the
number of quarks circulating in the loop, $2$ in our case.) This itself does not violate $U(1)_A$
because \eqref{anomaly} amounts a total divergency which in principle could be naively removed (as
it happens in quantum electrodynamics)
\be \f{\alpha_s}{8\pi}G_a^{\mu\nu}\widetilde G_{a\mu\nu} = \partial_\mu K^\mu\ee
%
%
But the surface integral of the Bardeen current $K^\mu$ can not be set to zero. This is understood
when we realize that the QCD vacuum is degenerated being the multiple states gauge transformations
of the $G_a^\mu=0$ configuration. They can be classified by the so called topological winding
number $n$ which characterizes the way these configurations behave at spatial infinity. Notably,
the QCD lagrangian allows for non-perturvative solutions of the equations of motion, the so-called
instantons that mediate transitions between these vacuum states. But these instantons vanish very
slowly at spatial infinity ($\sim 1/r$) so they \textit{give finite contributions} to the integral
of the divergence of Bardeen's current. To be concrete, the contribution of
$G_{a}^{\mu\nu}(n_+,n_-,x)$, an instanton taking a vacuum state with winding number $n_-$ to one
with $n_+$ is
\be \int d^4x \partial_\mu K^\mu = \f{\alpha_s}{8\pi}\int d^4 x G_a^{\mu\nu}\widetilde G_{a\mu\nu}
= \nu \ee
where $\nu=n_+-n_-$, the diference in winding numbers. Clearly, $U(1)_A$ is violated by these
instantons.  This solves the Weinberg $U(1)$ puzzle but unfortunately (or not) it poses a new and
very interesting discussion remaining nowadays yet unsolved: \textit{the strong CP problem.}

Note that any physical amplitude must take these instanton solutions into account in the path
integral. For reasons of gauge invariance and cluster decomposition one choses a weight factor
$e^{-i\theta \nu}$ for any of those configurations. This is formally equivalent to include a new
term in the effective lagrangian
\be {\cal L}_\theta = \theta \f{\alpha_s}{8\pi}  G_a^{\mu\nu}\widetilde G_{a\mu\nu}
\label{thetaterm} \ \ .\ee
Such a term violates parity and time reversal invariance but conserves charge conjugation
invariance (and thus violates CP). The picture looks a bit more complicated if we take into account
that quarks gain a mass after the spontaneous symmetry breaking of the $SU(2)_I\otimes U(1)_Y$
gauge symmetry of the standard model. In general this induces a complex mass matrix
\be {\cal L}_{mass} = -\overline{q}_{iL} M_{ij} q_{jR} - \overline{q}_{iR} M^\dagger_{ij} q_{jL}
\ee
($q$ holds for up and down quarks here) that can be diagonalized by a biunitary transformation that
 involves precisely a $U(1)_A$ rotation \eqref{rotationsU2xU2} with
\be \lambda_{i=1,2,3}^{\small L,R}=0 \ \ ; \ \ \lambda_0^{\small R}= -\lambda_0^{\small
L}=\f{1}{4}\mathrm{Arg Det} M \label{diagonalizationQuarkMassMatrix}\ee
This rotation modifies the measure of the quark fields in the path integral of the theory
\cite{Fujikawa:1979ay} introducing another term like \eqref{thetaterm} in the
lagrangian\footnote{See the exposition in \cite{Weinberg:1996kr}.}, shifting the value of $\theta$
and allowing us to define an effective $\overline \theta$
\be \overline \theta = \theta + \mathrm{Arg Det} M \ee
so actually the CP violating term in the QCD sector of the standard model is the sum of two very
different contributions, $\theta$ coming from QCD vacuum dynamics and $\mathrm{Arg Det} M$ from the
electroweak breaking sector. There is no convincing reason why any of these quantities should be
zero, and it is even more unlikely that being both non-zero they cancel exactly\footnote{If
$U(1)_A$ would have been a true symmetry then this transformation would have had no consequences
and the phases of quark masses would be unobservable as well as the term \eqref{thetaterm} which
could be rotated away.} \cite{Peccei:2006as}.

However, as mentioned before, a nonzero value of $\overline\theta$ implies CP violating observables
related to the strong interactions and, remarkably, it induces an electric dipole moment for the
neutron which up to now has eluded detection. The latest measurements \cite{Baluni:1978rf} set the
impressive bound
\be \overline \theta \lesssim 10^{-10} \ \ \ ,\ee
which finally poses the strong CP problem:

\textit{why is $\overline\theta$ so small being the sum of two, in principle unrelated, quantities?
}

Trying to solve the strong CP problem has motivated a lot of work and attractive speculations
\cite{Peccei:2006as}. Here I focus on the proposal of Peccei and Quinn
\cite{Peccei:1977hh,Peccei:1977ur} which has a relevant phenomenological consequence for this work:
the introduction of a pseudo Goldstone boson, the axion \cite{Weinberg:1977ma,Wilczek:1977pj}.

Let us first assume a new axial symmetry $U(1)_\mathrm{PQ}$ having a color anomaly ${\cal N}^c$. If
the symmetry is spontaneously broken at a very high energy $f_\mathrm{PQ}$ (it has to be like this
since it is not apparent at low energies) then the low energy effective lagrangian would include a
massless Goldstone boson, $\xi$. Because of the anomaly, $\xi$ develops an interaction
term\footnote{We can read it in eq. \eqref{Nambu-Gold-theo:current-field-correlator} where we see
how Goldstone bosons couple their relative currents. This mechanism is the same that provides
$\pi^0$ an interaction $\f{\pi^0}{F_\pi}\f{\alpha}{2\pi} F^{\mu\nu}\widetilde F_{\mu\nu}$
responsible for the anomalous $\pi^0\go \gamma\gamma$ decay. For further details see
\cite{Weinberg:1996kr}.}
\be -\f{\alpha_s {\cal N}^c}{8\pi}\f{\xi}{f_\mathrm{PQ}} G_a^{\mu\nu}\widetilde G_{a\mu\nu}
\label{a.gluon.gluon-coupling} \ \ . \ee

In the rest of the low energy effective theory $\xi$ enters only through its derivatives as a
Goldstone boson should. Now, provided that the rest of the theory preserves CP\footnote{CP
violation coming from the electroweak sector apart from the quark mass matrix is just a small
perturbation to this situation.} except the $\theta$-term \eqref{thetaterm} then the effective
potential is even in $\overline \theta +{\cal N}^c\xi/f_\mathrm{PQ}$, having a minimum at
$\overline \theta + {\cal N}^c\xi/f_\mathrm{PQ}=0$ which conserves CP. The consequence is that the
$\xi$ field develops an expectation value $\langle \xi \rangle$ such that $\overline\theta$ is
canceled. We can say that the Peccei and Quinn mechanism relaxes $\overline\theta$
\textit{dynamically} to zero solving the strong CP problem.

Note that the PQ symmetry is almost completely similar to the QCD $U(1)_A$. The subtle point that
makes the difference is that now $f_\mathrm{PQ}$ is a scale much higher than $\Lambda_\mathrm{QCD}$
and therefore the instanton effect can be considered \textit{now} a small explicit breaking of
$U_\mathrm{PQ}(1)$ while it was a $0(1)$ breaking of $U_A(1)$. It is because of this reason that
the axion exist and the GB of the $U_A(1)$ does not. Due to this small explict breaking, however,
the axion acquires a small mass, and hence it is rather a pseudo-Goldstone boson.

Interestingly enough, many of the relevant properties of the axion are independent of the exact
realization of the PQ symmetry. For example the $\xi$ field mixes with the neutral pion $\pi^0$
resulting in two massive eigenstates\footnote{Corrections due to the other $0^-$ mesons are small
and irrelevant for this discussion.}, one very close to $\pi^0$ and the other called \textit{axion}
whose mass can be easily obtained to be
\be m_a = m_{\pi^0} {\cal N}^c \f{\sqrt{m_u m_d}}{m_u+m_d}\f{F_\pi}{f_\mathrm{PQ}} \simeq 1.3 \
\mathrm{meV} \ {\cal N}^c \f{10^9\ \mathrm{GeV}}{f_\mathrm{PQ}}  \ .\ee

Note also that being $U(1)_\mathrm{PQ}$ an axial symmetry, the axion must be a pseudoscalar
spin-zero boson.

Also, the axion field always appears in the lagrangian suppressed by the Peccei-Quinn scale,
$f_\mathrm{PQ}$. The original Peccei and Quinn model set $f_\mathrm{PQ}$ close to the electroweak
scale but such a low energy scale was soon excluded. The so-called invisible models
\cite{Dine:1981rt,Kim:1979if} propose a much higher value of $f_\mathrm{PQ}$ which makes the axion
 quite difficult (though not impossible) to detect. Remarkably, the Peccei-Quinn solution to the
strong CP problem works independently of the size of this parameter.

\begin{figure}\centering \vspace{-.3cm}
  \includegraphics[width=11cm]{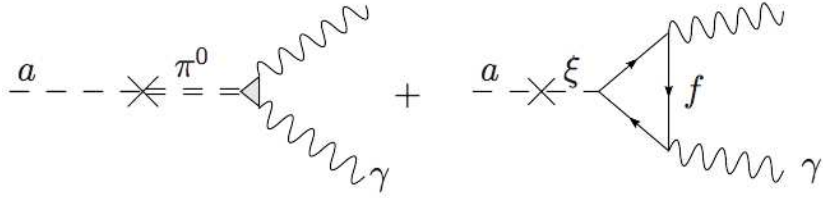}\vspace{-0.5cm}
  \caption{\small Feynman diagrams for the axion coupling to two photons through its components along the original
  $\pi^0$ and $\xi$. Here, $f$ stands for a sum over fermions carrying Peccei-Quinn charge. }\label{agammagamma.diagrams}
\end{figure}

Furthermore, it is quite generic that the axion develops a two photon coupling, denoted as
\be -\f{g_{a\gamma}}{4}\ a\  F^{\mu\nu}\widetilde F_{\mu\nu} \label{a-gamma-gamma.coupling}\ee
with $F^{\mu\nu}$ the field strength tensor of the electromagnetic field. The coupling
$g_{a\gamma}$ is the sum of two contributions, one coming from a plausible $U(1)_\mathrm{PQ}$
electromagnetic anomaly and the other coming from $a-\pi^0$ mixing (See Fig.
\ref{agammagamma.diagrams})
\be g_{a\gamma} = \f{\alpha}{2\pi f_\mathrm{PQ}}\left( {\cal N}^e -\f{2{\cal N}^c}{3}\f{ m_u +
4m_d}{m_u+m_d} \right) \ee
Where $\alpha$ is the fine structure constant and ${\cal N}^e$ is the electromagnetic anomaly of
$U(1)_\mathrm{PQ}$. A cancelation of this coupling is rather unlikely due again to the different
nature of the two contributions\footnote{An accidental tuning however is possible in some concrete
models. See \cite{Kaplan:1985dv} as a notorious example.}.
The axion has also couplings to quarks and leptons which have been extensively considered in the
literature but, as it will become clear, it is the electromagnetic coupling which has the richest
implications nowadays. The phenomenology of this coupling will be the main subject of Chapter
\ref{constraints}.

There are beautiful examples of Goldstone bosons other than pions and axions. For instance the
so-called \textit{Majorons} \cite{Chikashige:1980qk,Chikashige:1980ui} come from the hypothetical
spontaneous breaking of the ``Baryon minus Lepton" number $U(1)_{B-L}$ which was invoked to provide
a Majorana mass for neutrinos in a Higgs mechanism fashion. Other celebrated examples are
\textit{Familons} \cite{Wilczek:1982rv}, arising from the spontaneous breaking of family
symmetries, which are very much related to axions and the strong CP problem.

The general lesson to be learned from this section is that Goldstone bosons arise whenever there is
a global continuous symmetry which is spontaneously broken by the vacuum state. If the symmetry is
slightly explicitly broken the corresponding Goldstone excitations become massive. Interactions of
Goldstone bosons are always suppressed by the energy scale of the spontaneous breaking and involve
derivatives except in the case of the anomalous interactions \eqref{a.gluon.gluon-coupling} coming
from the axial anomalies of their corresponding symmetries. The parity and internal quantum numbers
of the Goldstone are those of the charge density $J^0$ associated to the broken symmetry.

Quite naturally these Goldstone Bosons can acquire a coupling to two photons through quantum
corrections like those giving the anomalies (See Fig. \ref{agammagamma.diagrams}). But there is one
important point to remark here. In what follows I will be considering the \textit{phenomenological
implications} of vertices like \eqref{a-gamma-gamma.coupling} not restricting myself to the axion
case. It happens that they are essentially similar (See Chapter \ref{constraints}) to those coming
from the corresponding parity-even gauge field structure
\be \f{g}{4}\ \phi\ F_{\mu\nu} F^{\mu\nu} \ \ , \label{phi-gamma-gamma.coupling}\ee
which I will consider altogether. There are however important theoretical differences between them.
$F_{\mu\nu} F^{\mu\nu}$ is CP and P conserving (See section \ref{Parity Arguments and parity-even
ALPs}) and thus \eqref{phi-gamma-gamma.coupling} respects these symmetries for a CP and P even
field $\phi$. This Goldstone mode is to be originated from the spontaneous breaking of a
vector-like symmetry (rather than axial). We know that this is not possible in theories with only
vector-like fermions (with non-chiral interactions) like QED or QCD because of the Vafa-Witten
theorem \cite{Vafa:1984xg}. However, this is certainly not a strong restriction; the standard model
itself is not subject to it, indeed.

However \eqref{phi-gamma-gamma.coupling} cannot have the same origin than
\eqref{a-gamma-gamma.coupling} since there is nothing like vector anomalies. Put it another way,
$F_{\mu\nu}F^{\mu\nu}$ is not a total derivative. As Goldstone bosons have derivative interactions,
\eqref{phi-gamma-gamma.coupling} \textit{cannot arise, even for a parity-even Goldstone boson}.

These theoretical considerations disfavor somehow parity-even \textit{massless} scalars coupled to
two photons. However (as we saw in the case of the axion) typically global symmetries are not exact
and the full effective description of particle physics at low energies will contain terms that
explicitly break the symmetry. We know that these terms \textit{would provide a Goldstone boson
mass} so in principle \textit{they can also naturally lead to interactions like}
\textit{\eqref{phi-gamma-gamma.coupling}}.


\newpage
\section{Paraphotons and millicharged particles}

Extensions of the standard model quite generally contain extra $U(1)$ gauge factors. These gauge
factors can be fundamental or sit within a non-abelian gauge group. Such gauge symmetries are
necessarily accompanied by their corresponding spin-1 massless gauge boson which is automatically a
new low energy candidate for new physics.

Gauge bosons can obtain mass from the spontaneous breaking of its related symmetry but purely
$U(1)$ symmetries allow also for a Stuckelberg mass-term
\cite{Stuckelberg:1937,Stuckelberg:1938a,Stuckelberg:1938b,Stuckelberg:1938c}. At the same time
there is at first glance no reason for their masses to be small, on the contrary, it is more
natural if they are of the order of the highest mass scale in the theory, which it is usually
assumed to be the Planck mass $m_{\mathrm{Pl}}\sim 10^{19}$ GeV. Of course, these arguments do not
rule out a small mass-scale, we know at least two completely ``unnatural" mass parameters in
nature, the electroweak scale ($\sim 100$ GeV), and the cosmological constant ($\sim 0.1$ meV).

There are many examples of such particles in the literature: mirror photons, $Z'$s, shadow photons,
U-bosons, etc. They differ in the interpretation that is given to the new $U(1)$ symmetry or the
way it links with the standard model sector. Here I will refer mainly to ``paraphotons", a term
first introduced by Okun \cite{Okun:1982xi}. To my knowledge he was the first one to consider these
particles, at least in the framework I will discuss here.

On a basic level, the impact of paraphoton models is to proportionate deviations from the
predictions of QED, at least at low energies. But also, paraphoton models include
\textit{naturally} millicharged particles, i.e. particles whose electrical charge is not a multiple
of the electron charge but a small number\footnote{See the discussion in \cite{Gies:2006ca} about
charge quantization.}. They are, together with paraphotons, necessary ingredients of our solution
to the PVLAS-astrophysics inconsistency. Interestingly enough, if the millicharged particles are
fermions, they can have naturally small masses in chiral theories like the standard model.

Quantum electrodynamics is usually said to be the most precise theory of all sciences so deviations
from their predictions are extremely restricted. This implies that paraphotons and particles
coupling to them have to be somehow isolated from our \textit{standard} particle physics. It has
become a fixed tendency to consider that paraphotons and other light paracharged particles coupled
to them inhabit a kind of \textit{hidden sector}. The connection between this and our sector would
be then driven by radiative corrections with very massive particles acting as \textit{mediators}
between the two worlds.

Consider a toy model in which we have quantum electrodynamics\footnote{For my purposes it is
sufficient to consider how paraphoton models alter the standard QED scenario. A wider scope would
consider the whole standard model \cite{Holdom:1990xp}.} in our sector (labeled $0$) and a similar
theory for the hidden sector (labeled $1$). The complete gauge symmetry is then $U(1)_0\times
U(1)_1$ and the gauge piece in the lagrangian is the sum of two similar parts
\bea {\cal L}_{QED} \equiv {\cal L}_{0} = -\f{1}{4}F_{0\mu\nu}F_0^{\mu\nu} +
e_0Q_0\overline{\psi}_0\slashed{A}_0\psi_0  \label{lagrangian-QED}\\
{\cal L}_{1} = -\f{1}{4}F_{1\mu\nu}F_1^{\mu\nu} + e_1Q_1\overline{\psi}_1\slashed{A}_1\psi_1
\nonumber\eea
plus a term that mixes kinetic terms of the gauge bosons
\be {\cal L}_{mix} = -\f{\epsilon}{2}F_{0\mu\nu}F_1^{\mu\nu} \label{mixing}\ee
where $F^{\mu\nu}_{0,1}=\partial^\mu A^\nu_{0,1}-\partial^\nu A^\mu_{0,1}$ are the field strength
tensors of the gauge bosons $A_{0,1}^\mu$, $e_{0,1}$ are the respective gauge couplings and
$Q_{0,1}$ are the $0,1$-charges of the different fermions $\psi_{0,1}$. We are interested in models
in which there is isolation of the 0 and 1 sectors so I put the tree level term \eqref{mixing} to
zero at zeroth order\footnote{There are two natural ways to obtain this: let $U(1)_1$ sit within a
non-abelian gauge group and the tracelessness of its generators will ensure $\epsilon=0$,
alternatively consider a discrete symmetry $A_0\leftrightarrow -A_1$ at high energy broken for
instance by fermion masses at lower scales. Both cases have been discussed in
\cite{Dienes:1996zr}.} and consider that the masses of mediators, fermions charged under both
$U(1)$'s ($\psi_{01}$'s) are very high so they can be removed from the low energy physics.

\begin{figure}   \centering
  \includegraphics[width=7cm]{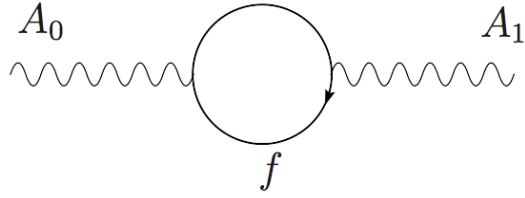}\vspace{-.6cm}
  \caption{\small One loop contribution to $\epsilon$. Only multicharged fermions $\psi_{01}$,
  acting as mediators, run in the loop.
  \label{0-1.mixing}}\vspace{-.4cm}
\end{figure}

Although decoupled, these fermions (of mass $m_j$) lead to a small correction\footnote{Other
radiative corrections produce the usual charge, mass and field strength renormalization which do
not interfere with my conclusions here.} to $\epsilon$ through the diagram depicted in Fig.
\ref{0-1.mixing}
\be \epsilon = \f{e_0 e_1}{6\pi^2} \sum_{\psi_{01}} Q_0^{\psi_{01}}Q_1^{\psi_{01}}\ \mathrm{Log}\ m_{\psi_{01}} \ \ \
.\ee
%
%
The dimensionful logarithm is understood to be regularized in any of the standard ways but a simple
choice of the charges of the heavy fermions can evade this complication. For instance, I can choose
two fermions, $\psi^a_{01}\in (1,1)$ and $\psi^b_{01}\in (-1,1)$, to get $\epsilon= \f{e_0
e_1}{6\pi^2} \mathrm{Log}\ \f{m^a}{m_b}$. If one does not assume a fine cancelation due to mass
degeneracy and use $e_1\sim e_0$, the obtained value is of order $10^{-3}$. However one expects
indeed higher degrees of symmetry and degeneracy as the energy scales involved (here the
$\psi_{01}$ masses) are increased and thus it is ``natural" to obtain much smaller values for
$\epsilon$ for example in supersymmetric models or in string theory \cite{Dienes:1996zr}.

We call $\{A_0,A_1\}$ the ``interaction basis" of the gauge bosons because the interactions with
their relative fermions are diagonal \textbf{by definition}. In opposition to this we can find a
``propagation basis" $\{\widetilde{A}_0,\widetilde{A}_1\}$ in which \eqref{mixing} is rotated away
and the gauge fields are eigenstates of propagation. In the course of diagonalizing the kinetic
gauge sector we are led with mixed charge assignments. Then, in our calculations, \textit{mixing
terms can be traded for off-diagonal charges}, a fact that I reviewed in
\cite{Masso:2006xg,Redondo:2006xx} in the context of the model proposed in \cite{Masso:2006gc}
solving the PVLAS-Astrophysics puzzle (Chapter \ref{Compatibility of CAST search with axion-like
interpretation of PVLAS results}). This can be shown in two ways, either by direct diagonalization
of the kinetic part of the lagrangian or by studying the $\psi_0 \psi_1$ elastic scattering
amplitude. Let me show both of them.

\subsection{Diagonalization of the kinetic matrix}

First write the kinetic part of the lagrangian in a matrix form, adding a general mass term that
will prove to be very convenient later on
\be {\cal L}_{kinetic} = -\f{1}{4}F^T{\cal M}_FF + \f{1}{2}A^T{\cal M}_A A \ . \ee
Where $F,A$ are vectors with $0,1$ components and Lorentz indices have been suppressed. I choose
the mass matrix to be diagonal in the interaction basis since it is more likely that this mass is
gained through interactions, then
\be {\cal M}_F = \left( \begin{array}{cc}
                                            1 & \epsilon  \\
                                            \epsilon & 1
 \end{array}\right)\ \  ;\ \   {\cal M}_A = \left( \begin{array}{cc}
                                           m_0^2 & 0 \\
                                           0   & m_1^2
 \end{array}\right) \ . \ee
To diagonalize ${\cal M}_F$ and express the result in canonical form we perform a rotation $R$ of
$45^\mathrm{o}$ followed by a contraction $C$ ($[C R]{\cal M}_F[CR]^T =1$). In the resulting basis,
${\cal M}_A$ is not diagonal but it is symmetric so a further rotation $R'$ is able to diagonalize
it also ($[R' C R] {\cal M}_A [R'CR]^T =$ Diag$\{\widetilde m_0^2,\widetilde m_1^2\}$). The matrix
transforming from the ``interaction" to the ``propagation" basis is therefore $U=R'CR$ and takes a
very simple form when we consider the interesting asymptotic case $\epsilon\ll \delta =
|m_0^2-m_1^2|/(m_0^2+m_1^2)$
\be U(\epsilon\ll\delta) = \left( \begin{array}{cc}
                                            1 & \epsilon\f{m_1^2}{m_0^2-m_1^2}  \\
                                            \epsilon\f{m_0^2}{m_1^2-m_0^2} & 1
 \end{array}\right) \label{U-epsilon<delta} \ee
The fermion interaction in this basis acquires the promised non-diagonal charges. For instance, we
get a $\epsilon$-sized electric charge for the paracharged fermion
\be e_1 Q_1\overline{\psi}_1\slashed{A_1}\psi_1 \rightarrow e_1
Q_1\overline{\psi}_1(\slashed{\widetilde A_1}+ \epsilon\f{m_0^2}{m_1^2-m_0^2}\slashed{\widetilde
A_0})\psi_1 \ \ , \label{parafermion-charge} \ee
and a small paracharge to standard model charged particles, for instance electrons $\psi_0^e$, with
$Q^e_0=-1$, implying
\be e_0 Q_0\overline{\psi^e}\slashed{A_0}\psi^e \rightarrow e_0 Q_0\overline{\psi^e}(\slashed{\widetilde A_0}+
\epsilon\f{m_1^2}{m_0^2-m_1^2}\slashed{\widetilde A_1})\psi^e \ \ \label{electron_paracharge} . \ee

The massless case $m_0=m_1=0$ is special. The freedom to choose the matrix $R'$ becomes an
arbitrariness of the definition of the propagation basis and thus of the charge assignments. It is
conventional then to \textit{define} the propagation photon $\widetilde A_0$ along the direction of
the interacting photon so standard model charged particles, $\psi_0$'s, do not interact with the
resulting paraphoton $\widetilde A_1$ but parafermions $\psi_1$ acquire an $\epsilon$-sized
electric charge. In this case we have
\be U(m_0=m_1=0) = \left( \begin{array}{cc}
                                            1 & 0  \\
                                            -\epsilon & 1
 \end{array}\right) \ \ \ . \ee
However, this is nothing but a definition without physical meaning. We could also have chosen a
situation in which standard model particles have paracharges, or even a mixed situation. I have
found that one can get the most general result from \eqref{U-epsilon<delta} performing the limit
$m_0=r m_1 \rightarrow 0$ ending up with
\be U(m_{0,1}=0) = \left( \begin{array}{cc}
                                            1 &  \epsilon\f{1}{r-1} \\
                                            \epsilon\f{r}{1-r} & 1
 \end{array}\right) \label{U-epsilon<delta-m=0} \ee
for arbitrary $r$ as long as $r\neq 1$ to ensure that the off-diagonal elements of \eqref{U-epsilon<delta} are always
small.

It is interesting to note that one cannot get this result by putting $m_0=m_1$ into the corresponding $U(\epsilon\gg
\delta)$ and letting $m_0\rightarrow 0$ . Although $m_0=m_1$ seem to imply also a rotational freedom, a small
$\epsilon$-sized mass difference arises in the propagation basis $\widetilde m_0 \neq \widetilde m_1$ breaking the
rotational invariance (This mass difference is proportional to $m_0=m_1$ so this does not spoil the freedom in the
massless case.).

In general $\widetilde e_0,\widetilde e_1, \widetilde m_0, \widetilde m_1$, etc. will differ from the interaction
values in small $\epsilon$-sized quantities.

\subsection{$f_0f_1$ scattering amplitude}

The charge assignments of the last subsection can also be obtained in a very simple calculation, a
procedure that turns out to be much simpler in models involving several paraphoton fields (I will
need two of them for my purposes). Consider the elastic scattering amplitude of a $0$-fermion with
a $1-$fermion, by using the mixing term in \eqref{mixing} as an interaction (It can be done
provided $\epsilon$ is small)
\be {\cal A}=(e_0Q_0 j^0_\mu) \f{i}{q^2-m_0^2} (\epsilon i q^2) \f{i}{q^2-m_1^2 }(e_1Q_1j^{1\mu})
\label{efef-mass-amplitude} \ee

($q$ is the 4-momentum carried by the bosons, $j_\mu^0,j_\mu^1$ the currents of $0,1$ particles and
I have used $q^\mu j_\mu^{0,1}=0$) which we can decompose using
\be
\f{-iq^2}{(q^2-m_0^2)(q^2-m_1^2)}=\f{m_1^2}{m_0^2-m_1^2}\f{i}{q^2-m_1^2}+\f{m_0^2}{m_1^2-m_0^2}\f{i}{q^2-m_0^2}
\label{propagator_division} \ee

This amplitude is equivalent to the sum of two single boson exchange diagrams as shown in
Fig.\ref{efef-mass}, which require that we assign an $\epsilon$-charge to $\psi_1$ and a
$\epsilon$-paracharge to $\psi_0$:
\begin{figure}[t]
\begin{centering}
  \includegraphics[width=15cm]{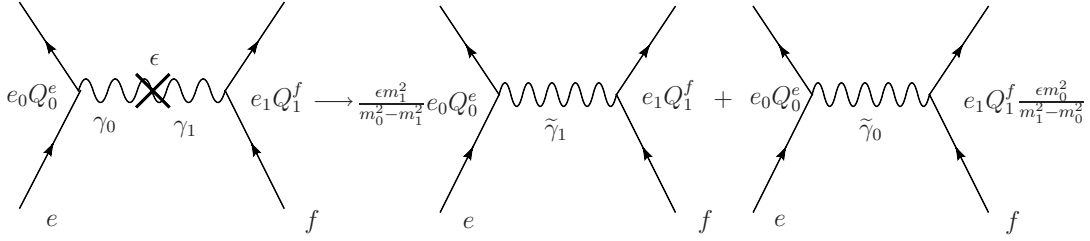}
  \caption{\small Mixing between massive photons is equivalent to non-diagonal $\epsilon$-charges for the fermions.
  \label{efef-mass}}
  \end{centering}
\end{figure}
\be Q_0^{\psi_1}=\epsilon Q_1^{\psi_0} \f{m_0^2}{m_1^2-m_0^2}\f{e_1}{e_0} \hspace{1cm}
Q_1^{\psi_0}=\epsilon Q_0^{\psi_1} \f{m_1^2}{m_0^2-m_1^2}\f{e_0}{e_1}  \label{efef-charges}\ee
We naturally associate these intermediate bosons with the propagation states we calculated before,
and we find (always at first order in the small quantity $\epsilon$) from
\eqref{lagrangian-QED},\eqref{U-epsilon<delta} and \eqref{efef-charges} that, as expected, both
approaches give identical charge assignments.

The diagram studied, however, does not give us information about the values of the order$-\epsilon$
corrections to $m_0,m_1,e_0,e_1$. To calculate them we shall study $e^-e^-$ or $f^-f^-$ elastic
scattering. In that case, beyond the $0^{th}$ order diagram in which the two electrons ($f$s)
interchange a $\gamma_0$($\gamma_1$) there is another contribution in which the
$\gamma_0$($\gamma_1$) intermediate state turns for a while into a $\gamma_1$($\gamma_0$) to
reconvert again into the original ``flavor". These diagrams provide the $o(\epsilon^2)$ corrections
to the masses and gauge couplings.

If $m_1\simeq m_0$ the expansion \eqref{propagator_division} would break down and this method probes to be not very
useful. However, again the massless case is special. Use again the limit $m_0=r m_1\rightarrow 0$ with $|r-1|>
\epsilon$ in \eqref{propagator_division} or \eqref{efef-charges} to get the same assignments
\eqref{U-epsilon<delta-m=0} we found within the previous diagonalization procedure.

\subsection{Some remarks} \label{Some remarks subsection}

First note that only particles charged under $U(1)_0$ will receive a $\epsilon$-sized charge of
type $1$ and viceversa. Neutral particles as the neutrino will continue being completely neutral if
photons mix with a new gauge boson from a given sector. Notably this holds for composite ``states"
as for instance atoms. Also, notice from \eqref{electron_paracharge} that for instance, electrons
will gain an opposite paracharge than protons because their charge is opposite.

\begin{figure}\centering\vspace{-.1cm}
  \includegraphics[width=9cm]{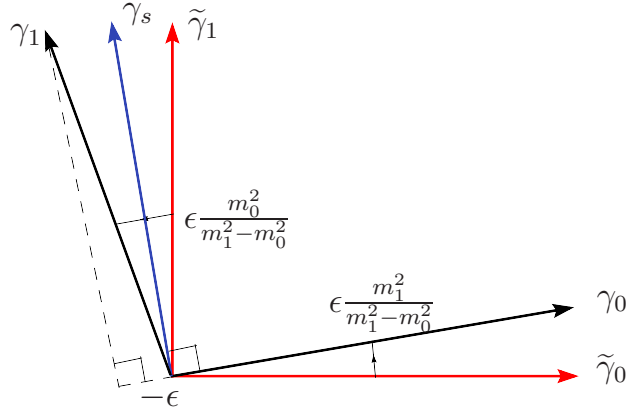}\vspace{-.3cm}
  \caption{\small The two basis of the $U(0)_0\times U(1)_1$ theory. The interaction basis is labeled $\gamma_0,\gamma_1$
  (and displayed in black) and the propagation basis $\widetilde \gamma_0,\widetilde\gamma_1$ (displayed in red).
  The first one is non orthogonal because of the small mixing of size $-\epsilon$.
  The two angles are assumed to be small (but magnified in the picture) and they also differ in the
  small quantity $\epsilon$. The sterile state $\gamma_s$ with respect to the $U(1)_0$ interactions of the standard
  model is displayed in blue.}\label{U-picture}
\end{figure}

Second, note that the matrix \eqref{U-epsilon<delta} is not unitary, reflecting the fact that
``interaction" states are not orthogonal, since they have a kinetic mixing term. Such a
complication does not make difficult the calculation of the $\gamma_0-\gamma_1$ oscillation
probability. If we prepare $|\gamma_{0,1}\rangle$ states using \eqref{U-epsilon<delta} their
temporal evolution will be easily expressed in the propagation basis
\bea |\gamma_0 (t)\rangle &=& |\widetilde \gamma_0\rangle e^{iE_0t} +
\epsilon\f{m_1^2}{m_0^2-m_1^2}|\widetilde
\gamma_1\rangle e^{iE_1t} \\
|\gamma_1(t)\rangle  &=& |\widetilde \gamma_1\rangle e^{iE_1t} +
\epsilon\f{m_0^2}{m_1^2-m_0^2}|\widetilde \gamma_0\rangle e^{iE_0t} \eea
where $E_i$ is the energy of the state $|\widetilde \gamma_i \rangle$ and I have not taken into
account the normalization, irrelevant for the following computation since it just provides
$O(\epsilon^3)$ terms. We find
\bea P(\gamma_0\rightarrow \gamma_1) &=& |\langle \gamma_1|\gamma_0 (t)\rangle |^2 = \\
&=& \epsilon^2\pa{1+\f{4m_0^2m_1^2}{(m_0^2-m_1^2)^2}\sin^2\pa{\f{\Delta E}{2}t}} + O(\epsilon^2)
\eea
Where $\Delta E = E_0-E_1 = (m_0^2-m_1^2)/2E$ for the relativistic limit $E>> m_{0,1}$. We see
clearly that in addition to the typical oscillation sinus, a $t=0$ contribution has arisen, due to
the fact that $\gamma_0$ has a small component along $\gamma_1$ (another way of looking at the
mixing term) of size $-\epsilon$. This is made clear in Fig. \ref{U-picture}.

Another interesting difference with neutrino oscillations is that in the case $m_0=0$, there are no
oscillations, even if $m_0^2=0\neq m_1^2$. This situation corresponds to one where $\cal U_{21}=0$
and then $\gamma_1=\widetilde \gamma_1$ so clearly no oscillation could happen since one of the
``interaction" states is a ``propagating" one, namely $\langle \gamma_1|\gamma_0 \rangle
(t)=-\epsilon \langle \gamma_1|\gamma_1\rangle e^{iE_1t}$.

It is common to consider experiments of disappearance of photons ($\gamma_0$), as for instance
light-shinning-through-walls experiments (See subsection \ref{Laser experiments subsection}). It
will prove useful then to define the sterile states $\gamma_{s0},\gamma_{s1}$, as the states
orthogonal to the interaction ones $\gamma_0,\gamma_1$. The probability of $\gamma_0-\gamma_{s0}$
oscillations is therefore derived in complete analogy to the neutrino oscillations case to be
\be P(\gamma_0\rightarrow \gamma_{s0})= |\langle \gamma_{s0}|\gamma_0 \rangle (t)|^2 = 4\pa{\f{\epsilon
m_1^2}{m_0^2-m_1^2}}^2\sin^2\pa{\f{\Delta E}{2}t}  \label{photon-sparaphoton oscillations}  \ee
These sterile states were called "paraphotons" by Okun \cite{Okun:1982xi} although here we have
used the name for a general abelian gauge boson. It is mandatory to compare this construction with
the original work of Okun \cite{Okun:1982xi}. There are two main differences. First, he did not
consider a current of type $1$. Second, he did not wonder where do the mass terms come from. In my
work I need particles charged under the hidden sector gauge group $U(1)_1$, but most importantly, I
assume that the mass terms come from interactions within the two gauge groups $U(1)_0,U(1)_1$
separately, i.e. the interaction states have a diagonal mass matrix.

\clearemptydoublepage


%% file: chapter3.tex
\chapter{The PVLAS experiment \label{PVLAS}}

\section{Introduction}
A photon can convert into a neutral boson $\phi$ in an electromagnetic field in the presence of a
$\phi\gamma\gamma$ interaction. This is the so-called Primakoff conversion first proposed by Henry
Primakoff to measure precisely\footnote{In this case they use the Coulomb potential generated by an
atom as the electromagnetic field.} the $\pi^0\go\gamma\gamma$ decay constant
\cite{Primakoff:1951ww}.

Moreover, if the $\phi$ particle is light enough, this vertex would lead to $\gamma-\phi$ ``mixing"
where a coherent superposition of the two particles arises, in complete analogy to the $K^0$-meson
or neutrino systems. Notably, in this case and due to the presence of the magnetic field, the
angular momentum needs not to be conserved, allowing the mixing of photons with particles of higher
(but integer) spin.

\begin{figure}[t]\centering \vspace{-.1cm}
  \includegraphics[width=17cm]{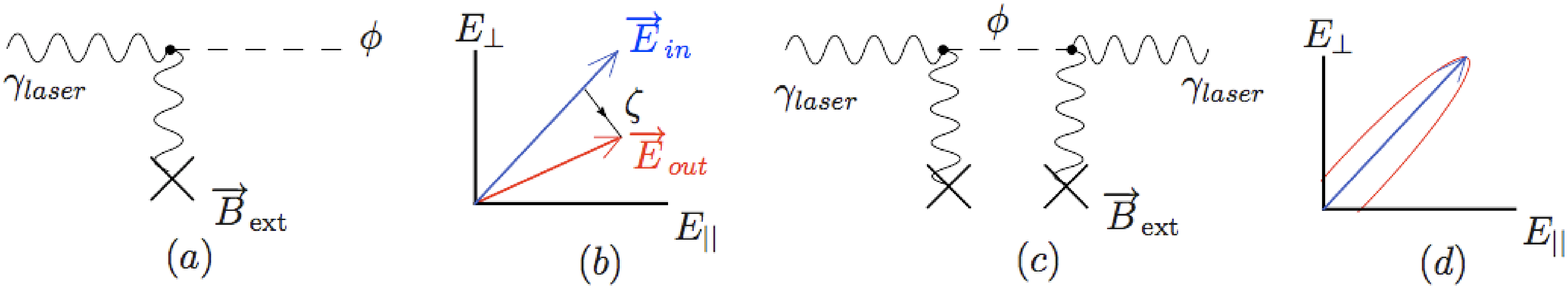}
  \caption{\small A laser beam traveling through a transverse magnetic field develops a small rotation
  of the polarization plane and a small ellipticity in the presence of a new boson $\phi$ coupled
  to two photons if this coupling is different for both photon polarizations.
  The real production of $\phi$ depicted in $(a)$ accounts for the dichroism, as shown in $(b)$
  where we see the plane transverse to the propagation and the two laser polarizations
  are labeled $E_{||}$ and $E_{\perp}$ (Electric field of the laser beam parallel and perpendicular
  to the external magnetic field). The $\phi$'s reconverted into photons $(c)$ are phase-shifted
  and account for the ellipticity shown in $(d)$.}\label{rotation-ellipticity-ALP}
\end{figure}

This immediately suggests a possibility for looking for novel $\phi$ bosons coupled to light in a
classical optical setup. Consider Fig. \ref{rotation-ellipticity-ALP}.$a$, a photon from a laser
beam interacts with an external magnetic field and it is converted into a $\phi$ boson. If this
absorption is selective, i.e. is different for photon polarizations parallel ($||$) and
perpendicular ($\perp$) to the magnetic field direction, the \emph{vacuum permeated by magnetic
fields would act exactly as a dichroic medium} since both polarizations will be depleted at
different rates. The net effect on a polarized beam will be a rotation of the polarization plane
which can be measured. This can be seen in \ref{rotation-ellipticity-ALP}.$b$, where I depicted a
view of the plane perpendicular to the laser direction showing \emph{in} and \emph{out}
polarizations of the incoming laser. It is straightforward to see that the sign of the rotation
will be positive if the coupling favors $\gamma_{||}-\phi$ transitions and negative in the opposite
case. As we will see, the sign of the rotation will be used to determine the parity\footnote{More
precisely the parity structure of the two-photon part of the $\phi\gamma\gamma$ interaction.} of
$\phi$.

Maiani, Petronzio and Zavattini were the first to realize that the two photon coupling of low mass
bosons (in fact they considered only spin zero bosons) also leads to birefringence for light
propagating in strong magnetic fields \cite{Maiani:1986md}. The $\phi$ boson can be produced and,
after some distance, reconverted\footnote{This can happen even off-shell, i.e. with 4-momentum
squared different to $m^2_\phi$.} as shown in Fig \ref{rotation-ellipticity-ALP}.$c$ resulting in a
phase delay of the photon. Again, if $\gamma_{||}-\phi-\gamma_{||}$ transitions are more likely
than $\gamma_\perp-\phi-\gamma_\perp$ or viceversa there would appear a phase difference between
the two polarizations and then the outgoing beam would show a counter-clockwise or clockwise
ellipticity (See Fig. \ref{rotation-ellipticity-ALP}.$d$). We can conclude then that the
\emph{vacuum permeated by magnetic fields would behave as if it were a dichroic and birefringent
medium}.

\begin{figure}\centering \vspace{-.3cm}
  \includegraphics[width=6cm]{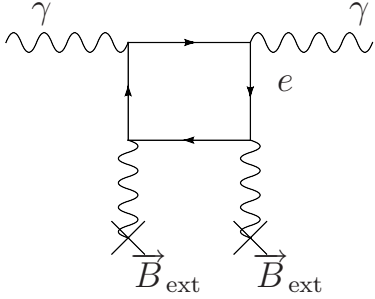}\vspace{-.4cm}
  \caption{\small QED Feynman diagram for the vacuum birefringence of light in a magnetic field.
  There are other diagrams differing from this in permutations of all photon lines.}\label{EH}
\end{figure}

%
Indeed it was known long ago (See \cite{Adler:1971wn} and references therein) that pure QED effects already produce
birefringence in a transverse magnetic field in vacuum (a vacuum Cotton-Mouton effect). This is due to virtual electron
loops like the one depicted in Fig. \ref{EH} which lead to the following indices of refraction \cite{Adler:1971wn}
\be n_{||}= 1+\f{7}{2}\xi \hspace{1cm};\hspace{1cm} n_{\perp}=1+\f{4}{2}\xi  \label{QED-birefringence}\ee
\be \mathrm{with}\ \ \ \xi = \f{\alpha}{45\pi}\pa{\f{B_\mathrm{ext}}{B_c}}^2 \ \ \ \mathrm{and}\ \
\ B_c = m_e^2/e \ee
%
%
It was T. Erber \cite{Erber:1961} the first to propose the use of a polarized laser beam
propagating in a magnetic field to measure the vacuum QED effects. However, he was wrong thinking
that it leads to a rotation of the polarization plane, not to an ellipticity. Some years after
Iacopini and Zavattini \cite{Iacopini:1979ci} wrote an interesting experimental
proposal\footnote{Apparently unaware of Erber's idea.} to detect this ellipticity at CERN. Time
passed by and the paper of Maiani, Petronzio and Zavattini \cite{Maiani:1986md} came out supporting
the idea of such an experiment\footnote{Little time after, Raffelt and Stodolsky
\cite{Raffelt:1988rx} showed that this paper contained some overestimated predictions. They had
claimed that they would be able to measure the QED birefringence effect and values of
$g_{a\gamma}\sim 10^{-10}$ GeV$^{-1}$, competitive with the strong astrophysical bounds.} with the
possibility of detecting low mass bosons coupled to light (at a time in which axions and other
light bosons like arions \cite{Anselm:1981aw}, majorons \cite{Chikashige:1980qk} or familons
\cite{Wilczek:1982rv} were very fashionable).

Finally, a pioneer experiment was performed in the Brookhaven laboratory by the
so-called\footnote{From Brookhaven, Rochester, Fermilab and Trieste, homeplaces of the members of
the collaboration.} BRFT collaboration. They found no signal of this kind of $\phi\gamma\gamma$
interactions\footnote{This experiment, as we will see, was also sensitive to the existence of
paraphotons.} \cite{Semertzidis:1990qc,Cameron:1993mr} so they were able to establish exclusion
bounds on the ALP parameter space. Most of the collaboration moved then to the National Laboratory
of Legnaro, near Padova (Italy), where, under the leadership of E.~Zavattini, they started to build
the PVLAS experiment \cite{Bakalov:1994} as a natural continuation of the BRFT apparatus and
technique. Next I will comment on its set up to end up with their latest results.


\section{PVLAS experimental setup and recent results}
\begin{figure}\centering \vspace{-.6cm}
  \includegraphics[width=16cm]{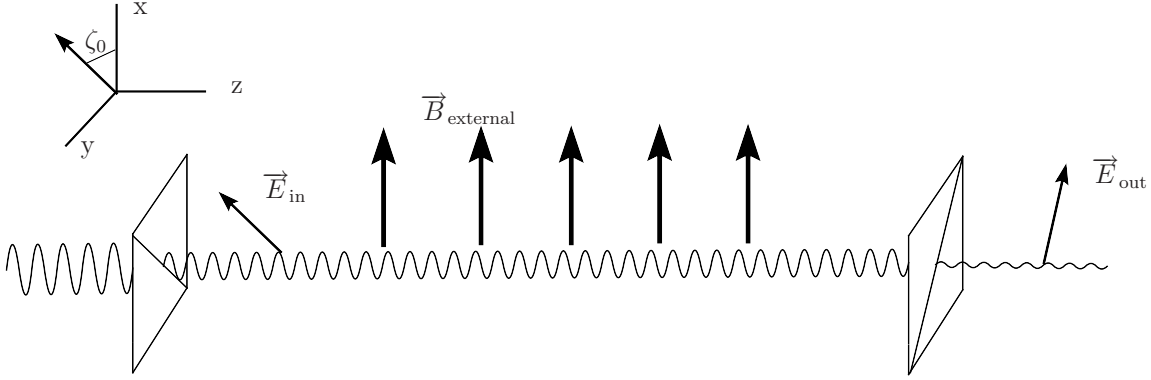}\vspace{-.7cm}
  \caption{\small Brief scheme of the PVLAS experiment essentials. A magnetic field between two crossed polarizers.
  }\label{PVLAS-basic-setup}
\end{figure}

The PVLAS experiment is as simple as beautiful. The basic idea consists in sending a laser beam
along a strong transverse magnetic $B_\mathrm{ext}$ field located between two crossed polarizers,
all in a cavity at high vacuum, see Fig. \ref{PVLAS-basic-setup}. It is clear that if light does
not interact with $B_\mathrm{ext}$, then no light would came out from the second polarizer.
Conversely if the magnetic field in vacuum has optical properties such as dichroism or
birefringence then the situation would change and some emerging light would ``enlighten" the
phenomenon.

As I mentioned this should be the case if one takes into account quantum fluctuations of the QED
vacuum structure, producing a net \textit{birefringence} but \textit{no dichroism}. Also, in the
case that a light boson with a two-photon coupling exists, the real and virtual production shown in
Fig. \ref{rotation-ellipticity-ALP} will lead to both a rotation and an ellipticity of the outgoing
beam.

In the PVLAS experiment the magnetic field intensity reaches $5.5$ T and the length of the magnet
is $1$ meter. With these parameters, the two perpendicular components of light emerging from the
magnetic field would have accumulated a QED-induced phase shift $ \sim 10^{-16}$ from
\eqref{QED-birefringence} implying an ellipticity\footnote{Ratio of the minor to the major axis of
the polarization ellipse shown in \ref{rotation-ellipticity-ALP}.d. See section
\ref{section_RotationEllipticity}.} of $\sim 10^{-16}$ which means that only a
ridiculous amount of light will traverse the two polarizers. Even using a MegaWatt laser (which
probably would destroy the polarizers), the emerging light would be far beyond the reach of any
present (and maybe future) photodetector.

For this reason, the PVLAS signal is enhanced in two different ways. First of all, the magnetic
field is located inside a pair of highly reflective mirrors (actually a Fabry-Perot cavity) which
increase the path of light through the magnet by a huge factor $N\sim 44000$. This might look quite
a cheap way of building a $44$ km length magnet, but it is not exactly the case as I will show
below. Second, the outgoing signal is enhanced by heterodyne combination with an ellipticity
modulator (called SOM for Stress Optical Modulator) before entering the detector. The principle of
heterodyne detection is very simple, two waves at frequency $\omega_1$ and $\omega_2$ traveling
through a non linear material will produce outgoing waves at frequencies $\omega_1\pm\omega_2$. The
amplitude of these waves depends linearly on both amplitudes of the incoming waves. This trick can
be used to amplify a very small amplitude by combining it with a larger one and look at any of the
interference waves. This procedure, however, demands that the small incoming wave to detect (with
our desired excess ellipticity) is beating at some frequency\footnote{The original setup in
\cite{Maiani:1986md} used a modulated Faraday cage. PVLAS is nowadays involved in the modifications
required to implement this other technique as a crosscheck of their results.} and in order to do
so, the PVLAS collaboration managed to make the magnet rotating at a small frequency\footnote{The
angle twisted in the photons-time-of-flight is a ridiculous $10^{-3}$ rad.} $\sim 0.5$ Hz.

\begin{figure}
  \centering \vspace{-.1cm}
  \includegraphics[width=15cm]{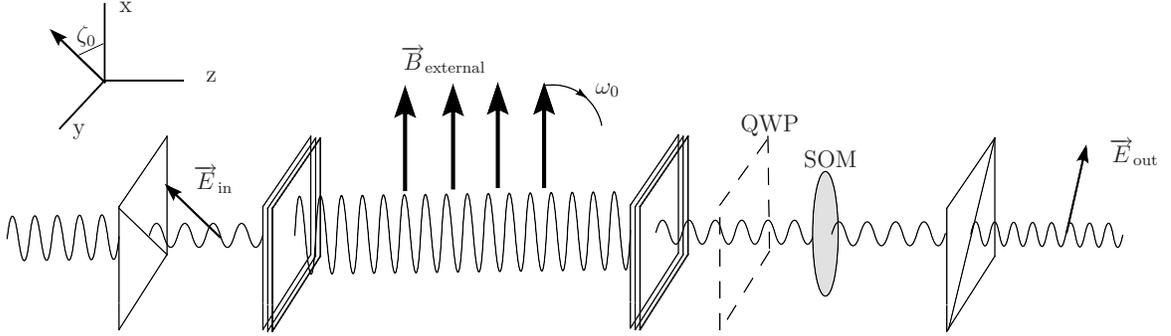}
  \caption{\small A closer look to the PVLAS experiment. The incoming light is linearly polarized and
  travels several times the interior of the Fabry-Perot cavity where a strong magnetic field is placed. The magnetic
  field is transverse to the propagation of the light and slowly rotates its direction at an angular frequency $\omega_0$.
  At the exit of the Fabry-Perot the light could have developed a small rotation and a small ellipticity
  due to mixing with ALPs or pure QED effects. A quarter wave plate can be used to convert
  rotations into ellipticities and viceversa. The SOM superposes a time dependent ellipticity to the incoming light. The
  small ellipticity (beating at $2\omega_0$) acquired at the magnet will beat with the SOM time-modulated ellipticity
  allowing heterodyne detection after the second crossed polarizer.
  The rotation entering the SOM will be left unchanged and because its smallness will pass undetected.
          }\label{PVLAS-complete-setup}
\end{figure}

The whole resulting setup is depicted in Fig. \ref{PVLAS-complete-setup} and is able to measure
ellipticities as small as $10^{-7}$ rad Hz$^{-\f{1}{2}}$. This setup can be adapted very easily to
measure rotations because a properly oriented quarter wave plate transforms small rotations into
ellipticities and viceversa \cite{Born:1980}, so if placed just before the SOM it will convert a
possible rotation generated by the magnetic field into an ellipticity which will be again enhanced
by heterodyne combination in the SOM and thus detected. Also, interchanging the slow and fast axis
of the quarter wave plate produces a sign flip in the outgoing ellipticity, providing a check of
the nature of the signal detected.

With this setup the PVLAS collaboration has recently reported \cite{Zavattini:2005tm} an excess
rotation of $3.9 \pm 0.5 \ \ 10^{-12}$ rad/pass for laser light of $1064$ nm traveling through a
$1$ m length, $5$ T, magnetic field. This signal has been reported after careful search of possible
systematic effects\footnote{For a review see \cite{Cantatore:IAS}} over several years. Also they
have reported in several conferences \cite{Cantatore:Patras,Gastaldi:QCD,Cantatore:IAS} positive
results for the search of birefringence, again at the $10^{-12}$ rad/pass level. This amazing
result exceeds by three/four orders of magnitude the QED expectations and nowadays \emph{remains
without a conclusive explanation within conventional physics}.

However, there are two non standard candidates that \textit{in principle} could account for the
PVLAS results within particle physics. The first one is a neutral boson weakly coupled to two
photons, one of the original motivations of PVLAS. Such a particle will be called from here on
axion-like particle (ALP). I devote next section to examine this possibility.

The second possibility has arisen very recently \cite{Gies:2006ca} and assumes the existence of very light millicharged
particles (MCP). Again, the reactions distinguish both photon polarizations and will lead to dichroism and
birefringence.

A very interesting paper has recently appeared comparing both ALP and MCP scenarios and it seems
that we will have to wait until more experimental data is released to decide which of them, if any,
is more likely to be the responsible for the PVLAS signal \cite{Ahlers:2006iz}.

Both explanations have many things in common however, the first one being that they are not
satisfactory at all because \textit{they are excluded by astrophysical arguments}, as I will show
in Chapter \ref{constraints}.

But moreover, they share a much more interesting feature, \emph{both scenarios can be made compatible with
astrophysical observations} within a concrete model I presented in \cite{Masso:2006gc} and constitutes the central work
of this thesis.

\newpage \section{Axionlike particle interpretation} \label{Axionlike particle interpretation-section}

The study of coherent $\gamma-\phi$ oscillations in presence of a long, \textit{static} homogeneous
magnetic field has been addressed by many authors
\cite{Maiani:1986md,Raffelt:1988rx,Das:2004qk,Gabrielli:2006im}. Here I follow the exposition of
Raffelt and Stodolsky \cite{Raffelt:1987im}.

I do not feel the necessity of including the effects of the QED vacuum since they are very small
compared with the PVLAS signal. But it is worth to recall that any birefringence will show up into
a rotation when the magnetic field rotates (as in the PVLAS setup)
\cite{Mendonca:2006pg,Adler:2006zs}. Nevertheless, it is quite intuitive (and has also been proved
rigorously in \cite{Adler:2006zs}) that these effects are negligible for the PVLAS setup\footnote{
The attempt made in \cite{Mendonca:2006pg} to explain the PVLAS signal in these terms is
mistaken.}. In what follows I neglect the rotation of the magnet.

Consider the following situation. A beam of laser light is shone against a region where there is a strong magnetic
field. I choose the beam direction as the z-axis and the magnetic field polarization to lie in the xz-plane. It is
convenient also to make the problem one-dimensional, in the sense that all fields we want to determine are functions of
time and $z$ only.
\be A^\mu(x^\nu) = A^\mu(t,z) \ \ \ ; \ \ \ \phi(x^\nu) = \phi(t,z) \ee
This is formally the case if the magnetic field would be infinite in the $x-y$ directions and the
laser is described as a plane wave. Effects from deviations of this simplistic picture are small.

Next, I consider that there is an interaction between two-photons and a new neutral boson, called
$\phi$, which is very light (we will see soon how light this ``light" means). The most discussed
case is that $\phi$ represents a spin-zero particle, but also one might consider that $\phi$ is a
spin-1 particle like a paraphoton or a spin-2 particle like the graviton.

In Chapter \ref{general concerns} we have found motivations for two different couplings associated
with opposite parity structure \eqref{a-gamma-gamma.coupling} and \eqref{phi-gamma-gamma.coupling}
which I rewrite here for completeness
\be \f{1}{4}g^+ F^{\mu\nu}F_{\mu\nu} \phi \hspace{1.5cm} ;  \hspace{1.5cm} \f{1}{4}g^- F_{\mu\nu}\widetilde F^{\mu\nu}
\phi \label{ALP-couplings} \ee
We know that parity is not exactly conserved by the weak interactions so there is no strong reason
a priori to think that it should be conserved by new additional interactions. This suggests not to
either assign a definite parity to $\phi$ nor to remove one of these two interactions. However, if
$\phi$ is a true Goldstone boson then it could easily happen that it has a definite parity.
Explicit breaking terms can modify this statement, but corrections should be small and then one of
the $g$'s will be much smaller than the other. Moreover, unless $g^+$ and $g^-$ are
fine-tuned\footnote{Maximal parity violation in a hidden sector like this might not propagate to
the SM sector and thus it should not be excluded at first glance.} these considerations will not
change our main conclusions.  This holds also for the case in which there is more than one neutral
boson $\phi$.

It is then enough to consider only one of the two interactions. Following criteria of theoretical
and historical importance I choose the parity-odd interaction which arises for axions and other
celebrated pseudo Goldstone bosons and drop the $-$ sign of the coupling for simplicity.

The lagrangian for the $\phi-\gamma$ system is\footnote{In all this thesis I am using the metric
$\eta_{\mu\nu}=\mathrm{Diag}\{1,-1,-1,-1\}$.}
\be {\cal L} = -\f{1}{4}F^{\mu\nu}F_{\mu\nu} + J_\mu A^\mu +\f{1}{2}\partial_\mu \phi \partial^\mu\phi
-\f{1}{2}m^2\phi^2 +\f{1}{4}g F^{\mu\nu}\widetilde F_{\mu\nu} \phi \ee
whose equations of motion are
\bea  \partial_\mu F^{\mu\nu} &=& J^\nu + g\widetilde F^{\mu\nu} \partial_\mu \phi \\
 (\partial^2 +m_\phi^2)\phi &=& -\f{1}{4} g F_{\mu\nu}\widetilde F^{\mu\nu}  \eea
where I have included the electromagnetic current $J^\nu$. We can divide the field strength into
two parts corresponding to the external magnetic field (with the definition\footnote{This is not a
``local" statement. A far away $J^\nu$ generates the magnetic field at the region where the laser
is shinning by contour conditions.} $\partial_\mu F_{ext}^{\mu\nu}=J^\nu$) and the laser beam,
\be F^{\mu\nu} = F^{\mu\nu}_{ext} + F_\gamma^{\mu\nu} \ \ . \ee
If we consider that the external magnetic field is much more intense that the laser beam we can
neglect terms involving two fields, $\phi F_\gamma$ and $F^{\mu\nu}_\gamma F_{\gamma\mu\nu}$ to
find a \emph{linear} system of two coupled second order differential equations
\bea \partial_\mu F_\gamma^{\mu\nu} &=& g\widetilde F_{ext}^{\mu\nu} \partial_\mu \phi \\
(\partial^2 +m_\phi^2)\phi &=& -\f{1}{2}g  F_{\gamma\mu\nu}\widetilde F_{ext}^{\mu\nu} \eea
Note that this is the case in the PVLAS experiment, since $|\overrightarrow{B}_\mathrm{ext}|\sim
5T\sim 1000\ \mathrm{eV}^2$ and the laser power is $\sim 1$ mWatt in a $2$ mm diameter gaussian
beam which means $|\overrightarrow{E}_\gamma|=|\overrightarrow{B}_\gamma|\sim 0.1\ \mathrm{eV}^2$
(taking into account the Fabry-Perot cavity which enhances $|\overrightarrow{B}|$ by a factor $\sim
10^4$ because the coherence of the reflections.) Then the wave equation for the time-varying
potential $\overrightarrow{A}$ (in Coulomb gauge $\overrightarrow{\nabla} \cdot
\overrightarrow{A}=0$) is
\bea  \partial^2 \overrightarrow{A} &=& g \overrightarrow{B}_T\, \partial_t \phi \label{ALP-phi-mixing-equations} \\
 (\partial^2 + m_\phi^2)\phi &=& -g \overrightarrow{B}_T\cdot \partial_t \overrightarrow{A} \nonumber \eea
Where $B_T$ is the transverse part of the external magnetic field ($\overrightarrow{B}_T\cdot
\widehat{z}=0$). As the coefficients of the unknown fields $A,\phi$ are time independent it is
natural to guess a solution of the form
\bea \overrightarrow{A}(t,z) &=& e^{i\omega t} \overrightarrow{A}(z) \\
\phi(t,z) &=& e^{i\omega t} \phi(z) \eea
which implies
\bea &&(\omega^2 +\partial^2_z)A_\perp(z)=0  \label{a-gamma mixing equations(Aperp)}\\
   &&\left[\partial^2_z + \left( \begin{array}{cc}
                                            \omega^2 & igB_T\omega \\
                                       -igB_T\omega & \omega^2-m_\phi^2 \end{array}
 \right)\right] \left(  \begin{array}{c}  A_{||} \\ \phi \end{array}\right)(z)=0
 \label{a-gamma mixing equations}\eea
Recall that the electric field will be $\overrightarrow{E}=\partial_t \overrightarrow{A}=i\omega
\overrightarrow{A}$. For simplicity in further expressions I will evade imaginary quantities
redefining $\overrightarrow{A}\rightarrow i\overrightarrow{A}$.

The solution of the first equation is trivially a plane wave with momentum $k_\perp=\omega$. In the coupled set we can
made a further approximation if the photons and $\phi$ are very relativistic, in this case
$\omega^2+\partial_z^2=(\omega+i\partial_z)(\omega-i\partial_z)\sim 2\omega(\omega-i\partial_z)$. The resulting
equation is therefore of first order and easier to solve
\be \left[\omega -i\partial_z + \left( \begin{array}{cc}
                                            0 & \f{gB_T}{2} \\
                                       \f{gB_T}{2} & -\f{m_\phi^2}{2\omega} \end{array}
 \right)\right] \left(  \begin{array}{c}  A_{||} \\ \phi \end{array}\right)(z)=0 \label{mixing-equation for agamma_par}
 \ \ .
\ee
The solution reads
\be
 \left(  \begin{array}{c}  A_{||} \\ \phi
\end{array}\right)(z) = \mathrm{Exp}\left(-i\omega z \right)\mathrm{Exp}\left(-i{\cal M}z \right)\left(  \begin{array}{c}
A_{||} \\ \phi
\end{array}\right)(0)\ee
being $\cal M$ the $2\times 2$ mixing matrix in \eqref{mixing-equation for agamma_par}. To get a useful expression for
$A_{||}(z)$ we must diagonalize $\cal M$ with a rotation $R$ of an angle $\theta$ defined by
\be R= \left( \begin{array}{cc}
                             \cos\theta & -\sin\theta \\
                             \sin\theta & \cos\theta \end{array} \right) \hspace{1cm} ; \hspace{1cm}
\f{1}{2}\tan{2\theta} = -\f{gB_T\omega}{m_\phi^2} \ee
The resulting states $(A_1,A_2) = (A_{||},\phi)R^T$ are eigenstates of propagation because for them the evolution is
decoupled
\bea
 \left(  \begin{array}{c}  A_1 \\ A_2
\end{array}\right)(z) = \mathrm{Exp}\left(-i\omega z\right)\mathrm{Exp}\left(-i{\cal D}z \right)\left(  \begin{array}{c}
A_1 \\ A_2
\end{array}\right)(0) \\
\mathrm{with} \ \ \ \ \ \ \ {\cal D}=\left( \begin{array}{cc}
                                           \f{-m^2_\phi + \sqrt{m^4_\phi+(2gB_T\omega)^2}}{4\omega}  & 0  \\
                                       0 & \f{-m^2_\phi - \sqrt{m^4_\phi+(2gB_T\omega)^2}}{4\omega}  \end{array}
 \right)
 \eea
These solutions carry momentum $p_1=\omega+\cal D_{11}$ and $p_2=\omega+\cal D_{22}$ and invariant
masses $m_1^2=\omega^2-p_1^2$ and $m_2^2=\omega^2-p_2^2$. When we specify the initial conditions
for \linebreak $A_{||}(z=0)=1, \phi(z=0)=0$ we find
\bea A_{||}(z)= & [R^T \mathrm{Exp}\left(i{\cal D}z\right)R]_{11} = &e^{-i\omega z}\left(e^{-i{\cal
D}_{11}z}\cos^2\theta+e^{-i{\cal D}_{22}z}\sin^2\theta\right)
 \label{ALP-phi-mixing-solutions1}\\  \phi(z)= & [R^T \mathrm{Exp}\left(i{\cal
D}z\right)R]_{21} = & e^{-i\omega z}\left(e^{-i{\cal D}_{11}z}-e^{-i{\cal D}_{22}z}\right)\sin\theta\cos\theta
\label{ALP-phi-mixing-solutions2}\eea
In order to get formulas for the probability of appearance of $\phi$ particles and disappearance of photons, we
interpret these $A_{||}(t,z),\phi(t,z)$ fields as probability wave functions. Then we find the probability of
$\gamma_{||}-\phi$ oscillations at a distance $z$ and time $t$
\bea \mathrm{P}(\gamma_{||}\rightarrow \phi) = & |\phi(z)^*A_{||}(0)|^2 & =\sin^2 2\theta\
\sin^2\left({\f{1}{2}}\Delta_{osc}z\right) \label{A_par_solution} \\
\mathrm{P}(\gamma_{||}\rightarrow \gamma_{||}) = & |A_{||}(z)^*A_{||}(0)|^2 & = 1-\mathrm{P}(\gamma_{||}\rightarrow
\phi) \label{phi_solution} \eea
with
\be \Delta_{osc}={\cal D}_{11}-{\cal D}_{22}=\f{\sqrt{m_\phi^4+(2gB_T\omega)^2}}{2\omega} \label{Delta_osc}\ee

\begin{figure}\centering
\vspace{-.5cm}
  \includegraphics[width=8cm]{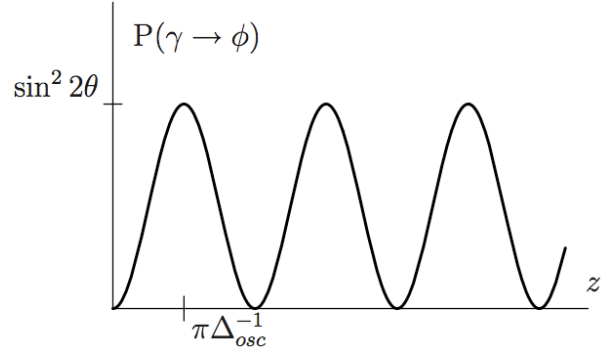} \vspace{-.5cm}
  \caption{\small Oscillation probability of photons into $\phi$ bosons as function of the distance $z$.}
  l\label{gamma-phi-oscillations}\vspace{-.5cm}
\end{figure}

The typical oscillation pattern arising from these formulae is shown in Fig.
\ref{gamma-phi-oscillations}. In order to maximize the signal there is an optimum distance at which
to place the detector at $z=n\pi/\Delta_{osc}$, with $n$ an integer.

There are two very interesting limits of these expressions, the weak and the maximal mixing limits.

\textit{Weak mixing limit, $\theta \sim 0$} .- \\
In this case $\tan 2\theta\simeq 2\theta \sim -2gB_T\omega/m_\phi^2\lll 1$ and the
corresponding eigenstates of propagation are very close to the original $A_{||},\phi$
\bea A_1 = A_{||}-\f{gB_T\omega}{m_\phi^2}\phi  \sim A_{||}  \\
     A_2 = \phi + \f{gB_T\omega}{m_\phi^2}A_{||}\sim \phi    \eea
We also can expand the square root of the diagonal elements of $\cal D$ to find that at first order
\bea p_1 = \omega + \f{g^2B^2_T\omega}{2m_\phi^2} +0(.)\ \ \ &\mathrm{and}& \ \ \ m_1^2 = -\f{g^2B^2_T\omega^2}{m_\phi^2} + 0(.) \\
     p_2 = \omega - \f{m_\phi^2}{2\omega} -\f{g^2B^2_T\omega}{2m_\phi^2}+ 0(.)\ \ \ &\mathrm{and}& \ \ \
     m_2^2 = m_\phi^2 +\f{g^2B^2_T\omega^2}{m_\phi^2}+0(.) \eea
also very close to the original photon and $\phi$. To get the results usually quoted as the ``weak
mixing" limit, one has to assume also that\footnote{An observation which it is not usually done.}
$\omega z \tan 2\theta  \ll 1$, then we obtain
\bea A_{||}(z)&=& e^{-i\omega z}\br{1-\f{g^2B^2\omega^2}{m_\phi^4}
\pa{2\sin^2\f{m_\phi^2z}{4\omega}+i\pa{\f{m_\phi^2z}{2\omega}-\sin\f{m_\phi^2z}{2\omega}}}+...}  \label{A-weakmixing}\\
\mathrm{P}\pa{\gamma_{||}-\phi}&=&\f{4g^2B_T^2\omega^2}{m_\phi^4}\sin^2\pa{\f{m_\phi^2z}{4\omega}}
\label{gamma-phi.oscillation_weakmixing}\eea

Interestingly enough this weak mixing limit applies in the case of the PVLAS ALP parameters since
for $B_T\sim 5$ T, $\omega= 1.2$ eV, $g^{-1}\sim 10^{6}$ GeV, $m_\phi\sim 1$ meV we find
\be \tan 2\theta \sim 10 ^{-6} \ee

\textit{Maximal mixing limit $\theta\sim \pi/4$} .-   \\
In this case $\tan^{-1} 2\theta \sim \cos 2\theta = \sin (\pi/2-2\theta) \sim
\pi/2-2\theta \sim -m_\phi^2/2gB_T\omega \sim 0$, implying $2\theta=\pi/2+m_\phi^2/2gB_T\omega $. The corresponding
eigenstates and eigenvalues are
\bea A_1 = \f{1}{\sqrt{2}}\left[A_{||}+\phi - \f{m_\phi^2}{4gB_T\omega}\left(A_{||}-\phi\right)\right]\\
     A_2 = \f{1}{\sqrt{2}}\left[\phi-A_{||} - \f{m_\phi^2}{4gB_T\omega}\left(A_{||}+\phi\right)\right] \eea
\bea p_1 = \omega + \f{gB_T}{2}+0(.) \ \ \ &\mathrm{and}& \ \ \ m_1^2 = -gB_T\omega + 0(.) \label{max_mix_mass1}\\
     p_2 = \omega - \f{gB_T}{2}+0(.) \ \ \ &\mathrm{and}& \ \ \ m_2^2 = gB_T\omega + 0(.) \label{max_mix_mass2}\eea
\bea A_{||}(z)&=& e^{-i\omega z}\br{1- 2\sin^2\f{gBz}{4} - i\f{m_\phi^2}{gB\omega}\sin\f{gBz}{2}} \label{A-maxmixing} \\
\mathrm{P}\pa{\gamma_{||}-\phi}&=& \sin^2\pa{\f{gBz}{2}} \label{gamma-phi.oscillation-maxmixing}
\eea
Note that in the  $m_\phi\rightarrow 0$ limit there is no imaginary part, as expected from the fact
that it comes from the difference in the propagation speed of photons and $\phi$'s of the same
energy, which vanishes for $m_\phi\rightarrow 0$. Still there is a real part that provides
$\gamma_{||}-\phi$ transitions. This is an opposite situation to neutrino oscillations because here
the oscillations are driven by the interaction $\phi\gamma\gamma$ and not by mass mixing.

Regarding the formulae for $\mathrm{P}\pa{\gamma_{||}-\phi}$ we see that the limits $z\rightarrow
0$ in maximal and weak mixing coincide. However, for long distances the maximal mixing formula
\eqref{gamma-phi.oscillation-maxmixing} can grow until $P=1$ while the weak mixing result
\eqref{gamma-phi.oscillation_weakmixing} has a small maximum value. These results will show their
interest later on, when I consider the possibility of measuring $\gamma-\phi$ oscillations in a low
pressure gas. There we will find that controlling the gas pressure we can effectively set
$m_\phi\go 0$ and get this maximal mixing situation.

\subsection{Parity arguments and parity-even ALPs}\label{Parity Arguments and parity-even ALPs}

\begin{figure}
  \centering \vspace{-.5cm}
  \includegraphics[width=12cm]{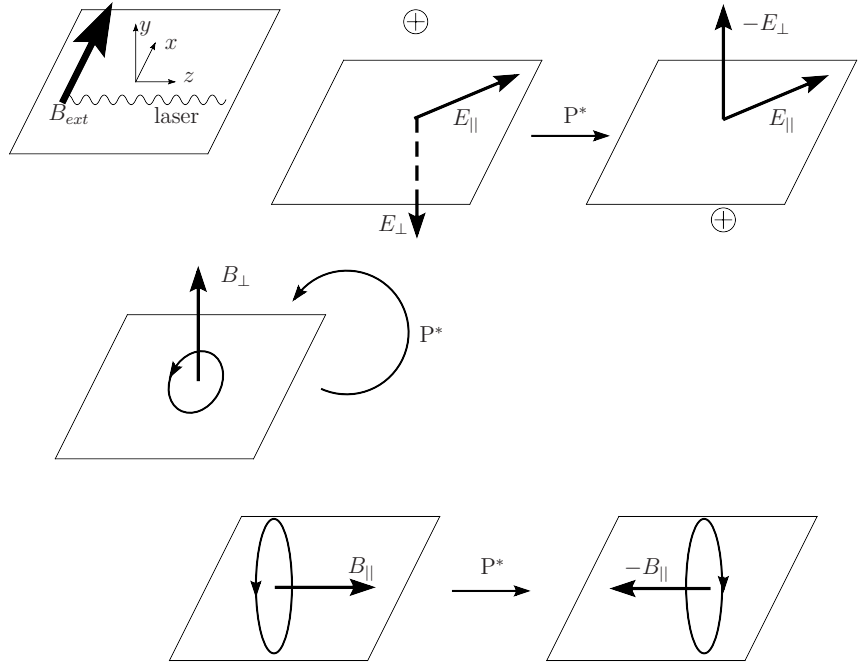}
  \caption{\small P$^*$ transformation properties for electric and magnetic fields contained
   in the $x-z$ plane ($||$) and orthogonal to it ($\perp$). The transformation acts on their relative
   sources (a positive charge for the electric field and a circular current for the magnetic field)
   which then determines the transformed fields.}\label{P-EB}
\end{figure}

It is quite remarkable that the set of equations \eqref{ALP-phi-mixing-equations} decouples into
two sub-sets, \eqref{a-gamma mixing equations(Aperp)} and \eqref{a-gamma mixing equations}. This
can be understood with parity arguments as shown in \cite{Raffelt:1988rx}. Instead of using
P$\overrightarrow{x}=-\overrightarrow{x}$ let us define the parity transformation\footnote{This is
nothing but the conventional parity P, with a further rotation of 180$^o$ in the $xz$ plane.}
$\mathrm{P}^*$ as
\be \mathrm{P}^*x_{||} =  x_{||} \ \ \ ; \ \ \ \mathrm{P}^*x_\perp = - x_\perp \ee
Where now I denote as $x_{||}$ the coordinates projected onto the plane containing the magnetic
field and the direction of the laser beam (the $xz$ plane) and $x_\perp$ the orthogonal direction
(the $y$ axis). Then, electric and magnetic fields transform according to
\bea \mathrm{P}^*E_{||}(x_\perp,x_{||})=E_{||}(-x_\perp,x_{||})\ \ \ ; \ \ \
\mathrm{P}^*E_\perp(x_\perp,x_{||})=-E_\perp(-x_\perp,x_{||}) \nonumber\\
\mathrm{P}^*B_{||}(x_\perp,x_{||})=-B_{||}(-x_\perp,x_{||})\ \ \ ; \ \ \
\mathrm{P}^*B_\perp(x_\perp,x_{||})=B_\perp(-x_\perp,x_{||}) \label{CP^* transformations on EB}\eea
where again $||,\perp$ refer to the vector components in the $xz$ plane and along the perpendicular
direction $y$. This can be easily seen if one considers how $\overrightarrow{E}$ and
$\overrightarrow{B}$ change if we apply this $\mathrm{P}^*$ transformation to a charge or a current
distribution (See Fig. \ref{P-EB}). Moreover the charge conjugation operation changes sign of both
electric and magnetic fields $\mathrm{C}\overrightarrow{E}= -\overrightarrow{E}$,
$\mathrm{C}\overrightarrow{B}=-\overrightarrow{B}$. The external magnetic field changes sign under
both C and $\mathrm{P}^*$ so is left invariant under the combined transformation $\mathrm{CP}^*$.
Finally, also the ALP interactions \eqref{ALP-couplings} can respect this symmetry. The
pseudoscalar interaction can be written in terms of the electric and magnetic fields
\be \f{1}{4}gF_{\mu\nu} \widetilde F^{\mu\nu}\phi= -g \overrightarrow{E}\cdot\overrightarrow{B}\phi
\label{parity_odd_ALP} \ee
which at the view of \eqref{CP^* transformations on EB} is $\mathrm{CP}^*$ invariant when $\phi(x)$ is CP$^*$-odd, i.e.
if $\phi(x)$ is P-odd\footnote{The $\phi$ field is neutral so C$\phi=\phi$ and scalar so invariant under the rotation
added to P$^*$ }. Conversely, the scalar interaction is written
\be \f{1}{4}gF_{\mu\nu} F^{\mu\nu}\phi= -\f{1}{2}g (\overrightarrow{E}^2-\overrightarrow{B}^2)\phi
\label{parity_even_ALP} \ee
and will be CP$^*$ invariant for a P-even $\phi(x)$ field.

\begin{figure}
  \centering \vspace{-.5cm}
  \includegraphics[width=12cm]{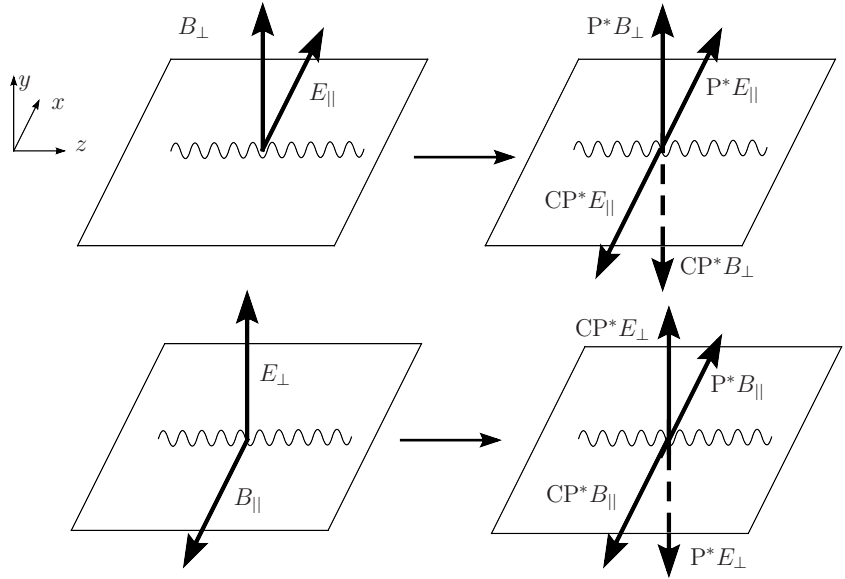}
  \caption{\small CP$^*$ transformation properties for plane waves.
  $||$-waves are CP$^*$-odd and $\perp$-waves are CP$^*$-even.}\label{CP-planewave}
\end{figure}

In any of these cases, the whole dynamics is invariant under CP$^*$ and thus the eigenstates of the system can be
classified by the eigenvalues of CP$^*$. Therefore, at the view of  Fig. \ref{CP-planewave}, we can conclude that:

\textit{${||}$-plane waves are} CP$^*$\textit{-odd and $\perp$-plane waves are}
CP$^*$\textit{-even}\footnote{This is true even if some small longitudinal component arises in a
medium.}. \textit{In the presence of Parity conserving interactions they can only ``mix" with
particles of the same parity. }

Splitting \eqref{parity_even_ALP} and \eqref{parity_odd_ALP} into the external magnetic field and
the laser contribution we get
:
\be \f{1}{4}gF_{\mu\nu} \widetilde F^{\mu\nu}\phi \simeq -g
\overrightarrow{E}\cdot\overrightarrow{B}_{ext}\phi= -g
 \overrightarrow{E}_{||}\cdot\overrightarrow{B}_{ext}\phi  \ee 
\be \f{1}{4}gF_{\mu\nu} F^{\mu\nu}\phi\simeq  g
\overrightarrow{B}\cdot\overrightarrow{B}_{ext}\phi= g
 \overrightarrow{B}_{||} \cdot\overrightarrow{B}_{ext} \phi = - g
 \overrightarrow{E}_\perp \cdot\overrightarrow{B}_{ext}\phi \ee
where for the last equality I have used $\overrightarrow{B}=\widehat{z}\times \overrightarrow{E}$
for a plane wave \textit{in vacuum}. At this point we understand that starting with the parity-even
structure \eqref{phi-gamma-gamma.coupling} we would have ended up with \textit{exactly the same
system of equations \eqref{ALP-phi-mixing-equations} interchanging the role of  $A_{||}$ and
$A_\perp$ states}. Finally, \eqref{ALP-phi-mixing-solutions1} and \eqref{ALP-phi-mixing-solutions2}
hold as the solutions for $A_\perp$ and a parity-even-$\phi$ mixing with the coupling
\eqref{parity_even_ALP}.

\subsection{Angular momentum conservation}

It is also is interesting to note that only \textit{transverse} magnetic fields are capable of
inducing transitions between photons and spin 0 particles. In order for these transitions to occur,
the ``mixing agent" (the external magnetic field) has to match the quantum numbers, in this case
the spin component $s_z$.

Formally we can write the external magnetic field as proportional to the angular momentum operator
$\overrightarrow{B}_{ext}\propto\overrightarrow{J}$ and consider the $\gamma-\phi$ transition
amplitude. Using $\gamma_\pm$ for the photon eigenstates of $J_z$ with eigenvalue $\pm 1$ we find
that
\be \langle \phi| J_z | \gamma_\pm \rangle = \pm \langle \phi| \gamma_\pm \rangle = 0\ee
by conservation of the total angular momentum. So a longitudinal magnetic field ($B_z$) cannot
mediate such transitions, although allows transitions between $\gamma_{||}$ and $\gamma_\perp$
states, as occurs in the Faraday effect to be discussed later on.

Finally, I have also considered together with E.~Mass{\'o} and C.~Biggio the case in which $\phi$
is an spin-2 particle \cite{Biggio:2006im}. In this case we have explicitly checked that a
transverse field can mediate transitions between photons and $s^\phi_z=\pm 2, 0$ states while a
longitudinal one will be responsible for the transitions to $s^\phi_z=\pm 1$.

\subsection{Rotation and ellipticity} \label{section_RotationEllipticity}

A general light wave propagating along the $z$-direction can be parameterized by means of the
Stokes parameters $S_{0,1,2,3}$ defined by
\bea S_0 = |E_x|^2 + |E_y|^2 \ \ \ ; \ \ \ S_1 =  |E_x|^2 - |E_y|^2 \\
     S_2 = 2 \mathrm{Re}\pa{E_x^* E_y} \ \ \ ; \ \ \ S_3 = 2\mathrm{Im}\pa{E_x^* E_y} \eea
Where the electric field is $\overrightarrow{E}= -\omega \overrightarrow{A}$.

\begin{figure}
  \centering \vspace{-.1cm}
  \includegraphics[width=10cm]{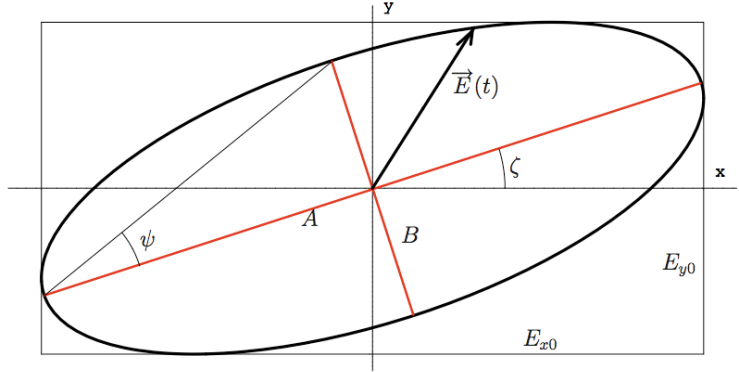}
  \caption{\small   Generic elliptical polarization. $\zeta$ is the angle between the mayor semiaxis
  and
  the $x$ direction (Parallel to $B_\mathrm{ext}$). The ellipticity $\varepsilon$ is the ratio of B and A, i.e. the minor
  over the major semiaxis. Understanding $\varepsilon$ as a tangent, the corresponding angle
  is called the ellipticity angle $\psi$.  } \label{rot-elip-def}
\end{figure}

From them it is easy to derive the formulae for the \textit{angle of polarization} $\zeta$, the
\textit{ellipticity angle} $\psi$, and the ellipticity $\varepsilon$, of a generically polarized
plane wave
\be \tan 2\zeta = \f{S_2}{S_1} \ \ \ ; \ \ \ \sin 2\psi =\f{S_3}{S_0} \ \ \ ; \ \ \ \varepsilon =
\arctan \psi \ee
The physical interpretation of these quantities is made clear in Fig. \ref{rot-elip-def}.

We can parameterize a small change in amplitude and in phase as
\be E_{||,\perp}\pa{z} = E_{||,\perp}(z=0)\pa{1-\eta_{_{||,\perp}}-i\varphi_{_{||,\perp}}} \ee
and let $\eta,\varphi$ be small real quantities, for which we can use $1-\eta-i\varphi\simeq(1-\eta)e^{-i\varphi}$.

Then, if the initial beam is linearly polarized at angle $\zeta_0$ (See Fig.
\ref{rotation-ellipticity-ALP}.a ), after passing the magnetic field it shows a rotation of the
polarization plane\footnote{We see here how if the magnet is rotating ``adiabatically"
($\zeta_0=\omega_0 t$), as in the PVLAS setup, the rotation beats with twice the frequency of the
magnet's revolution $2\omega_0$.}
\be \delta\zeta = \f{\sin 2\zeta_0}{2} \pa{\eta_{_{||}}-\eta_{_{\perp}}} \label{rotation_def} \ee
and a small ellipticity or ellipticity angle
\be \varepsilon \simeq \psi \simeq \f{\sin 2\zeta_0}{2} \pa{\varphi_{_{||}}-\varphi_{_\perp}}
\label{ellipticity_def}\ee
The sign of the ellipticity (or the ellipticity angle) tells us if we have \textit{clock-wise}
(sign $\varepsilon > 0$) or \textit{anti-clockwise} (sign $\varepsilon < 0$) elliptical
polarization, and it is also measurable.

It is easy to get $\eta_{_{||}}$ and $\varphi_{_{||}}$ from equations \eqref{A-weakmixing} and
\eqref{A-maxmixing}. If we had selected a parity-even ALP we would have got the same solutions but
replacing $||$ for $\perp$. In this case, we see that the rotation \eqref{rotation_def} and
ellipticity \eqref{ellipticity_def} will change sign, as I have argued from Fig.
\ref{rotation-ellipticity-ALP}.

\subsection{Further comments on the ALP interpretation}

We have calculated the resulting rotation and ellipticity on an initially polarized laser beam that
traverses a strong transverse magnetic field \textit{once}. However, in the PVLAS setup, the magnet
is located inside a Fabry-Perot cavity so that the \textit{laser beam performs $N$ reflections}
before entering the detector. If the $\phi$ field does not interact with the cavity mirrors, then
at every reflection, the $\phi$ particles produced scape the cavity and the $\phi$ field is reset
to zero. The final effect is analogous to shinning a laser beam through $N$ copies of the magnet
separated by barriers that absorb the $\phi$-particles. The resulting amplitude for the photon
field is then
\be A_{||}(z)= A_{||}(0) \br{1-\eta(z)-i\phi(z)}^N \simeq A_{||}(0)\br{1-N\eta(z)-iN\phi(z)} \ee

In \cite{Maiani:1986md} this point was unnoticed and they got a much more optimistic
\be
A_{||}(0)\br{1-\eta(Nz)-i\phi(Nz)}\ee
which will hold in models in which the $\phi$ field is reflected from the cavity mirrors.
Incidentally I found such a model in \cite{Jaeckel:2006id}, studying general frameworks in which
the astrophysical bounds on these particles could be evaded.
%

Note that we have started with an interaction $g\phi \widetilde F_{\mu\nu}F^{\mu\nu}$ which will be
most probably the first order of a complete effective lagrangian. With the mixing approach in
\cite{Raffelt:1988rx} we have managed to obtain a solution that involves further powers of $g^2$.
As long as additional effective vertices are expected in the realistic lagrangian, we cannot trust
in principle our results further than the $g^2$ level. However, as long as we can neglect the terms
involving two or more fields in the equations of motion the further interactions to consider, will
give the same oscillation pattern, i.e. \textit{the same functions}
\eqref{A_par_solution},\eqref{phi_solution}  but with corrections $\propto g^4$  to the mixing
angle and $\Delta_{osc}$. The same will hold for corrections to the relativistic assumption $m_\phi
\ll \omega$. In this case the expansion in $g$ is to be understood in $\theta$ and $\Delta_{osc}$
and \textit{not} on $A_{||}(z),\phi(z)$.

\subsection{Gas measurements} \label{Gas measurements_subsection}

Up to know we have considered $\gamma-\phi$ oscillations in vacuum permeated by a strong magnetic
field. The PVLAS collaboration has also performed measurements filling the magnetic field region
with a low pressure gas that may complete and give more soundness to the ALP interpretation. It is
mandatory here to include the theoretical tools to interpret these results.

It is well known that a gas in a transverse magnetic field becomes birefringent. This is the so
called Cotton-Mouton effect I mentioned before. The physical idea is simple to explain. The forward
scattering of light in neutral matter does produce a phase delay in its propagation. This is
accounted for with an \textit{index of refraction} which in isotropic materials cannot depend on
the light polarization. A gas at a low temperature can be considered such an isotropic medium
because their spins are randomly oriented. However, when you provide an external magnetic field
these neutral atoms tend to align their spins with the magnetic field and produce an anisotropy of
the medium. The forward scattering is thus different for the two linear polarizations ($||$ and
$\perp$) and then we expect a different index of refraction for $E_{||}$ and $E_\perp$ waves, i.e.
the medium becomes birefringent.

The resulting indices of refraction are quadratic in the external transverse magnetic field, $B_T$,
and are given by
\be (n_{||}-n_{\perp})_{\mathrm{C.M.}} = C\lambda B_T^2 \ee
where $\lambda$ is the light wavelength and $C$ is the so called Cotton-Mouton constant, that depends \textit{linearly}
on the pressure for low pressures.

Remarkably, if there is a \textit{longitudinal} component of the magnetic field also
$\gamma_{||}-\gamma_{\perp}$ transitions are expected, a phenomenon called \textit{Faraday effect}.
Hence, this effect also produces a rotation of the polarization plane for initially polarized
light. In this case, however, the effect is \textit{linear} in $B_L$.

Both magneto-optical effects can be included in our mixing formalism by including the corresponding forward-scattering
terms to \eqref{a-gamma mixing equations} getting
\be \left[\omega^2+\partial^2_z + 2\omega^2\left( \begin{array}{ccc}
                               (n_\perp-1)  & n_R       &   0                    \\
                                  n_R       & (n_{||}-1)& \f{gB_T}{2\omega} \\
                                    0  &   \f{gB_T}{2\omega} &-\f{m_\phi^2}{2\omega^2} \end{array}
 \right)\right] \left(  \begin{array}{c} A_\perp \\ A_{||} \\ \phi \end{array}\right)(z)=0
\ee
When only a transverse magnetic field is present our results get modified in a very interesting and simple way. Again
the mixing matrix splits in two parts ($||,\perp$). The rotation that diagonalizes the resulting mixing matrix in the
($\perp$) sector becomes
\be \f{1}{2}\tan 2\theta = \f{gB_T\omega}{2\omega^2(n_{||}-1)-m_\phi^2}\equiv
-\f{gB_T\omega}{\Delta m^2}\ee
and the nontrivial gas results follow from the substitution $m_\phi^2\rightarrow \Delta m^2 =
m_\phi^2-2\omega^2(n_{||}-1)$ in \eqref{A_par_solution}, \eqref{phi_solution} and in
\eqref{Delta_osc} which gives an oscillation length
\be \Delta_{osc}=\frac{\sqrt{(m_\phi^2-2\omega^2(n_{||}-1))^2+(2gB_T\omega)^2}}{2\omega} \ee

In particular, \textit{maximal mixing can be achieved easily} by adjusting the gas pressure (and
thus $n_{||}$) to get $\Delta m^2=0$. This is very interesting because one can benefit of using not
only strong, but also long magnets, like those used for bending particle beams at HERA or LHC,
without fear that the length could produce de-coherence of the beam.

Notably PVLAS can measure also ellipticity. This depends on the difference of the indices of
refraction of the two perpendicular components of light. Remarkably, it turns out that the sign of
$n_{||}-n_\perp$ is different for different gasses. For instance, it is positive for nobel gases
like Neon or Helium while for Nitrogen and Oxygen is negative. This provides an independent check
of the sign of the ALP coupling. Notice for instance the ellipticity \eqref{ellipticity_def} in
near maximal mixing regime\footnote{The conclusions hold for any situation but the formula is much
simpler in this case.} \eqref{A-maxmixing} for a \textit{parity-odd} ALP and in the limit $z \go 0$
\be \varepsilon = \f{N\sin2\zeta_0}{2}\pa{\f{m_\phi^2}{2\omega} - \omega(n_{||}-n_\perp) }z \ \ \ .
\label{ellipticity-sign-gas-measurements} \ee
We find that if we scan the values of $n_{||}-n_\perp$ by varying the pressure we will find a zero
of the ellipticity \textit{only} if $n_{||}-n_\perp$ is \textit{positive}, i.e. like Neon. It is
straightforward to see that in the \textit{parity-even} ALP case, the sign of the mass in
\eqref{ellipticity-sign-gas-measurements} is negative and thus the zero of the ellipticity requires
a negative value of $n_{||}-n_\perp$, i.e. like Nitrogen. This last possibility is the one that
PVLAS measurements in gas support empirically.

\newpage \subsection{Results and discussion}
\begin{figure}\centering
  \vspace{-.1cm}
  \includegraphics[width=13cm]{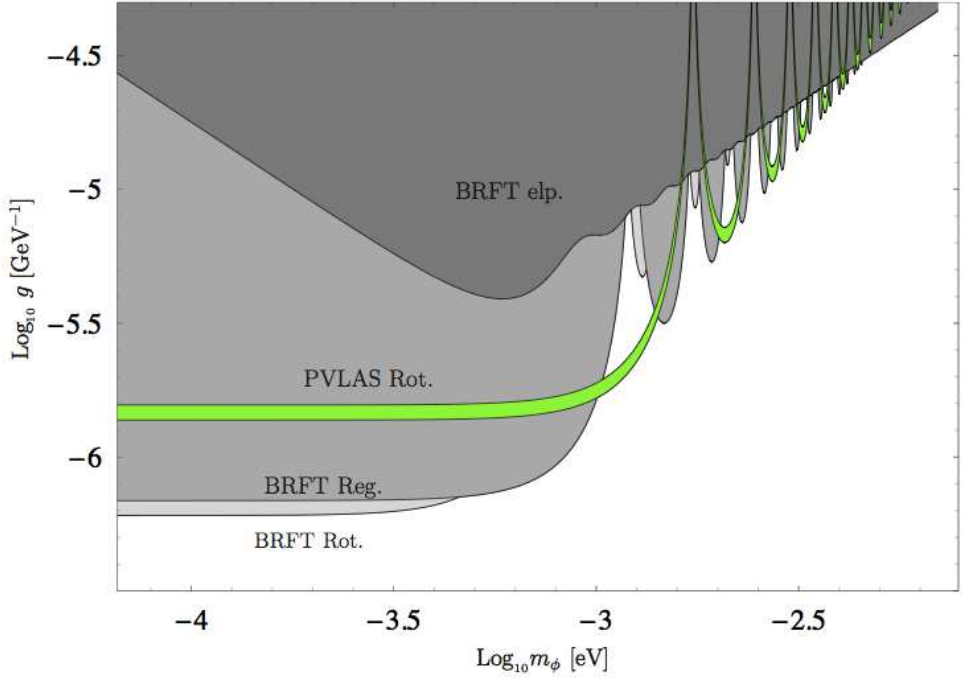}\vspace{-.3cm}
  \caption{Degeneracy in $g-m_\phi$ plane from the PVLAS rotation signal (green)
  \cite{Zavattini:2005tm}
  together with the excluded regions of the BRFT experiment \cite{Semertzidis:1990qc,Cameron:1993mr}
  .}
  \label{PVLAS-g-m-PLOT}\vspace{-.5cm}
\end{figure}

As was already advanced, the PVLAS collaboration has recently published \cite{Zavattini:2005tm} an excess rotation
signal of
\be |\delta \zeta| = 3.9 \pm 0.5 \ \ 10^{-12} \ \ \mathrm{rad/pass} \ . \label{PVLAS-signal} \ee
Note that, as the rotation depends on the two parameters of the ALP model, $g$ and $m_\phi$,
\eqref{PVLAS-signal} cannot be used alone to determine univocally their values. Using values of
$B=5$ T, $L=1$ m, and $\omega=1.18$ eV we find that the degeneracy fills the space between the
green curves plotted in Fig. \ref{PVLAS-g-m-PLOT}. We see that in any case
\be |g| > 1.16\ 10^{-6}\times \mathrm{GeV}^{-1} \label{PVLAS-signal-gmax} \ee
To provide finer values, the PVLAS collaboration combined \eqref{PVLAS-signal} with the exclusion
limits of the preceding BRFT experiment \cite{Semertzidis:1990qc,Cameron:1993mr} getting
\be 1\ \mathrm{meV} \lesssim m_\phi \lesssim 1.5\ \mathrm{meV} \ \ \ ; \ \ \ 2 \times 10^5\
\mathrm{GeV} \lesssim |M| \lesssim 6\times 10^5\ \mathrm{GeV} \ \ \ . \label{PVLAS_values for m and
g}\ee
We can see this preferred region and some other smaller islands in Fig. \ref{PVLAS-g-m-PLOT}. There
we see that the BRFT sensibility was higher at low masses. Still the BRFT detector seems to have
been placed near a minimum of the photon-ALP oscillation pattern. This can be considered a very
unlikely casuality. Indeed, it has generated several complaints in the community (See for instance
\cite{Melissinos:2007da}).

Other interesting possibility to extract these values is to combine results from rotation and
ellipticity measurements. The PVLAS collaboration has performed several measurements of ellipticity
during the last years and some of their results were available at the moment of the publication of
\cite{Zavattini:2005tm}. However these measurements are far more sensitive to systematics than
those of rotations and the final publication is suffering a considerable delay. However,
measurements of ellipticity have been released in many conferences (See for instance
\cite{Cantatore:Patras,Gastaldi:QCD,PVLASICHEP,Cantatore:IAS}). A very interesting and useful paper
\cite{Ahlers:2006iz} has recently appeared providing fits of the combined results of the PVLAS
rotation and ellipticity measurements together with the exclusion limits\footnote{Actually they
also include data from another ongoing experiment of the kind, Q\&A \cite{Chen:2003tp}, which has
already taken first results. However, at the moment their setup is not sensitive to the PVLAS
signal.} from the BRFT experiments. Their best fit to the ALP hypothesis narrows the PVLAS values
to
\be m_\phi \simeq 1\ \mathrm{meV} \ \ \ ; \ \ \ |M| \simeq 5 \times 10^5\ \mathrm{GeV} \ee
As mentioned in section \ref{Gas measurements_subsection}, the ellipticity measurements point to a
parity-even ALP.

Notice the absolute values of \eqref{PVLAS-signal} and \eqref{PVLAS-signal-gmax}. I explained
before that the sign of the rotation directly indicates the parity\footnote{In the case of an ALP
without definite parity and both type of couplings (schizon), the sign indicates the relative
importance of the two couplings $g^+,g^-$. See Appendix \ref{schizons}.} of the ALP. Unfortunately
the PVLAS results \cite{Zavattini:2005tm} did not include such a sign. Only very recently PVLAS has
made public new results that include the sign of the rotation, pointing to an \textit{odd-parity}
ALP \cite{Zavattini:2007}. Notice that if this claim and the ellipticity measurements are
confirmed, the bare ALP interpretation I have presented would be ruled out, since explaining the
two effects would require opposite ALP parities\footnote{This would be true even in the case of a
parity non-conserving interaction.}.

The real problem with this novel extremely weak interacting particle is that it is indeed
\textit{too strongly} interacting. Although it might seem that $g\sim 10^{-6}$ GeV$^{-1}$ is a
rather small coupling it has very important consequences, at least within two mature enough areas
of physics: stellar evolution and tests of gravitational interactions. The next Chapter is devoted
to show these and other constraints on ALPs.


\section{A further possibility: production of millicharged particles.}

There are other possibilities that have been proposed to account for the PVLAS signal. From my
point of view, the one deserving special attention here is presented in \cite{Gies:2006ca}. There,
the authors notice that a millicharged particle (MCP) of very small mass would be pair-produced
from laser light propagating in a strong transverse magnetic field.

This is a well-known non-perturbative phenomenon that depletes $||,\perp$ photon polarizations at
different rates producing dichroism and hence a rotation of the polarization plane of initially
polarized laser light. Further recombination of the pair will lead also to a birefringence effect
so both PVLAS measurements can be justified. In this case, however, the signs of the rotation and
the ellipticity do depend on the concrete mass and the millicharge of the MCP. Notice from Table
\ref{MCP-signs} that if the rotation and ellipticity measured by PVLAS are confirmed to be of
different sign the ALP interpretation would be ruled out but the MCP interpretation would have a
chance to fit the data.

The MCP hypothesis has been numerically analyzed in \cite{Ahlers:2006iz} and it requires a
millicharged particle whose electric charge and mass are very roughly\footnote{The concrete values
depend on the spin of the particle.}
\be   Q_f \sim 10^{-6}  \ \ \ ; \ \ \ m_f \lesssim 1\ \mathrm{eV}  \ \ \ .     \ee
Such couplings and mass are again strongly disfavored by astrophysical arguments, indeed the same
energy loss arguments leading to constraints on the ALP hypothesis. I will present these and other
constraints in Chapter \ref{constraints} but I can advance that observation on HB stars in globular
clusters still give the more demanding bound
\be  Q_f < 10^{-14} \ \ \  \mathrm{for}  \ \ \ m_f<10\ \mathrm{keV} \ee
As I will show, light millicharged particles are at the heart of our proposals
\cite{Masso:2005ym,Masso:2006gc}. However, the fact that their direct production leads to dichroism
and birefringence was unnoticed by us when we wrote \cite{Masso:2006gc}. Therefore, in spite of
using this source of rotation and ellipticity, we further required an ALP coupling to those
particles\footnote{This opens the possibility of the PVLAS signal to be a combination of ALP and
MCP real and virtual production, which is nowadays being investigated. }. From this point of view
the model in \cite{Masso:2006gc} is not minimal.

\renewcommand\arraystretch{1.5}
\begin{table}\centering
\begin{tabular}{|c|c|c|}
  \hline
                     &         $\psi >0$           &   $\psi <0$ \\  \hline
  $\delta\zeta > 0$  & MCP$0$($\chi\ll 1$) or ALP$^+$ &  MCP$0$($\chi\gg 1$)     \\ \hline
  $\delta\zeta < 0$  & MCP$\f{1}{2}$($\chi\gg 1$)     & MCP$\f{1}{2}$($\chi\ll 1$) or ALP$^-$ \\
  \hline
\end{tabular}
\caption{\label{MCP-signs} \small Signs of the rotation $\delta\zeta$ and birefringence angle
$\psi$ for the ALP and the MCP interpretations. Here $\chi = \f{3}{2}\f{\omega}{m_f}\f{\epsilon e
B}{m_f^2}$ is the relevant parameter for the MCP interpretation. $\omega$ is the energy of the
photons of the laser, $m_f,\epsilon$ the mass and electric charge of the MCP, $e$ is the electron
charge and $B$ the magnetic field, assumed to be transverse. MCP$0$ and MCP$\frac{1}{2}$ stand for
the case in which the MCP is a spin zero boson or a spin $1/2$ fermion. }
\end{table} 

\clearemptydoublepage

%% file: chapter2.tex
\def\baselinestretch{1}

\chapter{Constraints on novel low mass particles coupled to light\label{constraints}}

\def\baselinestretch{1.66}


This Chapter is devoted to a brief review of arguments that have been cast against the existence of
low mass ALPs, millicharged particles and paraphotons. I divide them into three broad classes,
Astrophysical, Cosmological and those based on laboratory direct or indirect searches. In some
cases, however, some of them could involve two of these environments.

Bounds on novel particles coupled to light have been studied by many authors and extensive reviews
can be found on MCPs \cite{Raffelt:2005mt}, axions or general ALPs
\cite{Raffelt:2006rj,Masso:1995tw,Masso:1997ru} and paraphotons \cite{Popov1999}. As we will see,
the astrophysical arguments provide the strongest bounds and typically laboratory bounds are only
quoted for particles with masses above the stellar temperatures, harmless in stellar dynamics.
However, our articles \cite{Masso:2005ym,Masso:2006gc} have pointed out the possibility that the
existence of new particles, of low mass for stellar standards, could help avoiding the
astrophysical bounds allowing a particle interpretation of the PVLAS experiment. Therefore, it is
important to provide laboratory alternatives to gain more information about these low mass
particles. Recently, many articles have been concerned about these revisited laboratory bounds, but
also new ideas have come into the field. In this section I try to review the most significant, a
hard task nowadays since the topic is already very hot and new ideas arise quite often.

\section{Astrophysical bounds}

Novel particles coupling to light modify the properties of stellar plasmas, specially if their mass
is light enough to allow kinematically their thermal production. These low mass particles, if
existing, must interact very weakly with normal matter (otherwise we would have discovered them) so
after their production they are likely to scape from the star, as neutrinos produced in the fusion
nuclear reactions do. In this way, novel particles provide a new, invisible, contribution to the
total stellar luminosity, a crucial parameter of stellar evolution. Another, less comfortable,
possibility is that the new particles interact strongly enough with the stellar medium that they
are reabsorbed soon after its production. In this case they do not contribute \textit{directly} to
the total luminosity but they accelerate the mechanisms of energy transfer inside the star
affecting notably the stellar structure.\def\baselinestretch{1}

In this section I briefly review how we can use the established knowledge about stellar evolution
to look for novel, low mass particles coupled to light.

\subsection{Introduction: stelar evolution}

Stars form by gravitational collapse of clouds of gas, mostly Hydrogen and Helium from the early
Nucleosynthesis.  The gravitationally bounded system losses energy by emission of electromagnetic
radiation. Both the resulting contraction and the rise of temperature, are explained by the Virial
theorem that relates the average gravitational and kinetic energy of a self-gravitating gas
supported by thermal pressure
\be \langle E_{\mathrm kin} \rangle = -\f{1}{2} \langle E_{\mathrm grav} \rangle \ee
We see that a decrease of the total energy implies that the potential energy becomes more negative
(the star contracts) and therefore the kinetic energy increases (the temperature
increases\footnote{Selfgravitating systems have a negative specific heat.}).

The configuration reaches a state of ``pseudo-equilibrium" when the contraction has increased the
temperature and density so much that allows the fusion of Hydrogen nuclei into Helium (typically
$T\sim 1.3$ keV and $\rho\sim$ 150 gcm$^{-3}$ in the center core of a star with the mass of the
Sun). The rates of nuclear reactions, depending steeply with the temperature, will produce a big
injection of total energy in response to an increase of temperature, therefore opposing to the
further contraction of the star.

At a smaller temperature the atoms have ionized and the electromagnetic radiation generated is not
free to escape from the star, but rather diffuses out following a random walk path. In the absence
of new physics, the total energy loss per unit time (total luminosity) of the star is due to
thermal electromagnetic radiation from the stellar surface and neutrino emission, mainly from the
center where the star is hotter.

A young star like our Sun burns Hydrogen until the temperature is so high at the center that also
Helium starts to ignite. Such resulting stars burn Helium in the central core (at a typical $T\sim
8.6$ keV and $\rho\sim 10^{6}$ gcm$^{-3}$) and Hydrogen in an outer shell and they are called
\textit{horizontal branch} (HB) stars. If the HB star is very massive (more than 7 times the Sun,
typically) it will reach an even higher temperature at which the Carbon and Oxygen, produced by the
Helium burning, will start to fuse producing heavier elements. Further increasing of the
temperature will cause new nuclear reactions leading to the production of heavier nuclei. The final
configuration of this kind has a degenerate Iron core\footnote{This is because the fusion of Iron
nuclei \textit{requires} energy.}, (a core supported by the degeneracy pressure of electrons) and
subsequent shells where the other light elements burn. At this point, the further heating of the
star, due to the outer shells burning, makes the Iron core growing until it exceeds its
Chandrasekhar limit (${\cal M}\sim1.4{\cal M}_\odot$), producing the collapse of the core. The
resulting implosion turns into the explosion that we see as Type-II Supernovae, leaving a black
hole or a neutron star surrounded by an expanding cloud of gas as the final stage of stellar
evolution.

However, lighter stars like our Sun will end up as degenerate Carbon-Oxygen stars, cooling by
thermal radiation from the surface and neutrino emission from the interior (they can not release
more nuclear power) until they finally fade out. These stars are called ``white dwarfs" because
being degenerate are very compact and thus the temperature is relatively high (a typical
\textit{surface} temperature is $1$ keV and density $\rho\sim 2\ 10^{6}$ gcm$^{-3}$).

Between every new burning phase we find an interesting intermediate state. The fusion products of
the heaviest element store in the center of the star displacing the burning to a surrounding shell.
The mass of the core grows until  thermal pressure is not sufficient to support the gravitational
pull and the electrons inside become degenerate. As more products are added to the core it tends to
shrink, and the external layers of the star expand to a configuration in which the energy transport
is mainly convective. This expansion makes the external temperature smaller and thus the color of
the star appears to be redder, hence their name: \textit{red giants} (RG)\footnote{Actually the
name red giant is usually restricted to stars where the inner core is made of Helium. Further
developed stars are called ``Asymptotic Red Giants" (AGB).}.

In this phase the temperature of the heavier element burning shell is determined mainly by the
gravitational potential arising just from the mass of the inner core by means of the Virial
dynamics. As this shell burns, the core gains mass and shrinks more and thus by the Virial theorem
it rises the temperature of the burning shell. This feedback accelerates the shell consumption so
much that these RG stages last much less than the first hydrogen burning phase (which is called the
\textit{main sequence} (MS) because stars spend most of their lives on it). Even when the
temperature is such that the core itself starts to ignite, the degeneracy pressure still dominates
and this energy gain does not lead to structural changes. Therefore, the rise in temperature is
unchecked and keeps on feeding positively on the energy generation rate. A RG star becomes brighter
than ever, reaching the so called \textit{Helium-flash}. The core then expands nearly explosively
from $\rho \sim 10^{6}$ gcm$^{-3}$ until the HB configuration is reached at $\rho \sim 10^{4}$
gcm$^{-3}$, where thermal pressure ensures the self-regulated burning provided by the dynamics of
the Virial theorem.

\subsection{The energy loss argument}

A new energy loss channel will lead to opposite effects in a star whose burning is regulated by the
Virial negative specific heat (MS and HB) and in a star dominated by a degenerate core mass (RG).
In an RG, the exotic energy loss opposes the positive feedback between the core and the external
burning shell. As the energy is more easily evacuated, the temperature of the core raises slower
thus allowing the core to be heavier at the He-flash. The main effect of an exotic energy-loss in a
RG is a delay of the Helium-flash and thus a larger He-core mass at the following HB phase.

In a self-regulated burning star the effects are more subtle because due to the logic of the Virial
regulation, any additional energy loss will readjust the star into a hotter and more compact
configuration in which the nuclear reactions can provide this new energy. In practice, because of
the very steep dependence of the nuclear reaction rates with the temperature, only a \textit{minor
adjustment is required} and the most important implication is that the nuclear fuel is exhausted
faster. Thus in the presence of a novel, unaccounted energy loss channel

\textit{a star lives longer in a RG phase and less in a MS or HB phase.}

For a HB star the luminosity is almost constant for its entire life so we can argue than a small
exotic contribution will lead to a HB phase duration of
\be t_{HB} = t_{HB}^{std} \f{L_{HB}^{std}}{L_{HB}^{std}+L_{HB}^{x}} \ee
Where $L_{HB}$ and $t_{HB}$ are the luminosity and lifetime of a HB star and $std$ holds for models
without novel particle losses. In a white dwarf a novel energy loss will simply accelerate its
thermal cooling competing with neutrinos. As in the case of Supernovae, this competition with
neutrinos can lead to very interesting bounds. However, in practice these are evolved stars are
more complicated to modelate. Being hotter and more dense than younger stars they could be thought
to be more sensitive to exotic energy loss channels, but the models I present in this thesis
suppresses the impact of such channels more efficiently at higher densities and temperatures.
Therefore, for the sake of simplicity I will discuss none of these cases and focus only in the MS,
RG and HB constraints, which allow for cleaner predictions, more statistics and consequently
stronger bounds.

\begin{figure}[h]\centering
  \includegraphics[width=12cm]{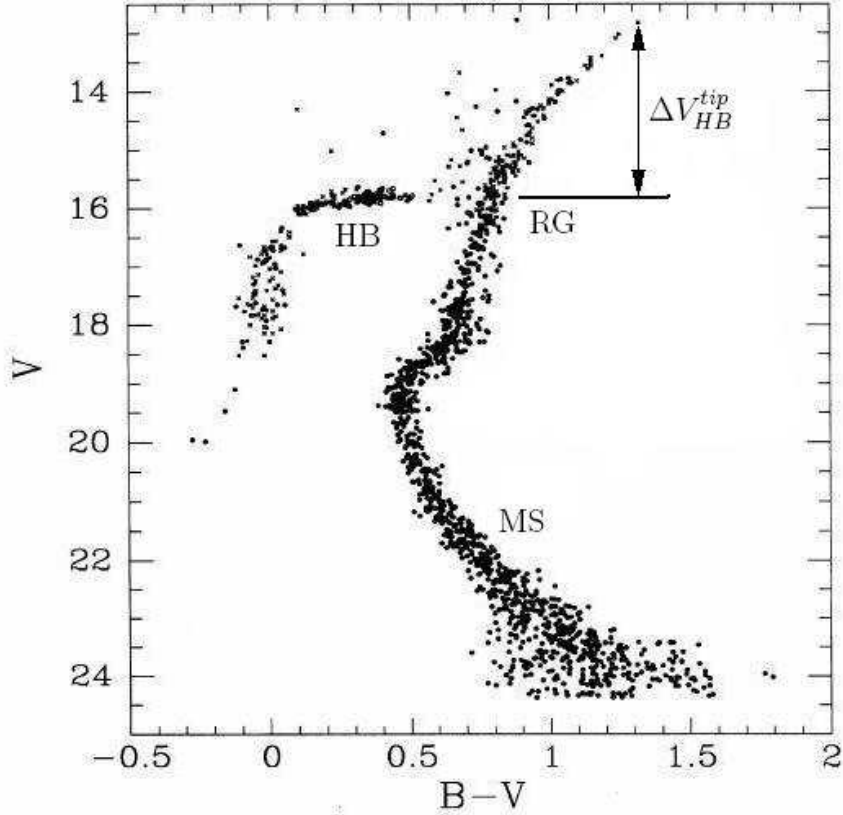}\vspace{-0.5cm}
  \caption{\small Color magnitude diagram for the globular cluster M15. V and B are the luminosities of the stars with
  filters in the visible and in blue. V is the apparent luminosity and $V-B$ is related to the temperature. Stars on the
  left side are bluer and thus hotter. The main stages of stellar evolution  are labeled: main sequence (MS),
  red giant (RG) and horizontal branch (HB). The ratio of the stars in the
  HB and RG stages and $\Delta V_\mathrm{HB}^\mathrm{tip}$, the difference in magnitude of the brightest RG
  and HB stars are sensitive to new energy loss channels. See the text for details.
  Figure taken from \cite{Durrel:1993} \label{m15}.}
\end{figure}

\newpage \subsection{Observations from globular clusters: HB and RG evolution.}

A globular cluster (GC) is a bound system of around $10^6$ first generation stars, i.e. composed
mainly by Hydrogen and Helium. All the stars have nearly the same small metallicity\footnote{Mass
fraction of elements heavier than Helium. This is a key point to produce precise statistics,
because typically $Z$ influences notably the star luminosity.}, because of their old age and only
differ by their initial mass. In the Milky Way there are around 150 GCs distributed in a spherical
halo which extends beyond the luminous disk. The color-magnitude diagram of such clusters reflects
very nicely the star evolution outlined before. In Fig. \ref{m15} we can see such a diagram for the
GC M15. Every point represents a star whose coordinates are \textit{color} and \textit{magnitude},
tightly related to the external temperature and the apparent luminosity.

The branch called MS englobes stars, like our Sun, burning Hydrogen and increasing the temperature
and luminosity. A star in the MS goes up and left in the color magnitude diagram. The red giant
phase is labeled RG and it is characterized by an increasing luminosity and a progressive reddening
due to expansion. At the tip of the RG branch, Helium starts to ignite and a self-regulated Helium
burning star (HB) arises. They are less brighter but much smaller than late RGs so they appear
bluer. The luminosity of HB stars is independent of their total mass, and a sub group called
RR-Lyrae is used as a standard candle to determine absolute luminosities. After the Helium burning
phase stars became white dwarfs (below the displayed region in Fig. \ref{m15}) or proceed to
heavier element burning (and eventually explode as Supernovae) and disappear from the diagram as
black holes or neutron stars.

More massive stars evolve faster so they enter sooner the RG and HB phases. Interestingly enough,
the RG and HB phases are much faster than the MS, so we can consider that stars in the RG and HB
branches have approximately the same mass, i.e. RG and HB branches are an isochrone of the
evolution of a unique star.

To study the impact of novel particles in stars of GCs  is not an easy task. One identifies
observables in the color-magnitude diagram that can be related to the duration of the MS, RG and HB
phases and compares these observables with numerical simulations. The difficult task is to relate
these mensurable quantities with parameters like masses, metallicities, etc.

The most important observables for our purposes are $\Delta V_{HB}^{tip}$ and the ratio $R$ of
stars in the HB and RG branches. Clearly $R$ is related to the relative velocities of the RG and HB
phases of a single star. As a new energy loss channel enlarges one and shortens the other, a
negative deviation from the standard value of $R$ could be the smoking gun of a new particle. On
the other hand $\Delta V_{HB}^{tip}$ measures the maximum brightness of a RG (the brightness at the
He-flash), which is expected to exceed a standard value if the degenerated core has grown larger
than its standard value.

High statistics have been collected showing the compatibility of these measurements with their
standard values from numerical simulations
up to a typical $10\%$. Raffelt has converted this agreement into a limit on the novel energy loss
channels in a HB, and in the RG core just before Helium-Ignition \cite{Raffelt:1996wa}. Curiously,
both observables, when confronted with simulations, lead to the same exclusion bound for the
average\footnote{ The spatial average over the HB interior or the RG core.} energy production rate
in exotica per unit mass,
\be \langle \Xi \rangle < 10\ \mathrm{erg\ g}^{-1} \mathrm{s}^{-1}  \label{GB energy loss bound}
\ee
for an HB or RG core. These small losses can be considered a small perturbation of the standard
picture and therefore $\Xi$ can be calculated from the standard values of temperature and matter
density given above. However, to evaluate the contribution of ALPs, paraphotons and millicharged
particles to $\langle \Xi \rangle$ one has to take into account the properties of the stellar
plasmas. Next I address this task to end up deriving the corresponding bounds on the novel
particles we are interested in.

\newpage \subsection{Dispersion relations and screening in a plasma}

\textit{Dispersion relations}.- In a plasma, the presence of charged particles affects the
propagation of light and hence the electromagnetic interactions. The calculation of
photon-initiated processes in a plasma has to take into account these modifications. Concerning our
models two aspects will be crucial, the non-trivial dispersion relation and the presence of
screening.

In a linear medium\footnote{The response of the medium is the generation of an induced current,
that in linear media is a linear function of the electromagnetic fields, i.e.
$J^\mu_{ind}=\Pi^{\mu\nu}A_\nu$.} the Maxwell equations in Fourier space (in Lorentz gauge
$\partial_\mu A^\mu=0$) are
\be (-K^2\eta^{\mu\nu} + \Pi^{\mu\nu})A_\nu(K)=0 \ee
where $K_\mu$ is the wave 4-vector, $\eta^{\mu\nu}$ the Minkovski metric and $\Pi^{\mu\nu}(K_\mu)$
is the polarization tensor that accounts for the interactions of light with the medium. It has to
be transverse in the sense that $K_\mu \Pi^{\mu\nu}=0$, in order to respect gauge invariance. In a
homogeneous medium there is a preferred frame in which the average constituents are at rest. In
this frame I will study the propagation of electromagnetic waves along the $z$ direction and use
$K^\mu=(\omega,0,0,k)$, so $\omega$ and $k$ are Lorentz scalars given by $\omega=U^\mu K_\mu$ and
$k^2=\overrightarrow{k}^2=(U^\mu K_\mu)^2-K^2$, where $U^\mu$ is the four-velocity of the medium.

The restriction to the Lorentz gauge leaves three physical degrees of freedom of the vector
potential $A_\mu$, two transverse and one longitudinal, whose polarization vectors are
\be e^\mu_\pm = (0,1,\pm i,0)/\sqrt{2} \ \ \ ; \ \ \ e_L^\mu = \f{\omega
K^\mu-K^2U^\mu}{k\sqrt{K^2}} . \ee
satisfying\footnote{There are some subtleties here; when $K^2=0$ one has some freedom to define
$e_L^\mu$. Also $|e_L|^2=-1$ only for space-like $K^\mu$, while it changes sign for $K^2>0$.}
$|e_i|^2=-1$. The longitudinal excitations correspond to waves with no magnetic field and electric
field equal to $\overrightarrow{E}_L=
\partial^0 \overrightarrow{A}_L-\overrightarrow{\partial}A_L^0 \propto (0,0,1)$, i.e. pointing in the direction of propagation, hence the name
``longitudinal". In the static case in vacuum they account for the static Coulomb potential of a
distribution of electric charge, but in a plasma they generally describe ``sound" waves in the
electric field.

Because of its transverse nature (in a 4-D sense), the polarization vector will be
\be \Pi^{\mu\nu} = \sum_{i=\pm,L} \pi_i(\omega,k) P^{\mu\nu}_i \ee
with $P_i^{\mu\nu}=-e_i^\mu e_i^{*\nu}$ the projectors into each polarization and $\pi_i$ scalar
functions encoding the response of the medium to the $i$ polarization. The corresponding dispersion
relations will be
\be -\omega^2+k^2+\pi_i(\omega,k)=0 \ee
If the medium is invariant under parity we will find that it cannot differentiate between
transverse excitations so $\pi_+=\pi_-\equiv \pi_T$.

The polarization tensor is, from a quantum field theory viewpoint, nothing but the photon-self
energy in the medium. In a plasma composed by electrons and other ions $\Pi^{\mu\nu}$ can be
computed at first order from the Compton scattering only from electrons \cite{Braaten:1993jw} since
the result is inversely proportional to the scatterer mass. This calculation is complicated because
of the average over the thermal distribution of electrons which makes difficult to provide
expressions for a general plasma. Here I am interested only in a classical plasma, i.e. whose
electrons are not relativistic and hence $T\ll m_e$. In this case the calculation is very much
simplified and one gets
\bea \omega^2 &=& k^2 +\omega_P^2\pa{1+\f{k^2}{\omega^2}\f{T}{m_e}} \ \ \ \ \ \mathrm{Transverse\
Plasmons,}
\label{dispersionrelation-Transverse}\\
\omega^2 &=& \omega_P^2\pa{1+3\f{k^2}{\omega^2}\f{T}{m_e}} \ \ \ \ \ \ \ \ \mathrm{Longitudinal\
Plasmons.} \label{dispersionrelation-Longitudinal}\eea
where $T$ is the temperature and $m_e$ the mass of the electrons. The plasma frequency $\omega_P$
for a classical plasma (in which the charged particles move slowly i.e. the temperature is much
smaller than particle masses) is given by
\be \omega_P^2 = \f{4\pi \alpha n_e}{m_e}\pa{1-\f{5}{2}\f{T}{m_e}}  \label{plasma frequancy classical}\\
\ee
where $n_e$ is the electron density. We recognize immediately that transverse plasmons have a
particle-like dispersion relation with an ``effective mass" $m_T\simeq \omega_P$ while the
longitudinal excitations have an almost fixed frequency $\omega\simeq \omega_P$.

\textit{Screening effects}.- In a plasma, the presence of an electron at a given point enhances the
probability of finding ions near this point and, inversely, it diminishes the possibility of
finding electrons nearby. Such polarization effect makes the medium to ``screen" the electric
charges of electrons and ions. We can compute screening effects from the Maxwell equations of a
plasma in the static limit $\omega=0$
\bea \pa{k^2+\pi_L(0,k)}A_0(k)          =  \rho(k)\\
     \pa{k^2+\pi_T(0,k)}\overrightarrow{A}(k) =  \overrightarrow{J}(k) \eea
One easily finds that $\pi_T(0,k)=0$, so stationary currents (and thus stationary magnetic fields)
are not screened. Conversely $\pi_L(0,k)\equiv k_s^2 \neq 0$ and, notably, it does not depend on
$k$ so the potential generated by an localized electron \textbf{($\rho(k)\sim \delta^3(k)$)} is
\be \pa{k^2+k_s^2}A_0(k)=\rho(k)\ \ \  \rightarrow \ \ \ A_0(r)\propto \f{1}{r}e^{-k_s r} \ .
\label{screened_Yukawa_potential}\ee
We find a potential of the Yukawa type which ``screens" the charge at distances higher than
$k_s^{-1}$. A direct evaluation of $\pi_L(0,k)$ shows that in the classical limit the screening
scale gives the well-known Debye-H\"uckel scale
\be k^2_s= k^2_D= \f{4\pi\alpha n_e}{T}=\f{m_e}{T}\omega^2_P \ \ \ .\ee
In contrast to their humble contribution to the dispersion relations, heavy ions contribute as much
as electrons to the screening since $k_s$ is independent of the mass. Indeed, the screening scale
is
\be k^2_S = \sum_i \f{4\pi\alpha Z_i n_i}{T}\ee
where the sum runs over all charged species in the plasma.

If, as an example, we calculate Coulomb scattering of electrons in a plasma, we can take into
account screening effect by using the modified potential \eqref{screened_Yukawa_potential} with
only a minor subtlety. This potential assumes that the particles entering the Coulomb potential
\textit{feel} an \textit{average charge cloud} around the target. This could be the situation if
the probes were slow particles because the screening particles have time to move around in the
probe's time of flight, but certainly this is not the case when the incoming particles are photons.
To compute such a cross section one must think that the photon feels \textit{one charge
configuration at a time} and then the average must be made after the amplitude of the process is
squared. This is equivalent to modifying the static Coulomb propagator in the cross section
\cite{Raffelt:1996wa}
\be  k^{-4} \rightarrow \left\{
\begin{array}{c}
  ({k}^2+k_s^2)^{-4}               \ \ \     \text{Slow\ probes} \\
  {k}^{-2}(k^2+k_s^2)^{-2} \ \ \ \text{Fast\ probes}
\end{array} \right.
\ee
where $k=|\overrightarrow{k}|$ is the momentum transferred.

\subsection{Selected processes} \label{selected processes subsection}

Armed with the tools for calculating reactions in a plasma I shall use them to discuss the main
energy loss channels in MS, RG and HB stars due to ALPs, paraphotons and millicharged particles.

\begin{figure}\centering
  \includegraphics[width=15cm]{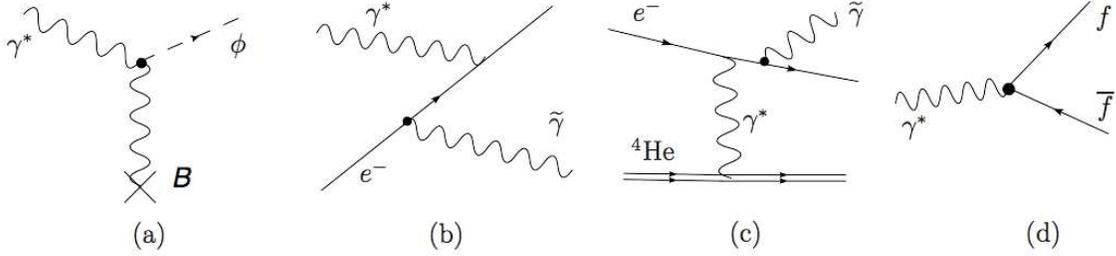}\vspace{-0.5cm}
  \caption{\small Feynman diagrams for the main channels for novel particle emission from stars.
  $(a)$ Primakoff emission of ALPs,
  $(b)$ Compton production of paraphotons, $(c)$ Bremsstrahlung of paraphotons in electron-$^4$He collisions and
  $(d)$ plasmon decay ($\gamma^*$) into millicharged particles. }\label{energylosschannels}
\end{figure}

\textit{Primakoff effect for ALPs} .- The main channel for low mass ALPs is their Primakoff
production \cite{Dicus:1978fp} from transverse plasmons (See Fig. \ref{energylosschannels}.a). This
calculation is addressed in detail for the case of the Sun in Chapter \ref{The Need for Purely
Laboratory-Based Axion-Like Particle Searches} to be integrated over a detailed solar model but it
is valid also for HB and RG stars. However, the latter have a much more dense core where electron
degeneracy effects turn out to be important, suppressing the production. On the other hand, the
Sun's interior is typically much colder. Therefore the strongest bound for Primakoff emission of
ALPs comes from HB stars. One finds from \eqref{gammaunsuppressed} that the energy loss rate per
unit mass is
\be \Xi_\phi = \f{g^2}{4\pi} \f{T^7}{\rho} F(x^2) \ee
where $x=k_s/2T$ and $F(x^2)$ is a smooth function of $x^2$ in the cases of interest. For a HB core
one can use $F\sim 1$ and $\langle T^7/\rho \rangle \sim 0.3 (8.6 \mathrm{keV})^7/(10^4$gcm$^{-3})$
leading to $\langle \Xi_\phi \rangle \sim 30 \pa{g\ 10^{10}\ \mathrm{GeV}}^2 $ erg g$^{-1}$
s$^{-1}$ and thus to the bound \cite{Raffelt:1996wa}
\be g_\mathrm{HB} < 0.6 \ 10^{-10}\ \ \mathrm{GeV}^{-1}   \label{HB-bound_for_ALPS}   \ee
%


%
\textit{Emission of Paraphotons:  Compton scattering and Bremsstrahlung} .- Massive paraphotons
like the ones considered in Chapter \ref{general concerns} couple to standard model charged
particles like electrons through a small paracharge \eqref{electron_paracharge} that depends on a
possible photon mass $\omega_P$. In the presence of a plasma, and for small values of the
paraphoton mass $\widetilde m$ one has
\be Q_1^i \simeq Q_0^i \epsilon \f{\widetilde m^2}{\omega_P^2} \ee
and therefore there is a strong suppression of paraphoton interactions in a plasma if typically
$\widetilde m \ll  \omega_P \sim$ keV. The most important energy loss channels for a non-degenerate
plasma (like MS and HB stars) are Compton scattering off electrons and bremsstrahlung in
electron-ion collisions because they involve the minimum number of gauge boson vertices (and thus
of powers of $\alpha$). In spite of the relative suppression of $\alpha\simeq 1/137$,
bremsstrahlung is competitive because it is not suppressed at high dense plasmas while Compton
scattering is.

The energy loss rate due to Compton emission can be extracted from the results in
\cite{Grifols:1986fc} for emission of scalar particles coupled to electrons since the total cross
section (after average on polarizations) for vector and scalar particles only differs in a factor
of two, accounting for the two photon polarizations.
\be \epsilon_{\widetilde \gamma} = \f{16 \alpha^2 \epsilon^2}{\pi}\f{\widetilde
m^4}{\omega_P^4}\f{Y_e T^4}{m_u m_e^2} \ee
where $Y_e$ is the number of electrons per baryon ($\sim 0.5$ in a Helium core) and $m_u,m_e$ the
atomic mass unit and the electron mass. The derived bound is $\widetilde m$ dependent. Using
$\omega_P \sim 2$ keV for transverse plasmons in typical HB conditions of density $\rho\sim 10^4$
gcm$^{-3}$ and chemical density I find\footnote{Note that a more careful bound should consider the
dependence of $m_\gamma$ with the density.}
\be \epsilon < 0.3 \pa{\f{\mathrm{meV}}{\widetilde m}}^2 \ \ .  \label{HB_compton_paraphotonBound}
\ee
The emission by bremsstrahlung in HB stars can be calculated again from the bremsstrahlung emission
of novel scalar particles derived in \cite{Grifols:1988fv,Grifols:1988fw} and gives a slightly
stronger bound
\be \epsilon < 0.1 \pa{\f{\mathrm{meV}}{\widetilde m}}^2 \ \ .
\label{HB_Bremsstrahlung_paraphotonBound}\ee
Clearly none of these bounds impose a serious restriction on the existence of
paraphotons\footnote{At least for those in the framework outlined in Chapter \ref{general
concerns}.}. Note that if the ``paraphoton" had an \textit{intrinsic} coupling to electrons (and
not driven by mixing effects) the derived bounds would not be valid, since no suppression would be
expected in this case. To recover the bounds in this case use $\widetilde m= 2$ keV in
\ref{HB_Bremsstrahlung_paraphotonBound} and the restriction reads typically $\epsilon < 10^{-14}$,
as it is found in \cite{Raffelt:2005mt}.

\textit{Millicharged particles and plasmon decay} .- Contrarily to what happens with the coupling
of charged particles with paraphotons, the small electric charge of a novel ``para-particle" $f$
would not change in a stellar environment and their production will be not suppressed.

The lowest order (in $\alpha$) production channel is plasmon decay into a pair of millicharged
particles (See Fig. \ref{energylosschannels}.d). This is possible because plasmons have a non
trivial dispersion relation. Transverse plasmons in a ``classical" plasma have an ``effective mass"
given by the plasma frequency \eqref{plasma frequancy classical} and thus can decay into a pair
$\overline{f}f$ as long as $\omega_P>2 m_f$. The dispersion relation of longitudinal plasmons in a
similar plasma fixes the energy $\omega\sim\omega_P$ independently of the value of $k$, so the
decay is kinematically allowed only for $\omega_P^2-k^2>(2m_f)^2$. Because of this ``momentum
cut-off" the longitudinal modes contribute much less than transverse ones to the energy loss and
can be neglected here.

The decay width of a plasmon into a pair of $\epsilon$-charged particles is  therefore
$\Gamma=\epsilon^2\alpha\omega_P^2/3\omega$ (in the star reference frame). After a thermal average
of initial energies one finds that the energy loss in MCPs per unit mass is
$\epsilon_\mathrm{MCP}\simeq 5 \epsilon^2$ erg s$^{-1}$ g$^{-1}$ in a HB core. In a RG core before
the He-flash the plasma frequency is four times larger $\omega_P\sim 8.6$ keV but the limit does
not improve substantially. Accordingly, one must impose
\be \epsilon < 2\ 10^{-14} \ \ \ .\ee
%


\subsection{The case of the Sun}

Main sequence stars are cooler and less dense than HB and RG stars so generally the production of
new particles is smaller and hence the bounds weaker. This is not an exception for the case of the
Sun, despite the fact that we can measure precisely its mass, luminosity, radius, metal contents,
etc. Additional information from the Sun's interior is obtained by Helioseismology and the study of
the neutrino flux. This detailed knowledge has provided a deep understanding of the solar dynamics
and the possibility of very detailed models of the present inner structure of the Sun, namely the
solar models of Bahcall \textit{et al}. . The solar bounds for a weakly interacting ALP are
reviewed and restated in Chapter \ref{The Need for Purely Laboratory-Based Axion-Like Particle
Searches} so there is no need for a further development here.

The Sun shows two further and unique characteristics that we can use to obtain information about
novel particles coupled to light. First, the energy transfer from the Hydrogen burning core to the
surface is thought to be radiative for almost the whole Sun. The existence of a new ``not so weakly
interacting" particle that avoids the energy loss argument being trapped in the solar interior can
affect this transfer mechanism and hence can also be constrained. Second, we can try to detect on
Earth any novel weakly interacting particle produced in the Sun the same we detect neutrinos. I
discuss these two possibilities in the next sections.

\subsection{The energy transfer argument}

Up to now I have focused on the possibility that new particles are weakly interacting so that they
leave the stellar interior after being produced. There might be a way of avoiding the resulting
cooling mechanism (and the corresponding bounds) if the new particle is reabsorbed inside the star.
However, even in this case, the particle contributes again to the stellar dynamics, leaving its
imprint in this case, in the speed of the energy transfer from the hot core to the outside.

There are three fundamental ways of transporting heat from the inner core to the stellar surface,
namely radiation, conduction and convection. In the first case, thermal photons from hotter parts
of the plasma carry the energy to the colder parts, in the second case, electrons are the carriers
of heat. Finally, in the third one, the plasma itself moves from the place where it gets hot to the
place in which it can release the energy. Convection is the most effective energy transfer
mechanism of all three so, wherever it takes place, the impact of a novel energy transfer mechanism
is likely to be smaller than in environments in which radiation or conduction dominates.

In the Sun, the energy transfer is thought to be completely radiative except for a small region
near the surface, so a novel energy transfer can be limited very well there. RG and HB stars are
brighter and thus have to resort using convection somewhere. A RG star is formed by a small Helium
core surrounded by a Hydrogen burning shell and a huge convective envelope and a HB star has a
Helium burning core dominated by convection and further alternate layers where radiation and
convection dominates. The complicated structure of these developed stars obscures the response to a
novel energy transfer mechanism, although some qualitative behavior has been discussed for RGs
\cite{Raffelt:1988rx}. I shall focus on the very much controlled case of the Sun.

Consider a plasma in local thermodynamic equilibrium. The flux of energy due to radiation of
particles of energy $\omega$ is proportional to their velocity $\beta(\omega)$, their mean free
path $\lambda(\omega)$, and, in the diffusion approximation, to the gradient of their energy
density $\cal E(\omega)$, i.e.
\be F(\omega)= -\f{\beta(\omega)}{3}\lambda(\omega) \nabla {\cal E}(\omega) =
-\f{\beta(\omega)}{3}\lambda(\omega) \f{\partial {\cal E}(\omega)}{\partial T} \nabla T\ee
Alternatively one commonly uses the \textit{opacity} $\kappa(\omega)$ to replace the mean free
path, both being related by $\lambda(\omega)=\pa{\kappa(\omega)\rho}^{-1}$ with $\rho$ the matter
density. If these particles are photons in local thermodynamic equilibrium, which is a good
approximation for the Sun's interior, we find $\beta(\omega)\sim 1$ and $\cal E(\omega) =
2/(e^{\omega/T}-1)$. Integrating over energies we find the total flux
\be F = -\f{4aT^3}{3\kappa \rho}\nabla T  \ee
with $a=\pi^2/15$. This last equation implicitly defines the effective or \textit{Rosseland mean
opacity} $\kappa$, an energy average over photon energies. We can easily write down the
contribution of a new particle to this quantity
\be \kappa^{-1}_x = \f{1}{4aT^3}\int_{m_x}^\infty d\omega \beta(\omega) \f{\partial {\cal E}(\omega)}{\partial T}\ee
so that $\kappa^{-1}_\mathrm{total}=\kappa^{-1}_\gamma+ \kappa^{-1}_x$. In the mean free path we
should include all processes that can lead to interchange of energy between the novel particles and
the Sun's thermal bath
.In the calculation of the opacity contribution one should include also the effects of
self-interactions of the novel particles that might slow down the energy transfer\footnote{In
\cite{Jain:2005nh} and \cite{Foot:2007cq} we find two models where self-interactions are intended
to evade the astrophysical limits of the ALP and MCP interpretations of the PVLAS particle.}.

Photon opacities are difficult to calculate and measure, mainly because of the atomic transitions
of Iron and heavy elements that, despite of its small abundance, have large contributions at
concrete frequencies. The OPAL opacity tables agree very accurately with the observed properties of
the Sun \cite{Bahcall:1987jc} from Helioseismology so one is led to the conservative\footnote{It is
worth noting that a detailed study of the consequences of the violation of this constraint lacks in
the literature.} constraint that any further exotic opacity has to be smaller than that of photons
\be \kappa^{-1}_x < \kappa^{-1}_\gamma \sim 1\ \mathrm{g\ cm}^{-2}  \ \ \ .\ee
where I have used the typical opacity in the center of the Sun.
\subsection{Helioscopes}

Any weakly interacting novel particle produced in the Sun could be detected at Earth if it
encounters an efficient enough detector. For instance, neutrinos from the inner Sun have been
detected in several underground experiments using different techniques (See the dedicated review in
\cite{Yao:2006px}) with the common characteristic of requiring a large target volume to compensate
the smallness of the interactions.

\begin{figure}[t] \centering
  \includegraphics[width=14cm]{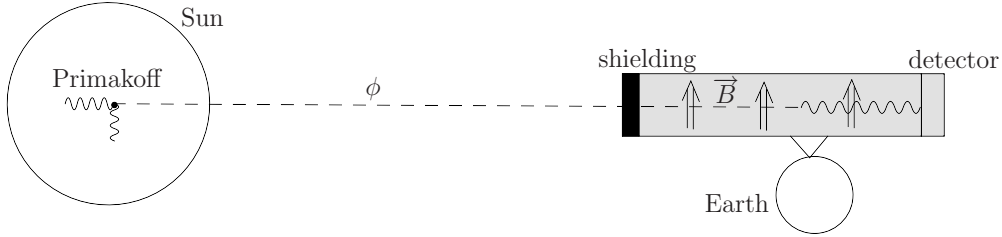} \vspace{-.5cm}
  \caption{\small Scheme of a Helioscope. The ALP $\phi$ (or any other particle capable of mixing with the photon
  inside a magnetic field) is produced in the solar core and leaves the Sun without further interactions. After a flight
  of roughly 8 minutes it enters a shielded long cavity permeated with a magnetic field in which it can oscillate into photons
  by inverse Primakoff effect, which finally can be detected.}\label{helioscope}
\end{figure}

In 1983, Pierre Sikivie proposed to use the coherent conversion of ALPs in a magnetic field
described in Chapter \ref{PVLAS} to detect a hypothetical flux of ALPs coming from the inner Sun
\cite{Sikivie:1983ip} (See Fig. \ref{helioscope}). This experiment would benefit from a coherent
enhancement if the particle is light enough or by filling the magnetic field region with buffer gas
to tune the photon index of refraction and restore coherence \cite{vanBibber:1988ge}, as was
explained in Chapter \ref{PVLAS}.
Three of such ``Helioscopes" have been built \cite{Moriyama:1998kd,Lazarus:1992ry,Zioutas:2004hi}
which have found no signal of solar ALPs. The CAST collaboration has recently published the latest
and more powerful results \cite{Zioutas:2004hi} using a $9.3$ m long decommissioned LHC test magnet
reaching $B=9$ T.

To derive bounds on the $\phi\gamma\gamma$ coupling one has first to calculate the expected flux of
$\phi$'s from the Sun. This is usually done by assuming that the emission of ALPs does not alter
the solar structure given by the precise solar models available, but just amounts to a small
perturbation. I have performed such a calculation in \cite{Jaeckel:2006xm} (see
Fig.\ref{figextra}) and my results agree with previous estimates. With an essentially similar
flux, the CAST collaboration has derived \cite{Zioutas:2004hi} the following bound
\be g < 1.16 \ 10^{-10}\ \mathrm{GeV}^{-1} \label{CAST_bound_for_ALPs}  \ \ \ .   \ee
Of course, these results assume a standard ALP history: one produced in the Sun by Primakoff effect
with an energy of some keV, the ALP goes through the Sun's outer shells, several km of ``not so
empty" space and finally the CAST shielding without any change to end up converting into an X-ray
detectable by any of the 3 CAST detectors. The failure of any of these three hypotheses could
easily evade the bounds\footnote{Remarkably, I have found and discussed examples where these three
hypothesis could fail. If the ALP strongly interacts with the solar medium it can be reabsorbed or
redshifted by interactions and simply keV ALPs do not scape the Sun
\cite{Masso:2005ym,Jain:2005nh}. The ALP can have also an invisible decay channel so it might not
survive the travel from the Sun to the Earth \cite{Masso:2005ym}. Finally if the ALPs have a
matter-density dependent mass $m_\phi=m_\phi(\rho)$ they could be not only produced below
expectations, but also they might be deflected by the Helioscope shielding
\cite{Jaeckel:2006id,Jaeckel:2006xm}.}.

Notably, these Helioscopes are also sensitive to the existence of paraphotons even in the case in
which the regenerating magnetic field is turned off. In \cite{Popov1999} the results of the
experiment \cite{Lazarus:1992ry} are analyzed and some prospects are made for further experiments.
The resulting bounds worsen as the mass of the paraphoton decreases because then the Compton
production in the Sun is increasingly suppressed by the photon plasma mass (See section
\ref{selected processes subsection}) or, in the oscillation formalism used in \cite{Popov1999}, the
oscillation $\gamma-\gamma_s$ is suppressed in presence of a large value of $\omega_P$. They get
\be \epsilon \lesssim 10^{-8} \f{\mathrm{eV}}{\widetilde m}
\label{lazarus_popov_bound_epsilon_paraphotons}\ee
valid for $10^{-4}$eV $<\widetilde m<100$ eV. In the range $100$ eV$<\widetilde m< 3$ keV the bound
saturates at $\epsilon < 10^{-10}$ and disappears abruptly for higher masses since in this cases
paraphotons cannot be created in the Sun (with a maximum temperature $\sim 1.3$ keV at the core).
Such a study has not been performed for the CAST telescope, but since the improvement on the ALP
coupling with respect to this experiment is roughly a factor of 10, we can
\textit{naively}\footnote{This bound will depend strongly on the CAST detection techniques, of
course, so a more detailed study is worth.} expect an increase in sensibility of the bound
\eqref{lazarus_popov_bound_epsilon_paraphotons} of the same order\footnote{A revision of these
bounds is in progress.}.

Interestingly enough, the Helioscope bounds are much more robust than stellar energy loss bounds if
we accept a suppression of the ALP emission from the Sun due to new physics. This is due to the
fact that an Helioscope signal is proportional to $g^4(\epsilon^4)$ (it involves production and
detection) while the stellar luminosity on exotica to $g^2(\epsilon^2)$. Therefore, in a
suppressing scenario the \textit{CAST bound is the more demanding constraint for ALPs and
paraphotons}.


\section{Bounds from cosmology}

The early universe also provides a sensitive laboratory for testing the existence of novel
particles coupled to light. In fact novel ``weakly interacting" particles are possibly required by
the current standard cosmological model which requires a rough $23\%$ of the critical density of
dark matter and a $67\%$ of dark energy \cite{Yao:2006px}.

The time of Big Bang Nucleosynthesis is commonly considered the frontier between the precision
cosmology and ``terra incognita". At temperatures of $0.1-1$ MeV the synthesis of light nuclei
takes place and we are provided with the oldest check of our knowledge of nuclear physics. The
accurate predictions constrain new thermal degrees of freedom at this epoch so we must ensure that
novel particles have not been produced at early times.

Much later on, at a temperature around $1$ eV, nuclei and electrons recombine and the universe
becomes transparent to light. The satellites COBE and WMAP as well as a plethora of dedicated
balloon and Earth-ground experiments have been able to measure very precisely that ancient light,
called the cosmic microwave background. Any novel particle interacting feebly with light could
distort the thermal shape of the spectrum and would be strongly constrained.

\subsection{Big bang nucleosynthesis and active degrees of freedom}

The standard theory of big bang nucleosynthesis (BBN) predicts the relative abundances of light
nuclei (Deuterium, He$^4$, Li$^3$,etc.) formed in the early universe from protons and neutrons at a
temperature $T\sim $ MeV. These light elements are able to combine despite the opposition of an
expanding universe which is controlled by the radiation energy density. Hence, the number of
degrees of freedom of radiative type (whose mass is smaller than the temperature) driving the
expansion can be cross-checked with the observed abundances to limit possible exotic degrees of
freedom. The deviation from the standard number of effective degrees of freedom
$g_*^{\mathrm{std}}=10.75$ at that epoch is constrained to be\footnote{Stronger constraints can be
derived by using populations out of equilibrium as in \cite{Davidson:2000hf}.} \cite{Cyburt:2004yc}
\be g_*-g_*^{\mathrm{std}} =  \sum_{\mathrm{B}}g_i \pa{\f{T_i}{T}}^4 +\f{7}{8} \sum_{\mathrm{F}}
g_i \pa{\f{T_i}{T}}^4 < 2.8 \label{BBN-constraint-number of species}\ee
where I have included a sum over possible new bosons (B) and fermions (F) and $g_i$ is the number
of internal degrees of freedom of the particle $i$. Here $T_i$ is the temperature of the species
$i$, which can be lower than $T$ if it has decoupled before the reheating caused by annihilation of
heavier standard particles (like QCD resonances, $\mu^\pm,\tau^\pm$,etc.).

We see that a spinless boson ($g=1$) or a Weyl fermion ($g=2$) are compatible with the bound even
with $T_i=T$, but this is not the case for a Dirac fermion ($g=4$) or a massive paraphoton ($g=3$).
In this thesis I try to show that in order to solve the PVLAS-Astrophysics inconsistency we need
new physics at a low energy scale, typically new particles and interactions. In fact, my best model
includes at least two paraphotons and a Dirac millicharged fermion (although it could be a boson as
well) with typical masses smaller than $1$ eV.

There are two arguments that can be used to circumvent these bounds. On the one hand we can assume
that such particles are not produced in the primordial soup (coming form the decays of an inflaton,
for instance) so they don't exist at the BBN epoch. Of course, we have to worry about its possible
thermal production after inflation by the interactions with normal matter that I will introduce.
The requirement of the freezing of these interactions all through the early universe (up to the BBN
epoch) provides bounds for these new interactions that my models have to satisfy. On the other hand
we can assume that these particles were produced indeed in the early universe but their
interactions with standard model particles freezed out before BBN. Therefore, their temperature can
be lower than that of photons and they can satisfy the bound\footnote{In this case they can
contribute to the dark matter of the universe and eventually its energy density could exceed the
accepted value $\Omega\simeq 1$ and overclose the universe. This argument has been used to obtain
bounds on ALPs \cite{Masso:1995tw} and MCPs \cite{Dobroliubov:1989mr,Davidson:1991si} but they
apply typically for ALP and MCP masses above $10$ eV and therefore are not interesting for this
work.} \eqref{BBN-constraint-number of species}.

A particle species decouples when the rate of the interactions that keeps it in equilibrium with
photons, $\Gamma$, is smaller than the rate of the expansion of the universe, $H$. In the early
universe the rate of expansion is proportional to the squared temperature, $H\propto T^2$. However,
the typical interaction rates of ALPs on the one hand, and MCPs and paraphotons on the other,
behave very differently with $T$. ALP interactions proceed through the irrelevant couplings in
\eqref{ALP-couplings} and thus will have a strong dependence with temperature. From dimensional
grounds and neglecting the ALP mass $\Gamma_\mathrm{ALP}\propto g^2T^3$, while MCPs and paraphotons
interact through marginal operators what will typically provide $\Gamma_{MCP,\widetilde
\gamma}\propto T$. Therefore, the ratio $\Gamma/H$ behaves very differently for both cases: it
increases with temperature for ALPs and decreases for MCPs and paraphotons. Consequently, ALPs in
thermal equilibrium in the early universe could decouple before BBN but this is not the case for
MCPs and paraphotons.

Therefore in what follows I will focus on the first possibility for MCPs and paraphotons and in the
second for ALPs to derive the required constraints from \eqref{BBN-constraint-number of species}.
Notably most of my models in \cite{Masso:2005ym,Masso:2006gc} are constructed to suppress the
couplings of these particles to normal matter in the hot and dense stellar plasmas. In the more
extreme primordial plasma of the early universe at the epoch of BBN these interactions will be even
more suppressed, relaxing very much these bounds.

\textit{ALP}.- Although the current bound \eqref{BBN-constraint-number of species} allows for the
existence of a spin zero boson it is interesting to know what would be the limit on the coupling
$g$ if the limit strengthens to $g_*-g_*^{\mathrm{std}} \lesssim 1$. Indeed this was the case some
time ago when the general study \cite{Masso:1995tw} was performed. There we can find that ALPs with
a coupling
\be g > 2\ 10^{-7}\ \mathrm{GeV}^{-1} \ \ \ee
will decouple before the reheating of photons due to pion and muon annihilation ($T\lesssim 200$
MeV) and therefore will have a smaller temperature, evading the BBN bound. For this bound, the
value $g_* - g_*^{\mathrm{std}}< 0.5 $ was used \cite{Walker:1992iz}. See \cite{Masso:1995tw} for
details.

 \textit{Lonely Paraphoton Model}.- Consider an extension of the standard model based just on
a new gauge symmetry $U(1)_1$, with its corresponding paraphoton and some sector responsible for
its mass on the basis of Chapter \ref{general concerns}.

A background of paraphotons will form if they couple to electrically charged particles with a small
paracharge $\epsilon$, for instance, by Compton production $\gamma e \rightarrow \gamma' e$. To
avoid such a production, the rate of this interaction $\Gamma \sim \epsilon^2\alpha^2 T$ has to be
smaller than the expansion rate of the Universe $H\sim T^2/m_{\mathrm{Planck}}$ , but note that in
such a plasma photons have an effective mass $m_\gamma^2=\omega^2_P \sim 4\pi\alpha T^2/9$ and
thus, by means of \eqref{electron_paracharge} the coupling is suppressed
($\epsilon=\epsilon_{T=0}m^2/\omega_P^2$) with respect to its vacuum value. Finally one gets
\be \epsilon_{T=0} \lesssim \alpha^{-1} \sqrt{\f{T}{m_{\mathrm{Planck}}}}\f{T}{m} \lesssim 10^{-8}
\f{\mathrm{MeV}}{m} \label{No-background-paraphotons-constraint} \ee
where for the last equality I have used a minimum temperature for BBN of $0.1$ MeV and
$m_{\mathrm{Planck}}=10^{19}$ GeV. This plasma-mass suppression makes
\eqref{No-background-paraphotons-constraint} completely harmless for small $m$ as it happens with
the energy loss in stars \cite{Popov1999}, see equations \eqref{HB_compton_paraphotonBound} and
\eqref{HB_Bremsstrahlung_paraphotonBound}.

\textit{Paraphoton + Millicharged fermion Model}.- Consider the extension of the later situation
with a new Dirac fermion charged under a new symmetry $U(1)_1$. This particle will acquire an
$\epsilon$-sized electric charge given by \eqref{parafermion-charge}.

Not surprisingly, a background of millicharged fermions forms much easier than one of paraphotons
because the small electric charge of millicharged particles \eqref{parafermion-charge} is not
suppressed by a relatively large photon plasma mass. The rate of plasmon decay into a
fermion-parafermion pair is $\sim \epsilon_{T=0}^2 \alpha^2 T$ and then we have to demand
\be \epsilon_{T=0} \lesssim \alpha^{-1} \sqrt{\f{T}{m_{\mathrm{Planck}}}} \lesssim 10^{-8} \ee
to prevent the formation of a background of millicharged particles. In order to achieve some
suppression for this and other channels in stellar plasmas I invoked a model with two paraphotons
which is presented in \cite{Masso:2006gc}. There we again recover a $\omega_P$-dependent electric
charge of parafermions in a plasma, and the bound is essentially similar to
\eqref{No-background-paraphotons-constraint} (regardless of small factors of $\alpha,\pi$,etc.) and
again completely harmless.

\subsection{Cosmic microwave background}

Novel particles weakly coupled to light during and after the time of decoupling of matter and
radiation can distort the CMB Planckian spectrum which has been precisely measured. At temperatures
around $1$ eV, the electrons and protons recombine and the universe becomes transparent to light.
It seems therefore that the dispersion relation of photons is at most due to the neutral atoms and
the few still unpaired charged particles, which give a negligible contribution. The CMB photons
satisfy therefore the  vacuum dispersion relation $q^2=0$.

In this situation ALPs would be produced by photon coalescence ($\gamma\gamma\rightarrow \phi$) or
in pairs ($\gamma\gamma\rightarrow \phi\phi$); MCPs would be produced in pairs
($\gamma\gamma\rightarrow\overline{f}f$) and paraphotons could be produced easily from them by
Compton scattering ($\gamma f\rightarrow\widetilde\gamma f$).

Very recently these arguments have been used to rule out millicharged particles with mass $m<$ eV
and millicharge $\epsilon<10^{-7}$ \cite{Melchiorri:2007sq}, although the bounds can be much
stronger ($\epsilon<10^{-9}$) in models with a paraphoton, like the ones considered here. The same
article mentions that the value of the ALP coupling candidate for explaining PVLAS, namely eq.
\eqref{PVLAS_values for m and g}, is still allowed by current precision of the FIRAS experiment at
the COBE satellite.

There could be ways to circumvent these bounds. For instance, it seems that if the millicharged
particles (or paraphotons) had thermalized after BBN, but before the CMB epoch, there would be no
distortion of the Planckian spectrum at decoupling and therefore no bounds\footnote{The impact of
such possibility seems not to have been addressed by the moment.}.

\section{Laboratory bounds}

The powerful astrophysical arguments, specially the energy loss one, restrict very much the
parameter space of novel low mass particles coupled to light, but they are certainly model
dependent. The searches for these particles under the controlled conditions of an Earth-based
experiment have the disadvantage of much smaller statistics but they benefit from a better
precision, a controlled environment and therefore a higher reliability of the results. In this
section I very briefly review the most relevant laboratory experiments constraining the existence
of ALPs coupled to light, paraphotons and millicharged particles present in the literature. I
divide the experimental efforts into collider and low energy experiments.

Experiments at accelerators typically benefit of a high amount of energy available in the center of
mass so that very massive particles can be produced. At the same time, clean collisions are rare
and the lack of statistics makes impossible to look for very weakly interacting particles like the
ones considered here. A higher precision is typically achieved in rare decays or spectroscopy of
bound systems. We will see that Positronium decays and measurements of the Lamb shift give very
strong bounds for these weak interactions. Optical experiments with lasers offer high statistics
and the possibility of coherence enhancement if the novel particles mix with photons, but they are
performed at low energies where backgrounds can be difficult to control. Finally, in the cases in
which the novel particle can mediate long range forces between macroscopic bodies, experiments
testing the inverse square law of the Coulomb and Newton forces offer the most impressive bounds
due to the cumulative effect over a macroscopic amount of particles. Let me review briefly all of
them.

\subsection{Bounds from colliders}

\begin{figure}[t]\centering
  \includegraphics[width=5cm]{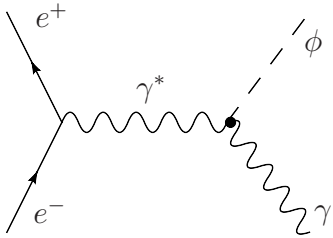}
  \caption{ALPs with an $\phi\gamma\gamma$ coupling are produced in $e^+e^-$ colliders.}\label{e+e-_to_phigamma}
\end{figure}

If novel weakly interacting particles take part in the products of a high energy collision, the
detectors will identify missing energy and momentum when reconstructing the standard particle
tracks. For instance in the SLAC storage ring PEP a dedicated detector (the ASP, for Anomalous
Single Photon detector) was built with this purpose. From the absence of $e^+ e^- \rightarrow
\gamma + \slashed{E}$ events one can  bound a $\phi\gamma\gamma$ interaction like
\eqref{a-gamma-gamma.coupling} (See Fig. \ref{e+e-_to_phigamma}) getting \cite{Masso:1995tw}
\be g < 5.5\ 10^{-4} \mathrm{GeV}^{-1} \ee
A recent paper \cite{Kleban:2005rj} has analyzed these results and derived new bounds from the
different LEP and KEKB experiments. Constraints as strong as $ g < 1.6\ 10^{-6} \mathrm{GeV}^{-1}$,
which will be sensitive to the PVLAS ALP, are suggested for LEP and KEKB experiments. However the
missing energy analysis \textit{has not been really performed} with LEP and KEK data, so these
bounds are derived under the assumption than no deviations from the standard model predictions are
to be found.

\begin{figure}[t]\centering
  \includegraphics[width=12cm]{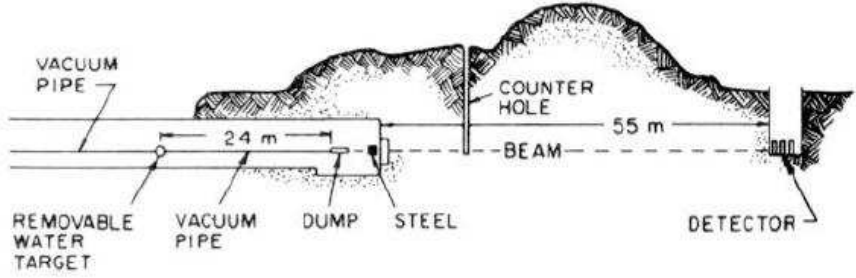}\\
  \caption{Schematics of the SLAC beam dump experiment. }\label{beamdump}
\end{figure}

Beam dump experiments have been used to limit the existence of millicharged particles. In such
experiments a beam of electrons is shone against a target and the products remaining after some
meters of shielding are analyzed (See Fig. \ref{beamdump}). In \cite{Davidson:1991si} the results
of the SLAC Beam Dump experiment were used to constrain millicharged particles with masses smaller
than $\sim 0.2$ GeV
\be \epsilon < 3\ 10^{-4} \ \ .\ee

Also, very recently it has been reported a search for millicharged particles in reactor neutrino
detectors \cite{Gninenko:2006fi}, yielding a slightly improved bound
\be \epsilon < 10^{-5} \ \ \ . \ee
\subsection{Accelerator cavities}
Particles coupled to light can be produced in accelerator cavities like the ones used in big colliders to accelerate
charged particles. In a recent publication \cite{Gies:2006hv} the Schwinger production of millicharged particles in
such cavities has been investigated concluding that low mass millicharged particles with $\epsilon> 10^{-6}$ would
leave a noticeable imprint on the quality factor of the studied cavities of the TESLA experiment \cite{Lilje:2004}. The
prospects for future cavities would lead to ameliorate this bound in an order of magnitude.

\subsection{Optical experiments} \label{Laser experiments subsection}

\begin{figure}
\begin{center}
\includegraphics[width=12cm]{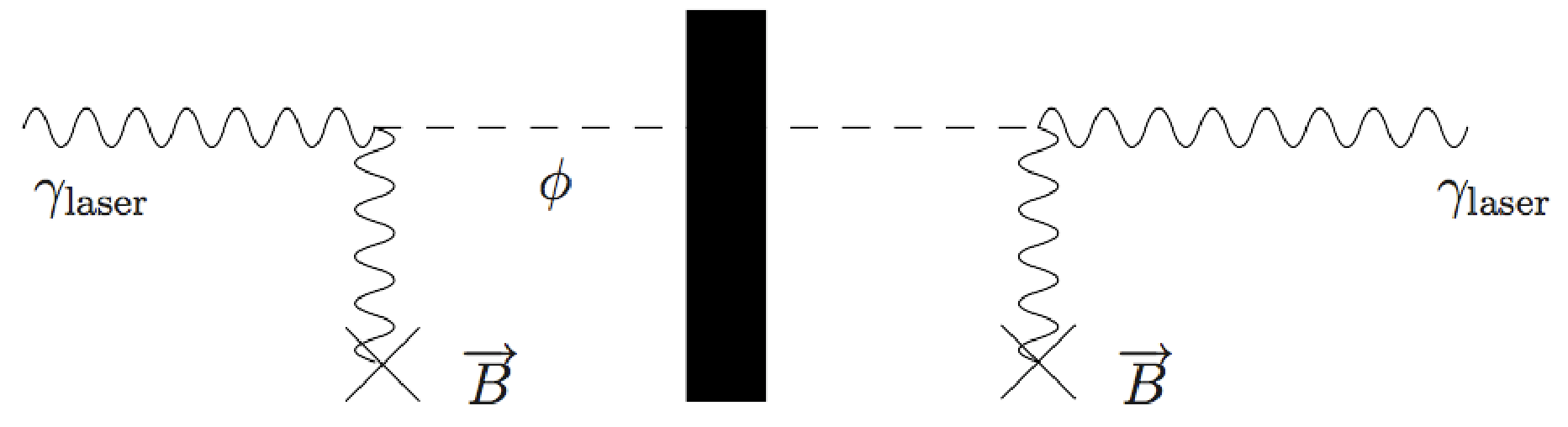}
\end{center}
\caption{Schematic view of a ``light shining through a wall'' experiment (LSW). ALP
production through photon conversion in a magnetic field (left), subsequent travel through an
opaque wall, and final detection through photon regeneration (right). The same scheme holds for a
sterile paraphoton $\gamma_s$ replacing $\phi$ without the external $B$ field insertions. \hfill
\label{photon-regeneration}}
\end{figure}

The PVLAS experiment, subject of Chapter \ref{PVLAS}, is only an example of the several experiments
that can be performed searching for low mass particles interacting with light with lasers in
optical setups. The simplest version of the family is the so called \textit{light shining through
walls} (LSW) experiment \cite{Anselm:1985bp,VanBibber:1987rq}. The basic idea is depicted in Fig.
\ref{photon-regeneration}. There we see how a photon from a laser oscillates into an ALP in the
presence of an external magnetic field $\overrightarrow{B}$, then this ALP crosses an opaque wall
to be further reconverted into a photon which can be easily detected at the end of the regenerating
magnet. In contrast to the PVLAS type of experiments the signal would be due to the appearance of
photons, so a positive result would provide a strong evidence for new particles. Because the right
side of the opaque wall can be strongly isolated, the backgrounds of such experiments can be very
much reduced. This is a key feature because in the PVLAS-type of experiments the disappearance of
photons is proportional to the $\phi\gamma\gamma$ coupling squared, while in LSW experiments the
signal goes with the fourth power of such small quantity.

The rate of regenerated/incoming photons can be easily calculated from the formulae derived for the
probability of $\gamma-\phi$ oscillation \eqref{phi_solution} in a magnetic field derived in
Chapter \ref{PVLAS}.
\bea \f{\dot{N}_\mathrm{counts}(\gamma-\phi-\gamma)}{\dot{N}_\mathrm{incoming}} &=& \
\mathrm{P}(\gamma\rightarrow\phi)\mathrm{P}(\phi\rightarrow\gamma) = \\
&=& \cos^2{\zeta}\  \sin^4{2\theta}\ \sin^2\f{\Delta_\mathrm{osc}L_1}{2}\
\sin^2\f{\Delta_\mathrm{osc}L_2}{2} \label{LSW-counting rate-ALP}\eea
With $\zeta$ the angle between the laser polarization and the magnetic field direction (both of
them assumed to be transverse to the beam direction). The $\gamma-\phi$ mixing angle, $\theta$, and
$\Delta_\mathrm{osc}$ are defined in section \ref{Axionlike particle interpretation-section} and
$L_{1,2}$ are the lengths of the first and second magnet, respectively. The outgoing photons will
have polarization parallel to the magnetic field if the $\phi\gamma\gamma$ coupling is odd, and
perpendicular if the coupling is even\footnote{If parity is not conserved by the $\phi\gamma\gamma$
interaction, $\zeta$  is the angle between the laser polarization and the schizon angle $\theta_s$
(See Appendix \ref{schizons}). The polarization of the outgoing photons will be along the
$\theta_s$ direction.}.

The BRFT collaboration performed such an experiment ending with no signal of novel particles
\cite{Cameron:1993mr}. This leads to the bound plotted in Fig. \ref{PVLAS-g-m-PLOT} which for $m\ll
1$ meV is
\be g < 3.6\ 10^{-7}\ \mathrm{GeV}^{-1}\ee
Notably, LSW experiments are also sensitive to paraphoton oscillations even without magnetic field.
From eq. \eqref{photon-sparaphoton oscillations} we find that the probability of photon
regeneration is
\bea \f{\dot{N}_\mathrm{counts}(\gamma-\gamma_s-\gamma)}{\dot{N}_\mathrm{incoming}} &=&
\mathrm{P}(\gamma\rightarrow\gamma_s)\mathrm{P}(\gamma_s\rightarrow\gamma) = \\
&=&16\epsilon^4 \pa{\frac{m_1^2}{m_0^2-m_1^2}}^4\sin^2\f{\Delta m^2 L_1}{4\omega}\ \sin^2\f{\Delta
m^2 L_2}{4\omega} \label{LSW-counting rate-ALP}\eea
Where now $L_{1,2}$ are the lengths from the photon source to the wall and from the wall to the
detector and $\Delta m^2$ is the square mass difference of the two propagating states. One might
wonder if $\gamma-\gamma_s$ oscillations can be enhanced in the presence of a magnetic field due to
low mass millicharged particles in loops \cite{Jaeckel:2007}.

The BRFT collaboration performed such a LSW experiment with null results \cite{Cameron:1993mr}.
Using $m_0\ll m_1$ their result can be quoted as
\be \epsilon < 1.7\ 10^{-4} \pa{\f{\mathrm{5.4\ 10^{-5} eV}}{m_1}}^2 \ee
for $m_1< 5.4\ 10^{-5}$ eV. From this mass up to $m_1<1.9\ 10^{-4}$ eV, the probability of
regeneration starts to oscillate and the bound (not quoted in \cite{Cameron:1993mr}) saturates at
$\epsilon \simeq 1.7\ 10^{-4}$. Beyond that mass the BRFT apparatus was not sensitive to paraphoton
oscillations.

Other optical experiments could be sensitive to the interactions of ALPs, MCPs and paraphotons with
light. Recently photon splitting has been proposed to test the PVLAS signal \cite{Gabrielli:2007zn}
and some experiments for $3\gamma\rightarrow\gamma$ coalescence have been proposed to measure the
QED vacuum effects (See \cite{Lundstrom:2005za} and references therein). Also, a new experimental
set up has been proposed to measure vacuum birefringence \cite{Heinzl:2006xc} and some technical
improvements have been suggested to increase the reach of the current LSW experiments
\cite{Sikivie:2007qm,Baker:2006ts}. It seems that optical experiments still have much to say about
ALPs, paraphotons and MCPs.

\subsection{Ortopositronium decays}

\begin{figure}[t]
  \centering
  \includegraphics[width=12cm]{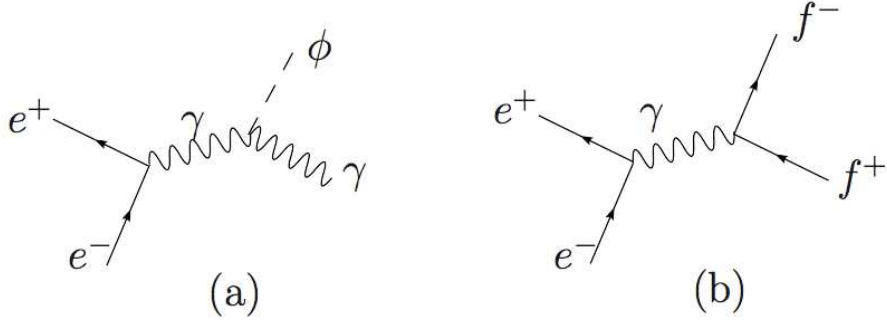}
  \caption{\small Contribution of ALPs (a) and millicharged particles (c)
   to the invisible and semiinvisible decays of ortoPositronium.}
  \label{orthopositronium diagrams}
\end{figure}

Ortopositronium (oPs) is the $1^3$S bound state of an electron and a positron. Being a spin 1
state, it cannot decay into a pair of photons because of the Yang-Landau theorem
\cite{Yang:1950,Landau:1948} and its main decay mode is oPs$\rightarrow 3\gamma$. This is a slow
process that favors the discrimination of additional decay channels into ALPs\footnote{This
reference does not consider the coupling $\phi\gamma\gamma$ but a direct Yukawa coupling of the ALP
to electrons.} \cite{Asai:1991rd} and millicharged particles \cite{Czarnecki:1999uk,Mitsui:1993ha}
(see the diagrams depicted in Fig. \ref{orthopositronium diagrams}). Experimentally one finds
\bea \mathrm{BR(oPs}\rightarrow \gamma+\mathrm{invisible})< 4.2 \ 10^{-7} \ \ \cite{Badertscher:2006fm}\\
     \mathrm{BR(oPs}\rightarrow \mathrm{invisible})< 8.6 \ 10^{-6} \ \ \cite{Mitsui:1993ha} \eea
and using $\Gamma (\mathrm{oPs})=7.05 \mu s^{-1}$ \cite{Czarnecki:1999uk} and the widths
\bea \Gamma\mathrm{(oPs}\rightarrow \overline{f}f)  &=& \f{\epsilon^2\alpha^5m_e}{6} \ \ \ \cite{Mitsui:1993ha}\\
     \Gamma\mathrm{(oPs}\rightarrow \phi^\pm \gamma)    &=& \f{\alpha^4 m_e^3}{8\pi}\times \left\{\begin{array}{c}
                                                                g^{+2} \\
                                                              \f{2}{3}g^{-2}
                                                            \end{array}  \right.
\eea
(where $\phi^\pm$ are the ALPs that couple to photons according to \eqref{a-gamma-gamma.coupling}
and \eqref{phi-gamma-gamma.coupling}) we find the following constraints on the relevant couplings
\be  g^+ < 1.5\ 10^{-3}\ \mathrm{GeV}^{-1} \ \ \ ; \ \ g^- < 1.9\ 10^{-3}\ \mathrm{GeV}^{-1} \ \ \
; \ \ \epsilon <
3.4\ 10^{-5} \ \ (\mathrm{MCP})
\ee
There are proposals for improving the limits \cite{Asai:1991rd,Mitsui:1993ha} by an order of
magnitude (See the references in \cite{Gies:2006ca}). Finally note that the new vertices are probed
at a momentum transfer squared $q^2=(2m_e)^2$, much higher than the energies related to the PVLAS
experiment. As our models \cite{Masso:2005ym,Masso:2006gc} are designed to decrease the novel
interactions at these high energies, these bounds would turn out to be negligible in such models.

\subsection{Radiative impact on QED precision measurements}

Apart from their real production (and possible detection) in accelerators, optical setups or oPs
decays, novel particles can affect QED precision observables radiatively, i.e. through its virtual
production. The most powerful measurements of QED are the electron anomalous magnetic
moment\footnote{The muon $g-2$ is much more sensitive to contributions of particles with masses
beyond the muon mass because in that case $g_i-2\sim m^2_i/M^2$ with $M$ the heavy mass. For
smaller masses one can benefit from the fact that $g_e-2$ is measured more precisely.} and the Lamb
shift, and both of them have been used to constrain novel particles coupled to light.

\textit{Anomalous magnetic moment}.-An ALP with a $\phi\gamma\gamma$ coupling will induce also a
contribution analogous to the pion-pole light by light one. This contribution can be extracted from
a number of papers studying the pion contribution (See for instance \cite{Knecht:2001qf}) but it
turns out to be a hard task for a poor bound. Therefore, and only for the sake of completeness, I
provide just an order of magnitude bound estimating that
\be \delta\pa{\f{g_e-2}{2}} = \f{\alpha}{\pi^2}\f{m_e^2}{M^2} \ \ . \ee
Similar expressions are obtained for the muon $g_\mu-2$.

\begin{figure}[t]\centering
  \vspace{-.3cm}
  \includegraphics[width=11cm]{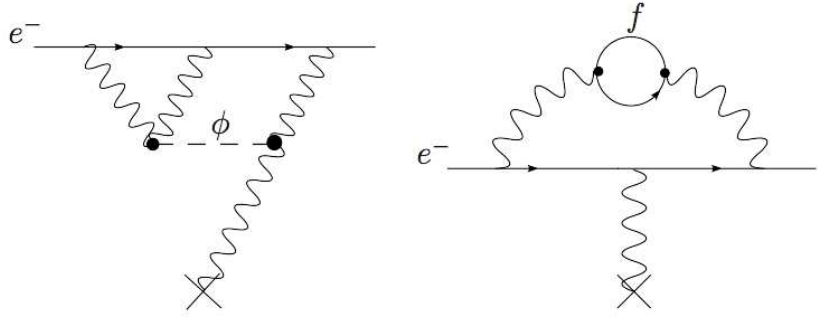}\vspace{-.5cm}
  \caption{\small Contributions to the electron $g_e-2$ from ALPs through
  the $\phi\gamma\gamma$ coupling (left) and from vacuum polarization
  produced by a millicharged particle $f$. (There are two additional diagrams to the ALP contribution
  differing in the way the photon lines are
  linked with the electron) }\label{g-2}
\end{figure}

A millicharged fermion $f$ will also induce a contribution (See Fig. \ref{g-2} for the relevant
diagram). Keeping only the leading logarithm from the general expression, that can be obtained from
\cite{Lautrup:1969fr} I find
\be \delta\pa{\f{g_e-2}{2}} = \f{\epsilon^2}{3}\pa{\f{\alpha}{\pi}}^2 \mathrm{Log}\f{m_e}{m_f} \ee
Experimentally, one finds the impressive result \cite{Mohr:2000ie}
\be \f{g_e-2}{2} = 0.0011596521859 (38)\ee
which allows to set the following constraints accepting that the experimental value corresponds to the SM
expectation\footnote{This argument is a bit too naive because in practice $(g-2)_e$ it is used to determine the
fine-structure constant $\alpha$ which is required to compute the SM prediction.  }
\be  g > 10^{-3}\ \mathrm{GeV}^{-1} \ \  (\mathrm{ALP}) \ \ \ ; \ \ \  \epsilon <
3\ 10^{-4} \ \ (\mathrm{MCP})
\ee

where I have used $m_f=1$ meV for the last bound. Constraints from the anomalous magnetic moment of
the muon are weaker. The contribution from paraphotons turns out to be negligible.

\textit{Lamb Shift}.- Novel particles coupled to light can contribute also to the energy levels of
atoms. Although it is not the most precisely measured shift\footnote{The 1S-2S shift is better
determined \cite{Bourzeix:1996}.}, the Lamb shift $\Delta E_{LS}=E_{2S_{1/2}}-E_{2P_{1/2}}$ of
Hydrogen has been used to look for massive \cite{Davidson:2000hf} ($m_f>m_e$) and light
millicharged particles, because disentangling their contribution is easier. The calculation of the
contribution from a $\phi\gamma\gamma$ coupling has not been addressed by the moment. After a mild
calculation the contribution for light millicharged fermions gives \cite{Gluck:2007ia}
\be \Delta E_{f} = - \f{\alpha^3m_e}{18\pi}\epsilon^2 \ \ . \ee
%
%
%
Assuming that this additional splitting lies between the experimental uncertainty of the most
precise measurement of the 2S$_{1/2}$-2P$_{1/2}$ Lamb shift in Hydrogen
\cite{Lundeen:1981,Hagley:1994,karshenboim:2005} $\Delta E_\mathrm{exp}/(2\pi\hbar)=(1075.85\pm
0.01) $MHz we find
\be 
\epsilon < \left\{
\begin{array}{l}
                    2.7 \ 10^{-4} \hspace{.3cm} (\mathrm{MCP}-0)\ \ \cite{Gluck:2007ia}   \\
                    1.1 \ 10^{-4} \hspace{.3cm} (\mathrm{MCP}-\f{1}{2})\ \ \cite{Gluck:2007ia}
                  \end{array}\right. \ee

\begin{figure}[t]\centering
  \includegraphics[width=4.5cm]{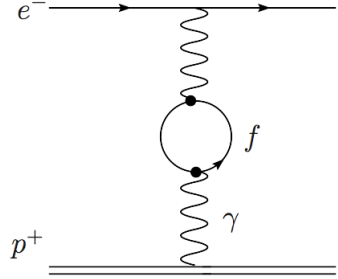}\vspace{-0.5cm}
  \caption{\small Contribution to the Lamb Shift from vacuum polarization from a novel millicharged
  fermion.}\label{LambShift}
\end{figure}

\subsection{Limits on a 5$^{\mathrm{th}}$ force}

The exchange of light bosons between matter fields leads to long range forces between macroscopic
bodies. The electrostatic Coulomb force and Newton's law of gravity are the only known forces of
such type and they are well understood in terms of photon and graviton exchange. Both forces are
extremely well tested experimentally and these tests can be used to constrain novel bosons coupled
to electrons and protons.

Goldstone bosons, however, only mediate spin-dependent forces at first order \cite{Gelmini:1982zz}
which are ineffective for tests using unpolarized bodies. The reason is that Goldstone bosons
couple \textit{derivatively }to fermions and then a parity-even structure gives $\partial_\mu\phi
\overline{\psi}\gamma^\mu\psi=0$ by means of the Dirac equation of motion of the
fermion\footnote{It is evident that this argument does not hold for flavor changing currents, but
then we will not expect a ``force" in the classical sense because the probes will change its
flavor. On the other hand we can not forget that non-derivative interactions can be generated by
terms that break explicitly the symmetry associated to the Goldstone.}. On the other hand, the
parity-odd structure $\partial_\mu \phi \overline{\psi}\gamma^\mu\gamma_5\psi = 2m_\psi
\overline{\psi}\gamma_5\psi$, which in the non-relativistic limit $\sim
2m_\psi\overrightarrow{\sigma}\cdot \overrightarrow{q}$, where $\overrightarrow{\sigma}$ are the
Pauli matrices representing the spin and $\overrightarrow{q}$ is the momentum transfer carried by
$\phi$. Although some work has been carried out to test these spin-dependent forces, the most
accurate measurements are still those testing unpolarized bodies\footnote{It is interesting to note
that exchange of two goldstones \cite{Ferrer:1998ue} or two fermions
\cite{Sikivie:1983ip,Grifols:1996fk} leads to a spin-independent force again, and thus tests of
spin independent forces lead to (weaker) bounds on spin dependent ones.}.

Regardless of its nature, I am concerned with ALPs that couple to light with both parity structures
\eqref{a-gamma-gamma.coupling} and \eqref{phi-gamma-gamma.coupling}. Chapter \ref{Light scalars
coupled to photons and non-newtonian forces} is devoted to examine the existing bounds on these
particles.

\begin{figure}[h]\centering
\includegraphics[width=12cm]{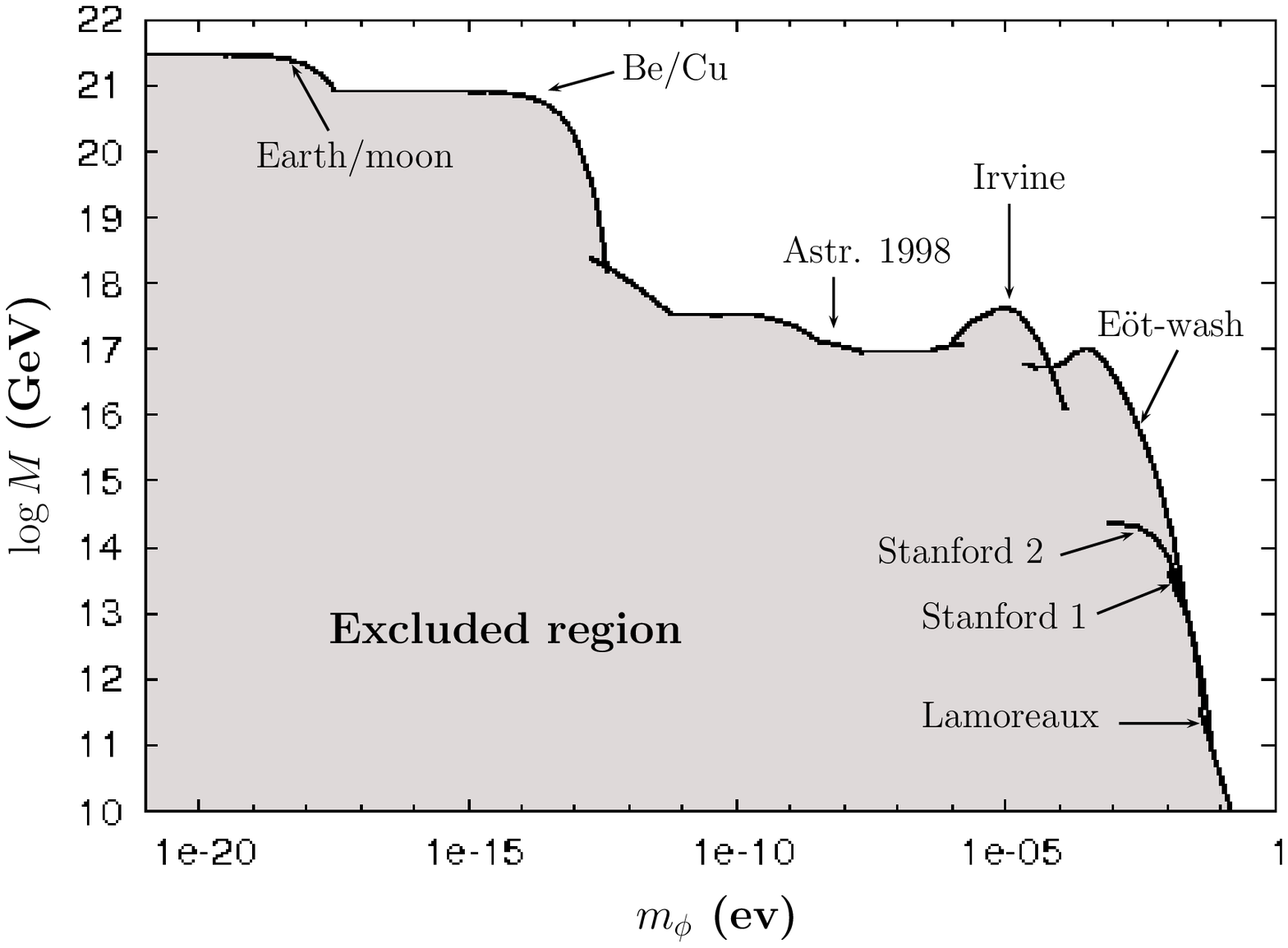}
  \caption{\small Constraints on an electromagnetic 5$^{\mathrm{th}}$ force. $\beta$ corresponds to our $\epsilon^2$
  while $\lambda=\widetilde m^{-1}$. Figure taken from \cite{Bartlett:1988yy}. Labels in the figure can be found
  in the references cited in \cite{Bartlett:1988yy}.}\label{5th_force_contraints_eps}
\end{figure}

Massive paraphotons will also couple to charged particles (recall Chapter \ref{general concerns})
as electrons and protons providing an electromagnetic type 5$^\mathrm{th}$ force. The ``paracharge"
of a standard fermion will be in this case proportional only to its electric charge, and hence
neutral bodies will also have zero paracharge. Then, paraphotons are to be looked for by providing
a net \emph{electric} charge to the test bodies. The reported negative results again place bounds
on the $\epsilon-\widetilde m$ plane that can be seen in Fig. \ref{5th_force_contraints_eps}.

\section{Summary plots}
\begin{figure}[t]\centering
\includegraphics[width=15cm]{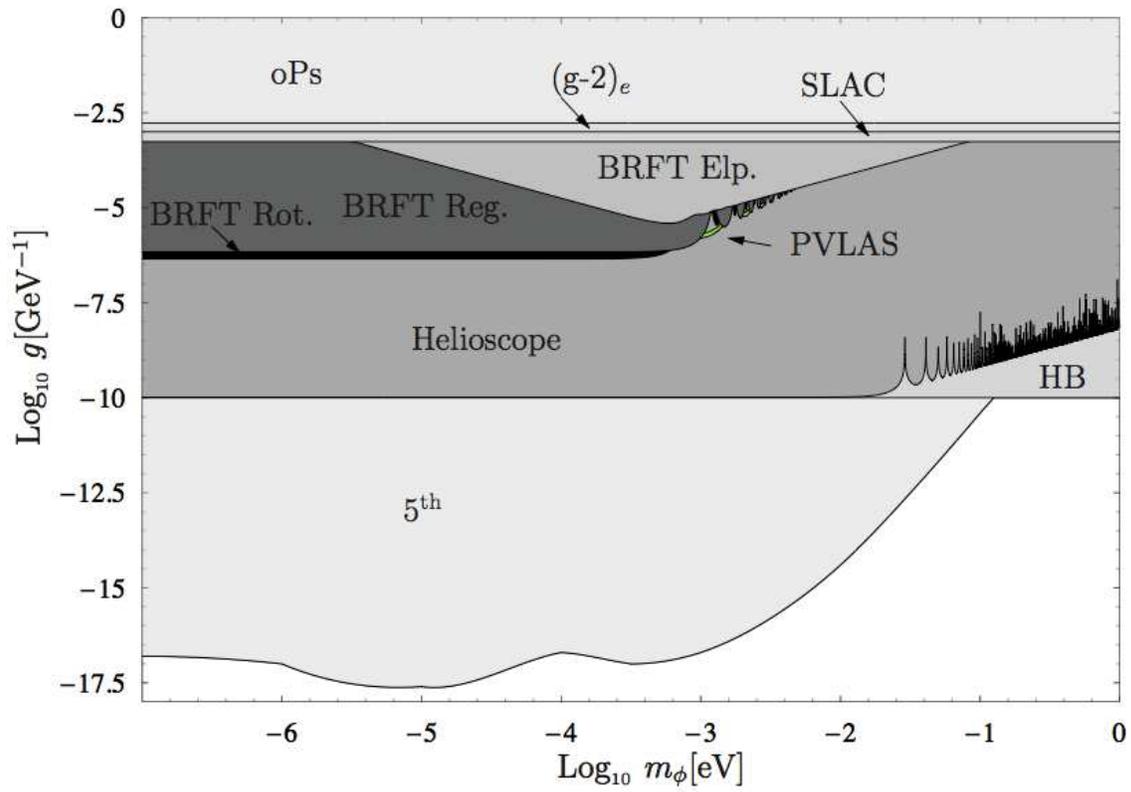}
  \caption{\small Summary plot for the constraints on ALPs. The 5$^{\mathrm{th}}$ bounds only apply for
  a parity-even ALP ($0^+$). The PVLAS signal is shown in green. \label{ALP bounds plane}}
\end{figure}

\begin{figure}[t]\centering
\includegraphics[width=13cm]{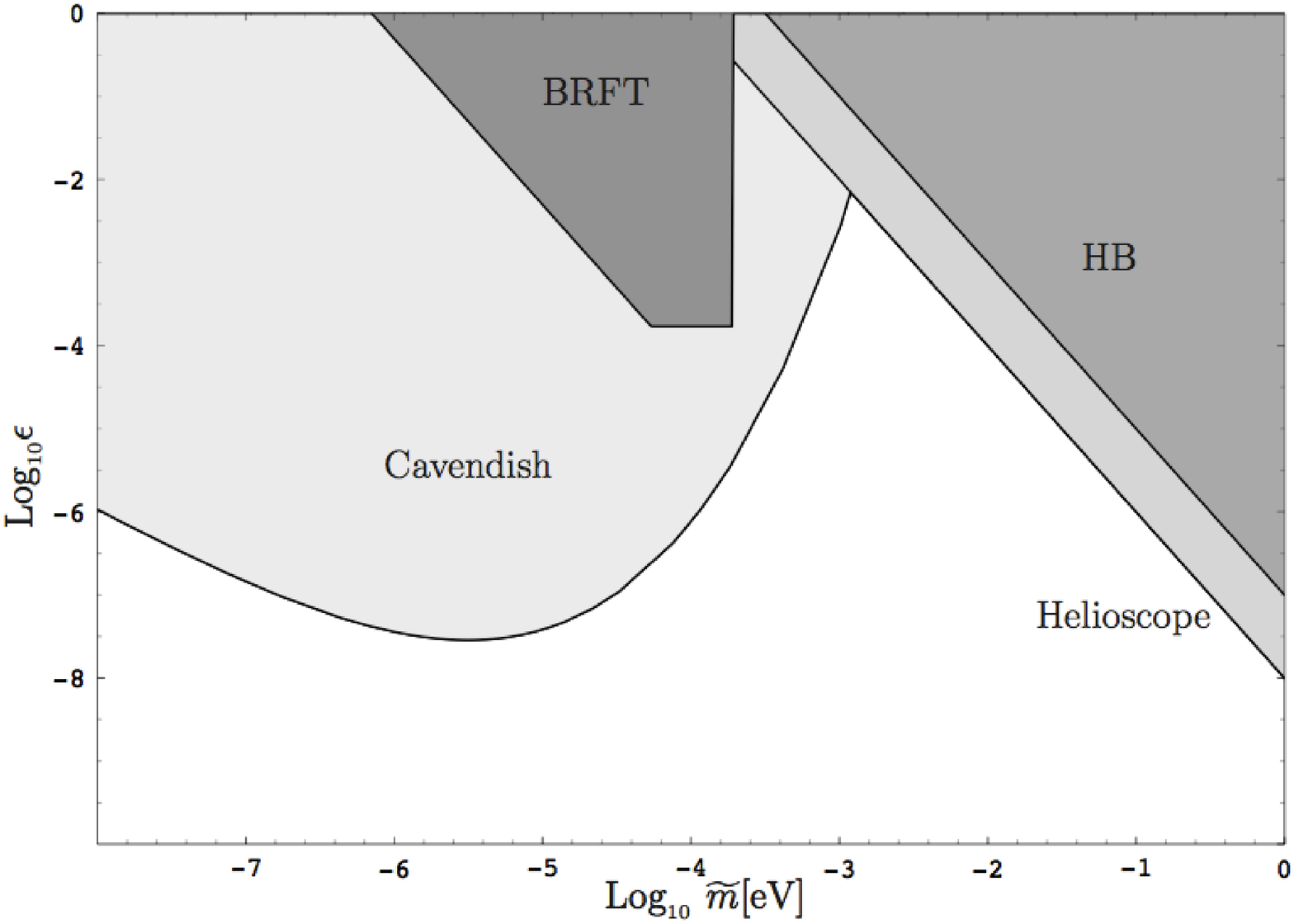}
\includegraphics[width=13cm]{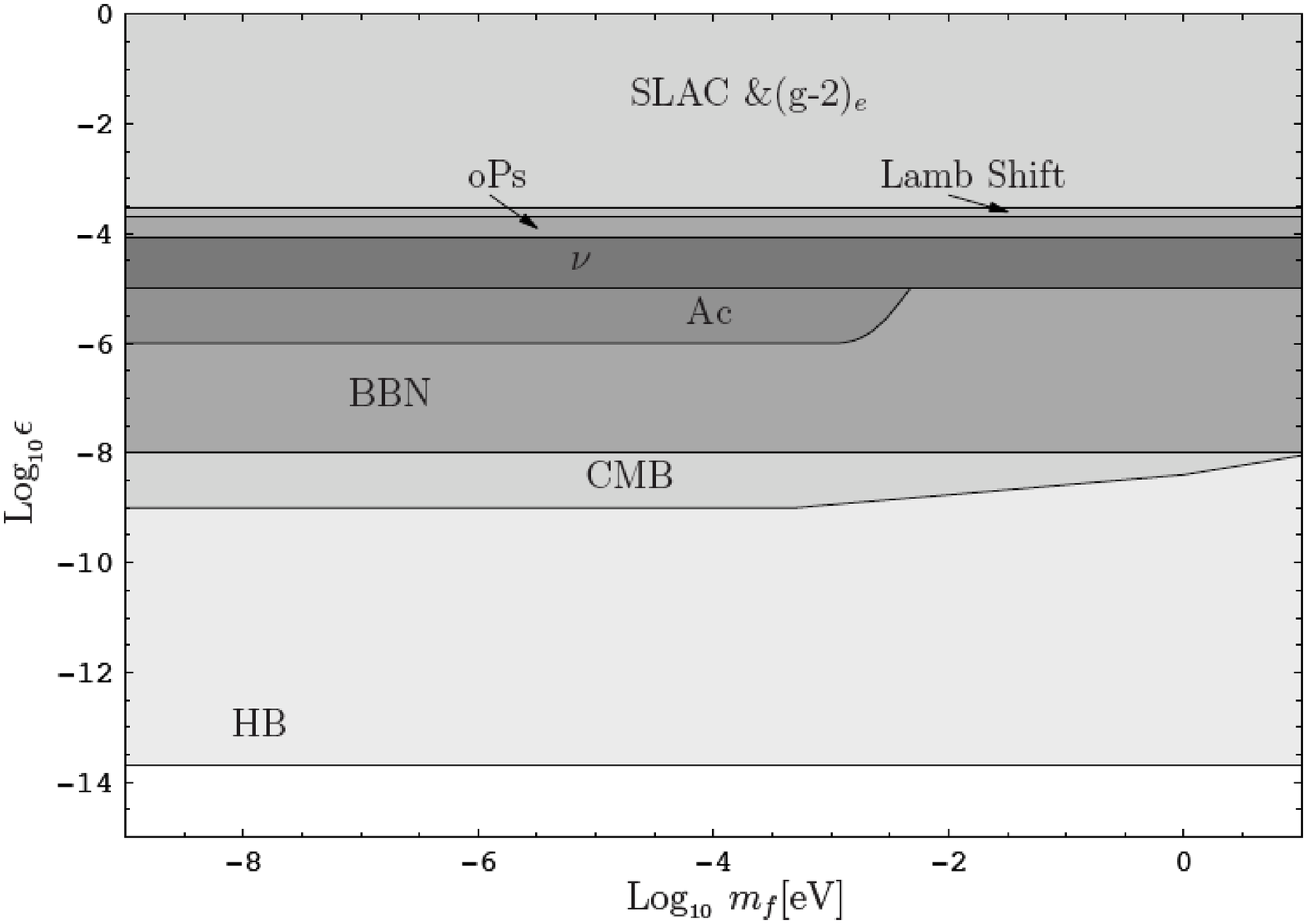}
  \caption{\small Excluded regions for the mass and coupling of paraphotons (up) and for
  spin-1/2 millicharged particles (down). For spin-0 MCPs the BBN bound disappears and the rest remain very close
  to the spin-1/2 case. See the text for details.  \label{paraphoton&MCP_bounds-plane}}
\end{figure}

In order to summarize this Chapter I display the bounds on ALPs, paraphotons and MCPs in Figures
\ref{ALP bounds plane} and \ref{paraphoton&MCP_bounds-plane}. In Fig. \ref{ALP bounds plane} we see
clearly that the hypothetical PVLAS ALP is strongly excluded by the HB lifetime argument,
Helioscope experiments and 5$^\mathrm{th}$ force searches (this last one holds only for a $0^+$
ALP). As we see in Fig. \ref{paraphoton&MCP_bounds-plane}, the MCP possibility ($m_\mathrm{MCP}<1$
eV and $\epsilon\gtrsim 10^{-6}$) is again very disfavored by the HB lifetime argument but also by
distortions in the CMB spectrum and the number of radiation species at the epoch of BBN. With the
possible exception of the 5$^\mathrm{th}$ force searches, laboratory searches are consistent with
both hypothesis, being the BRFT experiment the only one having competitive sensitivity with the
PVLAS apparatus.

\cleardoublepage
\thispagestyle{empty}


%% file: chapter4.tex
\def\baselinestretch{1}
\vspace{-1cm}
\chapter{Can the PVLAS particle be compatible with astrophysical bounds? \label{Evading}}
In Chapter \ref{constraints} I have shown that \textit{bare} ALP and MCP interpretations of the
PVLAS signal, involving each of them just a new particle and its corresponding interaction to light
are completely excluded by astrophysical arguments (a parity-even ALP is also excluded by 5$^{th}$
force experiments, as it will become clear in Chapter \ref{Light scalars coupled to photons and
non-newtonian forces}). In this thesis I have focused on the ALP interpretation and the general
question I want to answer is:

\textit{are there modifications of the ALP bare physics that avoid the astrophysical bounds?}

The answer is a modest ``yes" which I am intended to explain. However, I can advance that my
general strategy for invalidating the astrophysical bounds is to accept that these interactions are
not the end of the history. Let me state the number of hypothesis in the PVLAS-astrophysics puzzle:
\begin{itemize}\small
\item \textit{0-} $\phi$'s are elementary.
\item \textit{1-} $\phi$'s are produced by Primakoff effect in stars by the couplings \eqref{ALP-couplings}
\item \textit{2-} $m$ and $g$ have the same values in PVLAS than in stellar conditions.
\item \textit{3-} We use unperturbed standard stellar models to calculate ALP emission.
\item \textit{4-} $\phi$'s escape from the solar interior.
\item \textit{5-} We trust astrophysical calculations of evolutionary stellar timescales.
\item \textit{6-} $\phi$'s from the Sun enter into Helioscopes and convert into photons
                  by inverse Primakoff conversion.
\end{itemize}

Note that there are two kinds of hypothesis, the ones that affect our knowledge about the \textit{ALP physics}
(\textit{0,2}), completely unknown to us, (except for the existing bounds and maybe the PVLAS signal) and the ones
regarding aspects of \textit{established fields of physics} (\textit{1,3,4,5,6}).
As a natural strategy we will first have to try to adapt the new ALP to already understood physics
and \textit{not} the inverse, namely to adapt well established phenomena or theories to englobe the
bare ALP existence. This is well justified if one realizes that the ALP $\phi\gamma\gamma$ coupling
required by PVLAS \eqref{PVLAS_values for m and g} violates the astrophysical bounds by
\textit{five orders of magnitude!}. It is not very likely that a soft modification of our orthodox
ideas would leave room for the PVLAS ALP.

In this thesis I have worked and discussed two general ways of evading the astrophysical
constraints providing some new physics involving the PVLAS ALP, namely
\begin{itemize}
\item The ALPs produced in the Sun are \textit{trapped} inside,
\item The Primakoff production in stellar conditions is \textit{suppressed}.
\end{itemize}

\textit{Trapping ALPs} .- The idea consists in providing new (strong) interactions of the ALP with
some of the constituents of a stellar plasma\footnote{A further idea is that $\phi$ interacts with
dark matter accumulated in the Sun and other stars and not with the plasma. I regarded some models
without success. Another idea was presented in \cite{Jain:2005nh}.}. Instead of escaping, and
contributing largely to the stellar energy loss, the ALPs thermalize and slowly diffuse out to
contribute typically as much as photons, not compromising very seriously the evolutionary time
scales of stars.  The diffusion of novel particles is, however, also constrained by the energy
transfer argument and this imposes a new constraint on the feebleness of the novel strong
interactions, as I review in Chapter \ref{constraints}. This possibility demands reconsidering
hypotheses \textit{3} and \textit{6} so it is likely to generate some problems.

\textit{Suppression of stellar ALP production} .- This idea supposes that the $\phi\gamma\gamma$
interaction is suppressed in stellar environments with respect to the PVLAS experimental setup.
This can be achieved in several ways depending on the parameter that makes the difference. If we
accept that the ALP interaction is always of the type \eqref{ALP-couplings} we find that the
coupling $g$ in the Sun has to be roughly five orders of magnitude smaller than in PVLAS to evade
the HB bound \eqref{HB-bound_for_ALPS} and about 10 orders of magnitude smaller to evade the CAST
detection limit \eqref{CAST_bound_for_ALPs}.

I this thesis I have considered both possibilities and indeed I have built models for both of them.
However providing a strong coupling of $\phi$ to the stellar plasma constituents is not an easy
task if we examine its consequences out of the stellar environments. The second idea has showed up
to be more successful, and to my knowledge the model \cite{Masso:2006gc} provides the only known
way to reconcile PVLAS with particle physics and astrophysics. Next I dedicate a few words to
contextualize the work exposed in Chapters 5 to 8, corresponding to the four articles in which I
have published the main results of my PhD research.

The key point in all my work arises when we realize that the discrepancy between PVLAS and
astrophysics is really \textit{huge}. In order to reconcile them I believe we \textit{need} to
require that the new physics involving the PVLAS ALP behaves \textit{very differently} in the PVLAS
and the stellar environments.

If we consider $\phi$ and its coupling to two photons from an effective field theory point of view
we can interpret the scale $g^{-1}=M\sim 10^6$ GeV of the $\phi\gamma\gamma$ coupling as the
typical scale of the new physics. This is a very large scale, and the difference between the PVLAS
and stellar conditions is very small when compared to it. If this were the case the
$\phi\gamma\gamma$ interactions would be described fairly by \eqref{ALP-couplings} up to energies
of $10^6$ GeV and any chance of reconciling PVLAS and astrophysics would be lost.

From this viewpoint a reconciliation of the PVLAS ALP interpretation\footnote{Given in the terms of
Chapter \ref{PVLAS}} with the astrophysical constraints requires that:

\textit{the energy scale, $\Lambda_{NP}$, at which new physics appears giving the interactions
measured in PVLAS, \eqref{a-gamma-gamma.coupling} or \eqref{phi-gamma-gamma.coupling}, should lie
between the energies involved in the PVLAS experiment and those involved in the ALP emission from
the stars, i.e. } $\Lambda_{\mathrm PVLAS} <\Lambda_{\mathrm NP}< \Lambda_{\mathrm stars}$.
\textit{With $\Lambda_{\mathrm stars}$ of the order of keV it is clear that the new physics should
be at a very low energy scale.}

But, how to reconcile a small $\Lambda_{\mathrm NP}$ with the value from the PVLAS measurements
giving $M\sim 10^6$ GeV ? Up to now we have only found one solution that all models in this thesis
share. It requires the addition of a paraphoton $\gamma'$ that mixes kinetically with the
conventional photon $\gamma$ with a strength $\epsilon$. Then I made the hypothesis that the
effective interactions $\phi\gamma\gamma$ come from a more fundamental $\phi\gamma'\gamma'$ plus
kinetic mixing of the two gauge bosons. If the $\phi\gamma'\gamma'$ interactions come from the
equivalents to \eqref{ALP-couplings} with paraphoton field strength $F'_{\mu\nu}$ replacing the
photon field strength $F_{\mu\nu}$ and $g'$ is the coupling of the $\phi\gamma'\gamma'$ vertex then
we find that
\be  g' = \epsilon^{-2} g \sim \epsilon^{-2}\  10^{-6}\ \mathrm{GeV}^{-1} \ee
Using $\Lambda_{\mathrm NP} \sim g'^{-1} \sim 1$ eV we need to require a typical $\epsilon\sim
10^{-7}$.

\begin{figure}\centering
  \vspace{-.2cm}
  \includegraphics[width=9cm]{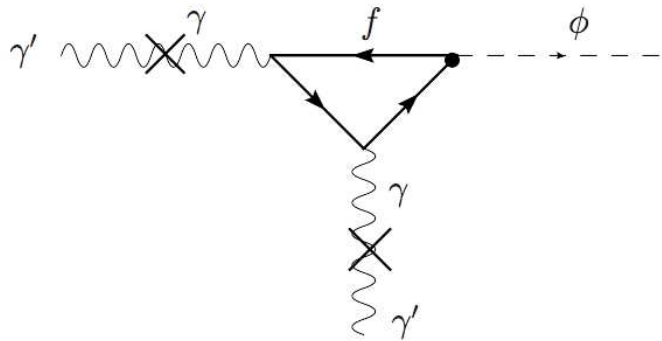}\vspace{-.2cm}
  \caption{\small The structure of the $\phi\gamma\gamma$ vertex responsible for the PVLAS signal
  in our models \cite{Masso:2005ym,Masso:2006gc}.}\label{triangle-mixing}
\end{figure}

The nature of the $\phi\gamma'\gamma'$ interaction is what makes the difference in all the models
proposed in \cite{Masso:2005ym,Masso:2006gc,Jaeckel:2006xm} (Chapters 5 to 7). While it is not
discussed in the first model of \cite{Masso:2005ym}, in the second we already introduce the
possibility that it is produced by a triangle loop in which a novel low mass fermion, $f$,
runs\footnote{We use a fermion for simplicity and because low mass fermions are more natural from
the effective field theory viewpoint.} (See Fig. \ref{triangle-mixing}). We couple $\phi$ directly
(and only) to this fermion, which is paracharged. This triangle loop  is also at the heart of the
model presented in \cite{Masso:2006gc}. Finally, in \cite{Jaeckel:2006xm} we perform a general
analysis without specifying the origin of the $\phi\gamma\gamma$ interaction.

All this reasoning holds if the important parameter is the \textit{energy} scale at which the
vertex $\phi\gamma\gamma$ is probed. But it might be as well that other environmental parameters
would be the responsible for the suppression. For instance in \cite{Mohapatra:2006pv} a model is
proposed in which at the high temperatures of the stellar interiors the new physics overcome a
phase transition that turns the $\phi\gamma\gamma$ coupling to zero. The temperature plays here the
role of the energy in the above discussion and we should require $T_{PVLAS}<T_0<T_{stars}$. Further
examples are provided for instance in Chameleon models \cite{Khoury:2003aq}. There, the mass of a
scalar field, the chameleon, turns out to be dependent of the energy density in this point, which
in a non-relativistic universe is nothing but the matter density. Let our $\phi\gamma\gamma$
coupling depend on such a field and we will get a matter density dependent coupling.

Moreover, the same model proposed in \cite{Masso:2006gc} (Chapter \ref{Compatibility of CAST search
with axion-like interpretation of PVLAS results}), albeit with some subtleties, produces a coupling
which depends on the plasma frequency $\omega_P$.

Needless to say, all the values of these parameters, temperature, matter density, photon plasma
mass, and others, are very different in the PVLAS and stellar conditions (See Table
\ref{PVLAS_SUN_HB}) so hopefully models can be built in which such a difference is the responsible
for the suppression of the production of ALPs in stars. These very interesting possibilities could
be studied altogether in a model independent way and I have performed such an analysis in
\cite{Jaeckel:2006xm} focusing on the case of the Sun. The reason has already been mentioned, given
a nonstandard suppression of the Primakoff ALP flux, the CAST bound scales with the fourth root
while energy loss arguments only with the square of the suppression. This work is presented in
Chapter \ref{The Need for Purely Laboratory-Based Axion-Like Particle Searches}.

Finally, in Chapter \ref{Light scalars coupled to photons and non-newtonian forces} we noticed that
if the PVLAS ALP is parity-even the $\phi\gamma\gamma$ vertex will induce radiatively long range
forces between macroscopic unpolarized bodies. In this case, our work gives the most demanding
bound for the ALP coupling, around 5 orders of magnitude more stringent than the best astrophysical
bounds for the mass suggested by the PVLAS experiment $m\sim 1$ meV. In the same paper we show that
our model \cite{Masso:2006gc} avoids this constraint.


%% file: JCAP.tex
\chapter{Evading astrophysical constraints on axion-like particles}
\label{Evading Astrophysical Constraints on Axion-Like Particles} \vspace{-.5cm}
\begin{flushright} {\footnotesize (In collaboration with
E.~Mass{\'o}. Published in JCAP \textbf{0509}, 015 (2005) \cite{Masso:2005ym}}) \end{flushright}

\section{Introduction}
An pseudoscalar axion-like particle $\phi$ coupled to photons
\begin{equation}
{\cal
L}_{\phi\gamma\gamma}=\f{1}{4M}F^{\mu\nu}\widetilde{F}_{\mu\nu}\p
\label{eq:1a}
\end{equation}
($F^{\mu\nu}$ is the electromagnetic field tensor,
$\widetilde{F}_{\mu\nu}$ its dual, and $\phi$ the axion-like
field) would be able to transform into photons when electromagnetic
fields are present
\begin{equation}
{\cal
L}_{\phi\gamma\gamma}=\f{1}{M}\overrightarrow{E}\cdot\overrightarrow{B}\,
\p\label{eq:1b}
\end{equation}
This is completely analogous to the well known Primakoff effect that involves the $\pi^0\gamma\gamma$ coupling. In this
paper we will be interested in the case that $\p$ is very light since then a number of interesting effects may happen.

When the $\p$ mass $m_\p<T_c$, with $T_c \simeq 1 - 10$ keV the
typical temperature of stellar cores as those of the Sun or
horizontal-branch stars, $\p$ particles are produced
by the Primakoff-like effect due to
the interaction (\ref{eq:1a},\ref{eq:1b}). If one further
assumes that the produced $\p$ flux escapes freely from the star
and thus constitutes a non-standard channel of energy-loss,
the strength of the interaction (\ref{eq:1a},\ref{eq:1b}) can be
bounded using observational data on stellar evolution time
scales \cite{Raffelt:1999tx}. For the Sun, one has the limit
\begin{equation}
M \gtrsim 4 \times \E{8}\ \mathrm{GeV} \label{sun}
\end{equation}
There is general agreement that these arguments applied to globular
clusters lead to an even stronger bound
\begin{equation}
M \gtrsim \E{10}\ \mathrm{GeV} \label{eq:2}
\end{equation}
(Here and thereafter, we understand that the bounds are on the
absolute value of $M$.)

In the allowed range for $M$, axion-like particles are still produced in the Sun and a calculable
flux reaches the Earth \cite{vanBibber:1988ge,Creswick:1997pg}. A proposal to detect this flux was
given in a pioneer paper by Sikivie \cite{Sikivie:1983ip}. The idea is that the interaction
(\ref{eq:1a},\ref{eq:1b}) allows solar $\phi$'s to transform back into X-ray photons in a cavity
with an external magnetic field. Such helioscopes have been built and limits on $M$ have been
obtained from the non-observation of this inverse Primakoff process in the cavity
\cite{Lazarus:1992ry,Moriyama:1998kd,Inoue:2002qy,Zioutas:2004hi}. The last result, from the CERN
Axion Solar Telescope (CAST) collaboration \cite{Zioutas:2004hi}, is
\begin{equation}
M > 0.9 \times\E{10}  \ \mathrm{GeV}   \label{eq:3}
\end{equation}
($95\%$CL) comparable to the bound based on the stellar energy loss arguments, eq.(\ref{eq:2}). The
strong limit (\ref{eq:3}) can be established when a coherent $\p\go\g$ signal is expected, which
happens for  $m_\p \lesssim 0.02$ eV. The CAST prospects \cite{Zioutas:2004hi} are to further
improve (\ref{eq:3}) and to extend the results to masses $m_\p$ up to 1 eV.

The research we present in this letter has been motivated by the observation  of a rotation of the
polarization plane of light propagating through a transverse, static, magnetic field by the PVLAS
collaboration \cite{Zavattini:2005tm}. A possible interpretation of this result is the existence of
a light axion-like particle $\p$ coupled to two  photons \cite{Maiani:1986md}. However, if
interpreted this way the scale appearing in (\ref{eq:1a},\ref{eq:1b}) must be
\begin{equation}
M \simeq 4 \times\E{5}  \ \mathrm{GeV}   \label{eq:pvlas}
\end{equation}
It results in such a strong coupling that it is in contradiction
with the bounds (\ref{eq:2}) and (\ref{eq:3}).
Yet, it is consistent with
the bounds coming from particle physics experiments  \cite{Masso:1997ru}.

Let us stress that if there exists a particle with the coupling $M^{-1}\simeq 2.5 \times\E{-6}  \,
\mathrm{GeV}^{-1}$ as given in (\ref{eq:pvlas}),
 it definitely cannot be the standard QCD
axion. The naming "axion-like" we use in the paper refers to the particle being
very light and to its pseudoscalar nature, reflected in the form of the  interaction
(\ref{eq:1a}).

Since at present there is no an alternative explanation of the PVLAS
data, we are faced to the challenge of finding a consistent model
that could explain the constraints
(\ref{eq:2}),  (\ref{eq:3}), and (\ref{eq:pvlas}) in terms of a light particle coupled
to two photons.
The route we have followed has been investigating ways to evade the
astrophysical bounds.
 We have worked out two
 possibilities that could solve the problem. The first is that
$\phi$-particles are indeed produced in the Sun but that they interact
so strongly
 that are trapped by the solar medium. Then, the energy of the
 emitted $\p$-particles is much lower than in the usual free-streaming
 regime and thus the CAST telescope is not able to detect them.
In Section \ref{trapping} we propose a simple model with paraphotons
that provides a way $\p$-particles are trapped. However, we
will see that it leads to too strong photon-paraphoton interactions that are
not consistent with other observations.
Even having this problem, we present the
model because, first of all, it remains to be seen
whether a sophistication of these ideas may lead to a consistent model.
Second, some of the issues we are faced are helpful in Section
\ref{suppression}.
The second possibility we examine is that
 the Primakoff process is suppressed when occurring
in a stellar medium.
 Then, there would be far less $\p$-particles emitted than expected.
We discuss in Section \ref{suppression} how a composite $\phi$ and
the corresponding form factor
of the $\p\g\g$ vertex could be responsible
for such a suppression of the
$\p$-flux. Finally, we present our conclusions and additional comments
in Section \ref{conclusions}.

 \section{Trapping regime} \label{trapping}

 Let us start analyzing the possibility that $\p$-particles are produced
in the solar core but that interact so strongly with the medium
that their fate is analogous to what happens to the stellar photons, namely,
they abandon the Sun after a lot of interactions, having
followed a random walk path. In this trapping regime, local
thermodynamic equilibrium applies and $\p$ would contribute to the
radiative energy transfer. The total opacity, including the exotic contribution,
\begin{equation}
k_{total}^{-1}=k_\g^{-1}+k_\p^{-1} \nonumber
\end{equation}
should not be much different from the standard solar opacity $k_\g^{-1} \approx 1$  g/cm$^2$, if we
do not want to ruin the standard solar model. Specifically, one imposes
\cite{Raffelt:1988rx,Carlson:1988jg}
\begin{equation}
k_\p^{-1} \lesssim 1 \f{\mathrm g}{{\mathrm cm}^2} \label{eq:5}
\end{equation}

The key point of course is to try to implement this possibility within a particle physics model.
The scenario we shall examine assumes paraphotons provide the trapping interaction. These vector
particles were proposed by Okun in \cite{Okun:1982xi} (see also \cite{Georgi:1983sy}) and further
developed by Holdom in \cite{Holdom:1986eq,Holdom:1985ag}. The basic idea is a modification of QED
that consists in adding an extra $U(1)$ abelian gauge symmetry. If $j_\mu$ is the electromagnetic
current involving charged particles $j_\mu \sim \bar{e}\g_\mu e + ...$ we start with the lagrangian
\begin{equation}
{\cal L}_0 =-\f{1}{4}F_1^{\mu\nu}F_{1\mu\nu}+ e_1j_\mu A_1^\mu
\end{equation}
This lagrangian has a $U(1)$ gauge symmetry group, and would be the
photon part of the QED lagrangian. The paraphoton
model assumes two groups $U(1)_1\times U(1)_2$ as the gauge symmetry, so that
one has two gauge fields $A_1$ and $A_2$.

In the line of \cite{Holdom:1986eq,Holdom:1985ag} we will assume that there are very massive
particles carrying charges under both $U_1$ and $U_2$ groups. At low energies, these massive
particles running in loops can be integrated out leaving the lagrangian
\begin{eqnarray}
&{\cal L}
=-\f{1}{4}(1+2\ep_{11})F_1^{\mu\nu}F_{1\mu\nu}-\f{1}{4}(1+2\ep_{22})F_2^{\mu\nu}F_{2\mu\nu} \nonumber\\
&+\f{1}{2}\ep_{12}F_1^{\mu\nu}F_{2\mu\nu}+ e_1j_\mu A_1^\mu
\label{eq:6}
\end{eqnarray}
The parameter $\ep_{12}$ is the induced mixing in the kinetic
terms, and $\ep_{11}$ and $\ep_{22}$ are also modifications due to these
loops. At first order in the small $\ep$-parameters,
we define new fields that
diagonalize and normalize  the kinetic terms,
\begin{eqnarray}
&A_\mu = (1+\ep_{11})A_{1\mu} \label{eq:7} \\ \nonumber &A'_\mu =
(1+\ep_{22})A_{2\mu} - \ep_{12}A_{1\mu}
\end{eqnarray}
We end up with the photon $A_\mu$ coupled to charged particles,  with
$e=e_1(1-\ep_{11})$, and with the paraphoton $A'_\mu$
\begin{equation}
{\cal L}
=-\f{1}{4}F^{\mu\nu}F_{\mu\nu}-\f{1}{4}F'^{\mu\nu}F'_{\mu\nu}+ej_\mu
A^\mu \label{eq:8}
\end{equation}

Different authors have added some physics to (\ref{eq:6}) and (\ref{eq:8}) so that phenomenological
consequences arise. In \cite{Okun:1982xi} and \cite{Georgi:1983sy} the effects of a paraphoton mass
were discussed. In \cite{Holdom:1986eq,Holdom:1985ag}, it was shown that the existence of light
particles having $U_2$ charge leads to these particles having an electric charge of size $\ep e$.
In \cite{Foot:2000vy} the paraphoton was identified with a mirror photon and some implications were
analyzed. The most recent work \cite{Dobrescu:2004wz} considers higher-order operators to describe
the interaction of the paraphoton with matter.

What we propose is to add to ${\cal L}$ in (\ref{eq:6}) the interaction
\begin{eqnarray}
{\cal L}_{\p\g_2\g_2}=\f{1}{4
M_2}F_2^{\mu\nu}\widetilde{F}_{2\mu\nu}\p
\end{eqnarray}
with $M_2$ a low energy scale. The axion-like particle is
therefore strongly coupled to the $U_2$ gauge boson. After
diagonalizing (\ref{eq:7}) we get a strong coupling of $\p$ to
paraphotons
\begin{eqnarray}
{\cal L}_{\p\g'\g'}=\f{1}{4
M_2}F'^{\mu\nu}\widetilde{F}'_{\mu\nu}\p
\end{eqnarray}
a weaker coupling with a mixed term
\begin{eqnarray}
{\cal L}_{\p\g\g'}=\f{\ep_{12}}{2
M_2}F^{\mu\nu}\widetilde{F}'_{\mu\nu}\p
\end{eqnarray}
and finally we get a term that couples $\p$ to photons, i.e., an
interaction as in (\ref{eq:1a}) with the identification
\begin{eqnarray}
\f{1}{M}=\f{\ep_{12}^2}{M_2}\label{eq:9}
\end{eqnarray}
with $M$ giving the strength (\ref{eq:pvlas}).

We now have the necessary ingredients to have a large opacity of
$\p$ in the solar medium. The dominant contribution to the opacity
comes from the process
\begin{eqnarray}
\p\g\go\p\g'\label{A}
\end{eqnarray}
where a virtual $\g '$ is exchanged.
The secondary paraphotons are further scattered
\begin{eqnarray}
\g'\g\go\g'\g'\label{B}
\end{eqnarray}
where now a $\p$ is exchanged.
In both reactions, (\ref{A}) and (\ref{B}), the initial $\g$ is
of course from the stellar plasma.

Having exposed the main idea, we proceed to the calculation
of the opacity, where
we shall content ourselves with order of magnitude estimates.
The head-on collision in (\ref{A}) has a total cross-section
\begin{eqnarray}
\sigma_\p=  \f{5}{384\pi} \left(\f{\ep_{12}}{M_2^2}\right)^2 s \label{eq:17}
\end{eqnarray}
with $s=(p_\p + p_\g)^2$. The total cross-section for (\ref{B}) is
\begin{eqnarray}
\sigma_{\g'}= \f{5}{768\pi} \left(\f{\ep_{12}}{M_2^2}\right)^2 s \label{eq:18}
\end{eqnarray}
with $s=(p_\g + p_{\g'})^2$. In (\ref{eq:17}) and (\ref{eq:18}) all particle masses are neglected in front of $s$.
To estimate the opacity we set
$s\simeq 4\langle E_\g^2\rangle $ and $\langle E_\g^2\rangle\simeq\,
10.3\,T^2$,
where $T$ is the temperature of the medium. We get
\begin{equation}
\langle \lambda_\p\rangle\simeq \f{1}{\sigma_\p n_\g}\simeq 5\times\e{7}\ep^{-2}_{12}\left(
\f{M_2}{\mathrm keV}\right)^4 \left( \f{T}{\mathrm keV}\right)^{-5} \,{\mathrm cm}
\label{lambdaphi}
\end{equation}
and
\begin{equation}
\langle \lambda_{\g'}\rangle\simeq \f{1}{\sigma_{\g'} n_\g}\simeq 1\times\e{6}\ep^{-2}_{12}\left(
\f{M_2}{\mathrm keV}\right)^4 \left( \f{T}{ \mathrm keV}\right)^{-5} \, {\mathrm cm}
\label{lambdaparaph}
\end{equation}

Requiring a large enough opacity, eq.(\ref{eq:5}), for the
conditions of the Sun core, $T\simeq 1\,$keV, $\rho\simeq
100\,$g$\,$cm$^{-3}$, we are lead to
\begin{equation}
\f{\ep_{12}}{M_2^2}\gtrsim \f{10^{-3}}{{\mathrm keV}^2} \label{C}
\end{equation}
This condition comes from the reaction (\ref{B}); the process
(\ref{A}) gives a weaker condition.

In our model, the Sun is a copious emitter of low energy axion-like particles and paraphotons.
However, there could be no axion-like particles reaching the Earth, because of the decay
$\p\go\g'\g'$. The lifetime of $\p$ with energy $E_\p\sim 3T_{\mathrm escape}$ is
\begin{equation}
\tau_\p = 1.3 \times 10^{-7} \left( \f{m_\p}{\mathrm eV} \right)^{-3}
 \left(\f{M_2}{\mathrm keV}  \right)^2  \left(\f{E_\p}{m_\p}  \right)\,
{\mathrm s}
\end{equation}
$\p$ would decay
before reaching the Earth when the parameters of our model are such that
$\tau_\p<500\,$s. In this case, only paraphotons, from emission or decay,
would survive the journey from the Sun to the Earth.

Using (\ref{eq:9}) and the experimental value (\ref{eq:pvlas})
 and then imposing condition (\ref{C}) we find the allowed
 values for  $M_2$ and $\epsilon_{12}$.
There is a maximum value for the mixing $\epsilon_{12} \lesssim  5\times10^{-7}$, and also a
maximum value for the scale $M_2\lesssim 25 \ {\mathrm eV}$.

Let us now discuss the cosmological constraints on the new
interactions. In the early universe, production of paraphotons
proceeds trough the reaction
\begin{equation}
\g\g\go\g'\g'\label{production}
\end{equation}
The interaction rate $\Gamma$ has to be compared to the expansion rate
$H$ of the universe to see whether the process (\ref{production}) is effective.
The calculation is similar to the one leading to (\ref{lambdaphi}) and
 (\ref{lambdaparaph}). Assuming a matter-dominated universe,
we have
\begin{equation}
\f{\Gamma}{H} \simeq \f{1}{200} \, \left( \f{\epsilon_{12}}{10^{-7}}\right)^4 \, \left( \f{\mathrm
eV}{M_2}\right)^4 \, \left( \f{T}{\mathrm eV}\right)^{7/2} \label{gh}
\end{equation}
Clearly, for the values of the parameters $M_2$ and $\epsilon_{12}$ discussed before and for $T>$ 1 eV, $\Gamma/H>1$ and thus a cosmic background of paraphotons will be born (when it is radiation that dominates, (\ref{gh}) has to be modified, but we reach the same conclusion). Once there is a $\g'$ population, the situation is catastrophic since the interaction (\ref{A}) is only
$\epsilon_{12}^2$-suppressed while (\ref{production}) is
$\epsilon_{12}^4$-suppressed.  As a consequence photons and paraphotons
would be in equilibrium for $T<$ 1 eV, in contradiction with the observation
of having a transparent universe for these low temperatures.

There might be other constraints on $\g-\g'$ interactions at high energies
coming from example from photon-photon interactions in accelerators.
However, here we should consider the issue that the vertex
could be subject to form factor effects. We will discuss about this topic in
the next Section.

\section{Suppression of the solar production} \label{suppression}

Let us investigate now a framework where the production in stellar cores is considerably
diminished. A first thing to notice is that  we should look at (\ref{eq:1a}) as an effective
lagrangian and consequently we should not expect it to be valid at arbitrarily high energies. The
well studied $\pi^0\g\g$ vertex is similar to (\ref{eq:1a}) and it is useful as a guideline. The
crucial point is that when one of the photons (or both) is off mass-shell the effects of the
$\pi^0$-photon transition form factor become manifest.

There are indeed a variety of measurements where the transition form factor of pseudoscalar mesons
can be observed, from moderate $q^2$ up to large momentum transfer
\cite{Dzhelyadin:1980kh,Aihara:1990nd,Behrend:1990sr,Gronberg:1997fj}. Let us emphasize that the
appearance of a form factor is expected on general grounds. From the theoretical point of view,
apart from the phenomenological VMD parameterization, one gets a form factor when using a
quark-triangle model \cite{Bramon:1981sw,Ametller:1983ec}, when calculating in  perturbative QCD
and when using some other methods
\cite{Lepage:1980fj,Ametller:1991jv,Dubnickova:2004ks,deMelo:2003uk,Xiao:2005af}.  All these
approaches are consistent among themselves and are able to fit the data. For example, when the
$\pi^0\g\g$ vertex is described by a quark triangle loop with off-shell photons, the explicit
calculation of the diagram leads to a form factor that can be identified with VMD provided one
assigns constituent masses to the internal up and down quarks \cite{Bramon:1981sw,Ametller:1983ec}.
Then, for high $q^2$ one has a suppression $M_\rho^2/q^2 \sim M_{u,d}^2/q^2 $.


These facts have encouraged us to postulate that the axion-like particle $\p$
is a confined bound-state of quark-like particles, that we will call preons in accordance with tradition. If for simplicity we consider one fermion $f$ as the only preon, $\p$ would be the $J^P=0^-$ $\bar f f$
bound state and the coupling to two photons would proceed through a triangle
loop with $f$ circulating in it. This would result in the appearance
of a form factor effect at high energies. When both photons are on-shell there is no suppression; these are the conditions in the PVLAS experiment and in the detection setup in CAST. However, in the solar medium there would be a suppression of the
$\p$ emission rate.

Let us calculate which is the required suppression $F$ in the Primakoff amplitude for having a
consistent scenario. If we call $M_{\mathrm pvlas}$ the value in (\ref{eq:pvlas}) and $M_{\mathrm
cast}$ the lower bound in  (\ref{eq:3}), we should have
\begin{equation}
\left[| F |^2  \f{1}{M_{\mathrm pvlas}^2} \right] \  \f{1}{M_{\mathrm pvlas}^2} \  < \ \left[
\f{1}{M_{\mathrm cast}^2} \right] \  \f{1}{M_{\mathrm cast}^2} \label{Fcondition}
\end{equation}
where in square brackets there is the relevant factor referred to production in the Sun and outside the brackets the factor corresponding to detection in CAST. In the lhs we assume there is suppression, while in the rhs we assume none because the CAST limit is obtained assuming no form  factor suppression in the solar production. Introducing numbers we obtain
\begin{equation}
| F |   \  < \ 2 \times 10^{-9}
\label{Fbound}
\end{equation}

We now turn our attention to the theoretical prediction for $F$, that we obtain from the
calculation of the preon-triangle diagram amplitude. For invariant masses $s_1$ and $s_2$ of the
photons, and values of the masses of $\p$, $m_\p$, and the internal fermion $f$, $M_f$, the
amplitude $F(s_1,s_2,m_\p;M_f)$ can be put in terms of dilogarithms
\cite{Bramon:1981sw,Ametller:1983ec}. Let us comment that $F$ is in general a complex quantity and
also that, as a form factor, we normalize it as $F(0,0,m_\p;M_f)=1$.

The values for $s_1$ and $s_2$ in the solar core will be in the keV range. Indeed, in the interior
of the Sun the Primakoff production is started by a photon of the  thermal bath with approximately
$ \omega_P^2 \simeq (0.4\, {\mathrm keV})^2 \simeq s_1$, with $\omega_P$ the  plasma frequency. The
virtual photon connecting the vertex to a proton (or to any charged particle) is subject to
screening effects, as discussed in \cite{Raffelt:1985nk}. These effects amount to cut the momenta
contributing to the Primakoff effect with the Debye-H{\"u}ckle scale $k_{DH}$, that in the solar
core is $k_{DH}^2 \simeq (9\, {\mathrm keV})^2 \simeq s_2$.

Provided the mass $M_f$ is much less than $s_1$ and $s_2$, we obtain  a strong suppression
compatible with (\ref{Fbound}). With the values of $s_{1,2}$ mentioned above and for $m_\p \lesssim
10^{-3}$ eV (these are the values for which a coherent effect in vacuum is expected  in the PVLAS
setup) we obtain numerically that $F$ satisfies (\ref{Fbound}) for
\begin{equation}
M_f  \lesssim  2 \times 10^{-2}\ {\mathrm eV} \label{Mf}
\end{equation}
To see a bit more clearly how the suppression arises, we have verified that the exact value for
$F$, in the limit $|s_2| \gg |s_1| \gg M_f \gg m_\p$ has the behavior
\begin{equation}
|F| \sim 10^2 \, \f{(2M_f)^2}{|s_2|}
\label{assy}
\end{equation}
Thus, $M_f$ plays the role of  the cut-off energy scale of the $\p\g\g$ vertex form factor. The scale of new physics is again a low energy scale.

Let us comment that, before, we have identified $k_{DH}^2$ with $s_2$ and that it is an approximation since the $t$-channel carries other momenta. However, $ |s_2| \gtrsim
k_{DH}^2$  always, so that, at the view of (\ref{assy}), the approximation is conservative.

There is a parameter that we have not discussed, namely
the electric charge $q_f e$ of the preon $f$. With the coupling
\begin{equation}
\f{1}{2 v}\, \bar f \gamma_\mu  \gamma_5 f \, \partial^\mu \p
\label{pff}
\end{equation}
the result of the calculation of the triangle for on-shell photons is
\begin{equation}
N\, \f{q_f ^2 \, \alpha}{\pi\, v} = \f{1}{M} \simeq  \f{1}{4\times 10^{5} \, \mathrm{GeV}}
\label{identify}
\end{equation}
where we have already identified the result with the coupling $1/M$ in
(\ref{eq:1a}) and with the experimental value in (\ref{eq:pvlas}). Also,
we have introduce a factor $N$ corresponding to having a confining $SU(N)$
gauge group in the preon sector.

Now we have to take into account the bound on light millicharged particles
\cite{Davidson:1991si,Davidson:2000hf} coming from BBN constraints,
\begin{equation}
q_f \lesssim 2 \times 10^{-9}
\label{qf}
\end{equation}
In a paraphoton model such a small electric charge could arise naturally
\cite{Holdom:1986eq,Holdom:1985ag}. Together with (\ref{identify}) it implies again low energy
scales
\begin{equation}
v  \lesssim 10^{-5} \, {\mathrm eV} \label{qf}
\end{equation}
(We have used $N=3$).

Let us point out that the $SU(N)$ gauge group should be totally independent of the color $SU(3)_c$ standard gauge group. Otherwise, among other undesired consequences, we would have in nature hadrons with charges near $\pm (2/3)e$, $\pm (1/3)e$,  and  $\pm (4/3)e$ that would form when binding a preon or antipreon with
quarks or antiquarks, of the kind $\bar u f$, $uuf$, etc.
Also, at tree level we should take $f$ as a singlet under the standard model,
with the small electric charge arising as a higher order effect.
With all these assumptions, we think the new force and particles could have been not noticed in other experiments. Yet, there are consequences, like the existence of bound states with higher spin, as for example a state with $J=1$ that would be unstable since it would decay into $\p \g$. Of course, from a phenomenological perspective,
it would be interesting to look for signals of the preon model. From a more theoretical point of view, here we do not
attempt to build a full model, for we think it would be premature.
Rather, we have shown a possible way to evade the astrophysical limits
on axion-like particles.

\section{Conclusions}\label{conclusions}

A recent review by Raffelt \cite{Raffelt:2005mt} emphasizes that the PVLAS result
(\ref{eq:pvlas}), interpreted in terms of a new light axion-particle coupled to photons,
would lead to the Sun burning much faster that what is actually observed.
There is the pressing issue of explaining the results in a consistent model.
Of course an independent check of the results would be most welcomed;
in fact there are interesting proposals for such type of laboratory
experiment where a high sensitivity would be reached
\cite{Ringwald:2003ns}.

In this paper, we report the work we have done trying to evade the astrophysical bounds and thus accommodating a light particle coupled to two photons.
The astrophysical limits assume 1) a flux calculated with the interaction
(\ref{eq:1a},\ref{eq:1b}) in the stellar core, and 2) that the produced particles escape the star without further interaction.

Our first attempt has been trying to find a model where 2) is not true. In our paraphoton model
with (\ref{C}), axion-like particles $\p$ are trapped in the stellar interior and so are the
paraphotons $\g'$ produced by $\p\g$ scattering. The large opacity makes $\p$ production not a
problem. It follows that the astrophysical limit (\ref{eq:2}), that assumes $\p$ freely escapes,
is no longer valid. These arguments have nothing to say about an axion-like interpretation of the
PVLAS result, because it is an Earthbound laboratory experiment, with $\p$ produced and detected in
the laboratory.

The model, however, leads to photons to interact
with paraphotons so strongly that it is excluded, at least in the simple framework
we have exposed where a cosmological background density of paraphotons
emerges.
Perhaps a more elaborated model with paraphotons, or another
model with a different
strong interaction can do the job of trapping particles
in the Sun without entering in conflict with other experiments.
Let us point out that
there are also astrophysical constraints on the $\p\g\g$ coupling
coming from red giants and from SN observations.
As far as the SN is concerned, the low
value (\ref{eq:pvlas}) makes $\p$ to be trapped in the SN core
in such a way that one does not need extra interactions \cite{Masso:1997ru}.
In any case, the exercise we have presented in Section \ref{suppression}
shows that it is not trivial to evade the astrophysical constraints.

The astrophysical bound could also be evaded if the $\p\g\g$ vertex, while fully operating at PVLAS
energies, is suppressed in the conditions of stellar interiors. In this case the condition 1) above
does not hold. We have explored the possibility that $\p$ is a composite particle and has a form
factor leading to a suppression of the production. We have shown that this scenario is able to
explain the puzzle. Our ideas are highly speculative since they involve preons with a new confining
force and probably a minuscule electric charge,  but notice that we have been inspired by the pion
and the $\pi^0\g\g$ vertex, that after all have the nice property of being real.  In any case, it
would be crucial to look for other phenomenological consequences of the preon model. We have not
found any that rules out it obviously.

Either in the case of a strong interaction leading to trapping or
of a suppression of the production,
the astrophysical bounds on axion-like particles could be evaded.
If indeed they are evaded,
there are drastic consequences for CAST, since then
the non-observation of X-rays does not imply a limit such as
(\ref{eq:3}).

Our main conclusion is that the explanation of the PVLAS result in terms of a light particle
coupled to photons is not necessarily in contradiction with other experiments and observations. Let
us emphasize that taking alone the PVLAS data, if interpreted in terms of new light particle
coupled to photons, it would already mean an interesting piece of new physics. But there is even
more. The result, taken together with the astrophysical limits and the CAST data, means that we
have to go beyond the "mere" existence of a new pseudoscalar $\p$ coupled to photons and even more
exotic physics has to be invoked. In the scenarios of the sort we have proposed the new physics
scale is at very low energies. If confirmed, it would be an exciting discovery. Otherwise, if
finally the models that try to evade the astrophysical constraints are shown not to be valid, the
situation will be no less exciting since an alternative explanation for the PVLAS laser experiment
result will be needed.

\section{Acknowledgments} We thank Carla Biggio, Zurab Berezhiani, Albert Bram{\'o}n, Giovanni
Cantatore, and Sergei Dubovsky for useful discussions. We are specially indebted to Georg Raffelt
for insightful remarks on a first version of the paper. We acknowledge support from the CICYT
Research Project FPA2002-00648, from the EU network on Supersymmetry and the Early Universe
(HPRN-CT-2000-00152), and from the \textit{Departament d'Universitats, Recerca i Societat de la
Informaci{\'o}} (DURSI), Project 2001SGR00188.

%% file: PRL.tex
\chapter{Compatibility of CAST search with axion-like interpretation of PVLAS results}
\label{Compatibility of CAST search with axion-like interpretation of PVLAS results}\vspace{-.5cm}
\begin{flushright} {\footnotesize (In collaboration with
E.~Mass{\'o}. Published in Phys. Rev. Lett. \textbf{97}, 151802 (2006) \cite{Masso:2006gc}})
\end{flushright}
\def\baselinestretch{1}

\section{ Introduction}
Very recently, the PVLAS collaboration has announced the observation of a rotation of the plane of
polarization of laser light propagating in a magnetic field \cite{Zavattini:2005tm}. This dichroism
of vacuum in magnetic fields may be explained as the oscillation of photons into very light
particles $\phi$. If true, this would be of course a revolutionary finding \cite{Lamoreaux:2006ei}.

The lagrangian that would describe the necessary $\phi\gamma\gamma$
coupling is
\begin{equation}
{\cal L}_{\phi\gamma\gamma}=\frac{1}{8M}\,
\epsilon_{\mu\nu\rho\sigma}F^{\mu\nu} {F}^{\rho\sigma}\, \phi
\label{L}
\end{equation}
when $\phi$ is a pseudoscalar, and when it is a scalar is
\begin{equation}
{\cal L}_{\phi\gamma\gamma}=\frac{1}{4M}\, F^{\mu\nu}F_{\mu\nu}\,
\phi \label{LS}
\end{equation}
with $F^{\mu\nu}$ the electromagnetic field tensor. We shall refer
to $\phi$ in both cases as an axion-like particle (ALP). Let us
remark that a transition to a spin-two particle contributes to the
polarization rotation negligibly \cite{Biggio:2006im}.

Either (\ref{L})  or (\ref{LS}) lead to $\gamma-\phi$ mixing in a
magnetic field and, if $\phi$ is light enough, to coherent
transitions that enhance the signal \cite{Maiani:1986md}.
Interpreted in these terms, the PVLAS observation
\cite{Zavattini:2005tm} leads to a mass for the ALP
\begin{equation}
1\ \mathrm{meV} \lesssim m_\phi \lesssim 1.5\ \mathrm{meV} \label{m_phi}
\end{equation}
and to a coupling strength corresponding to
\begin{equation}
2 \times 10^5\ \mathrm{GeV} \lesssim M \lesssim 6\times 10^5\ \mathrm{GeV} \ \ \ . \label{M_PVLAS}
\end{equation}
Of course we would like to have an independent test of such an interpretation. There are ongoing
projects that will in the near future probe $\gamma-\phi$ transitions \cite{Ringwaldpatras}. In the
meanwhile we should face the problem of the apparent inconsistency between the value
(\ref{M_PVLAS}) and other independent results, namely, the CAST observations \cite{Zioutas:2004hi}
on the one hand, and the astrophysical bounds on the coupling of ALPs to photons on the other hand
\cite{Raffelt:1996wa}.

The CAST collaboration has recently published  \cite{Zioutas:2004hi}
a limit on the strength of (\ref{L}) or (\ref{LS}). A light particle
coupled to two photons would be produced by Primakoff-like processes
in the solar core. CAST is an helioscope \cite{Sikivie:1983ip} that
tries to detect the $\phi$ flux coming from  the Sun,  by way of the
coherent transition of $\phi$'s to X-rays in a magnetic field. As no
signal is observed they set the bound
\begin{equation}
M > 0.87 \times 10^{10}\ \mathrm{GeV}  \ \ \ , \label{M_CAST}
\end{equation}
which is in strong disagreement with (\ref{M_PVLAS}).

Also, the production of $\phi$'s  in stars is constrained because
too much energy loss in exotic channels would lead to drastic
changes in the timescales of stellar evolution. Empirical
observations of globular clusters place a bound
\cite{Raffelt:1996wa}, again in contradiction with  (\ref{M_PVLAS}),
\begin{equation}
M > 1.7  \times 10^{10}\  \mathrm{GeV} \label{M_star}  \ \ \ .
\end{equation}

As it has been stressed in \cite{Masso:2005ym}, once we are able to
relax  (\ref{M_star}) we could also evade (\ref{M_CAST}). Indeed,
the CAST bound  assumes standard solar emission. From the moment we
alter the standard scenario we should revise (\ref{M_CAST}). In
\cite{Masso:2005ym,Jain:2005nh,Jaeckel:2006id} two ideas on how to
evade the astrophysical bound (\ref{M_star}) are presented. One
possibility is that the produced ALPs diffuse in the stellar medium,
so that  they are emitted with much less energy than originally
produced \cite{Masso:2005ym}. A second possibility is that the
production of ALPs is much less than expected because there is a
mechanism of suppression that acts in the stellar conditions. We
will present in this letter a paraphoton model with
 a low energy scale where the particle production in stars is
suppressed enough to accommodate both the CAST and the PVLAS
results.
\section{Triangle diagram and epsilon-charged particles \label{sectiontriangle}}
The physical idea beyond this letter is that to understand PVLAS and
CAST in an ALP framework we have to add some new physics structure
to the vertices (\ref{L}),(\ref{LS}). The scale of the new physics
should be much less than O(keV), the typical temperature in
astrophysical environments.

\begin{figure}[b]\centering
\includegraphics[width=7cm]{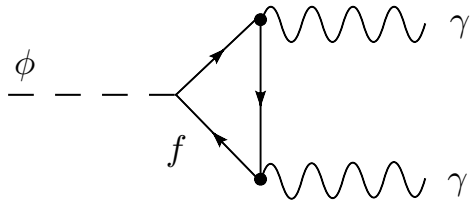}
  \caption{\it
    Triangle diagram for the $\phi\gamma\gamma$ vertex.
   \label{fig1} }
\end{figure}

We will assume that this structure is a simple loop where a new
fermion $f$ circulates; see Fig.(\ref{fig1}). The amplitude of the
$\phi\gamma\gamma$ diagram can be easily calculated and identified
with the coefficient in (\ref{L}) or (\ref{LS})
\begin{equation}
\frac{1}{M} = \frac{\alpha}{\pi} \frac{q_f^2}{v} \label{triangle}
\end{equation}
Here $\alpha=e^2/4\pi$, and the charge of the fermion $f$ is $eq_f$. The value of the mass-scale
$v$ depends on the $\phi\bar{f}f$ vertex. If $\phi$ is a pseudoscalar $v_{\mathrm
PS}=m_f/g_{\mathrm PS}$ while if is scalar $v_{\mathrm S}=f(m_f,m_\phi)$ not far from $v_{\mathrm
S}\sim m_f \sim m_\phi$ if $m_f\sim m_\phi$. Finally if $\phi$ is a Goldstone boson $v_{\mathrm
GB}$ is related to the scale of breaking of the related global symmetry.

From (\ref{triangle}) we see that $M$,  the high energy scale (\ref{M_PVLAS}), is connected to $v$.
As we need $v$  to be a low energy scale,  $q_f$ should be quite small.

Paraphoton models \cite{Holdom:1986eq,Holdom:1985ag} naturally incorporate
 small charges. These models are QED extensions with extra
$U(1)$ gauge bosons. A small  mixing among the kinetic terms of the gauge bosons leads to the
exciting possibility that paracharged exotic particles end up with a small induced electric charge
\cite{Holdom:1986eq,Holdom:1985ag}.

Getting a small charge for $f$ is not enough for our purpose since we need also production
suppression of exotic particles in stellar plasmas. With this objective, we will present a model
containing two paraphotons; if we allow for one of the paraphotons to have a mass, we will see we
can evade the astrophysical constraints and consequently the model will be able to accommodate all
experimental results. We describe it in what follows.
\section{A model with two paraphotons}
Let us start with the photon part of the QED lagrangian,
\begin{equation}
{\cal L}_0 =-\frac{1}{4}¼ F_0^{\mu\nu}F_{0\mu\nu}+ e\ j_{0\mu}
A_0^\mu
\end{equation}
where $j_{0\mu}$ is the electromagnetic current involving electrons, etc., $j_{0\mu} \sim
\bar{e}\gamma_\mu e + ...$. From the $U_0(1)$ gauge symmetry group, we give the step of assuming
$U_0(1)\times U_1(1)\times U_2(1)$ as the gauge symmetry group, with the corresponding gauge fields
$A_0$, $A_1$, and $A_2$. With all generality there will be off-diagonal kinetic terms in the
lagrangian, like $\epsilon_{01}\, F_0 F_1$ and $\epsilon_{02} F_0 F_2$ (Lorenz index contraction is
understood). We expect these mixings to be small if we follow the idea in
\cite{Holdom:1986eq,Holdom:1985ag} that ultramassive particles with 0,1,2 charges running in loops
are the responsibles for them. We will assume that these heavy particles are degenerate in mass and
have identical 1 and 2 charges so that they induce identical mixings
$\epsilon_{01}=\epsilon_{02}\equiv \epsilon$.

To write the complete lagrangian we use  the matrix notation $A
\equiv (A_0, A_1, A_2)^T$ and $F \equiv (F_0, F_1, F_2)^T$,
\begin{equation}
{\cal L} =-\frac{1}{4}\, F^T {\cal M}_F\,  F+ \frac{1}{2}\, A^T
{\cal M}_A\,  A + e\sum_i\, j_{i} A_i \label{L_complete}
\end{equation}
We call $A_0$, $A_1$, and $A_2$ interaction fields because the interaction term in
\eqref{L_complete} is diagonal, i.e. the interaction photon is defined to couple directly only to
standard model particles. Here the kinetic matrix contains the mixings,
\begin{equation}
 {\cal M}_F  = \left( \begin{array}{ccc}
                                            1 & \epsilon & \epsilon  \\
                                            \epsilon & 1 &   0          \\
                                            \epsilon & 0 & 1
 \end{array}\right)
 \label{MF}
\end{equation}
In general the diagonal terms are renormalized, $1\rightarrow 1 + \delta$, and  there are terms
${\cal M}_{F12}$. However, they  do not play any relevant role here and we omit them.

As said, we need one of the paraphotons to be massive but it will prove convenient to work with a
general ${\cal M}_A  = {\mathrm Diag}\, \{ m_0^2, m_1^2, m_2^2 \}$. Also, in the last term of
(\ref{L_complete}) we see the currents $j_1$ and $j_2$ containing the paracharged exotic particles.
To reduce the number of parameters we have set the unit paracharge equal to the unit of electric
charge, so that there is a common factor $e$.

Diagonalization  involves first a non-unitary reabsorption of the $\epsilon$ terms in (\ref{MF}) to
have the kinetic part in the lagrangian in the canonical  form, $(-1/4) F^T F$. After this, we
diagonalize the mass matrix with a unitary transformation that maintains the kinetic part canonical
ending up with the propagating field basis $\widetilde A$. We have $A={\cal U} \widetilde A$, with
\begin{equation}
{\cal U} =\   \left( \begin{array}{ccc}
              1                              & \epsilon\, \f{m_1^2}{m_0^2-m_1^2} &
              \epsilon\, \f{m_2^2}{m_0^2-m_2^2}  \\
              \epsilon\, \f{m_0^2}{m_1^2-m_0^2} & 1                              &   0          \\
              \epsilon\, \f{m_0^2}{m_2^2-m_0^2} & 0                              &     1
 \end{array}\right)
 \label{U}
\end{equation}
We see that the interacting and the propagating photon differ by little admixtures of
$O(\epsilon)$. (We work at first order in $\epsilon$).

We have developed a quite general two paraphoton model. The specific
model we adopt has the following characteristics. First, only one
paraphoton has a mass, say $m_1\equiv \mu \neq 0$, and $m_2=0$.
Second, in order to get the effects we desire we have to assign
opposite 1 and 2 paracharges  to $f$, so that the interaction for
$f$ appearing in the last term of (\ref{L_complete}) is
\begin{equation}
e\, \bar{f} \gamma_\mu f \, ( A_1^\mu -  A_2^\mu)
 \label{Lf}
\end{equation}

\begin{figure}[b]
\vspace{-.4cm}\centering
\includegraphics[width=10cm]{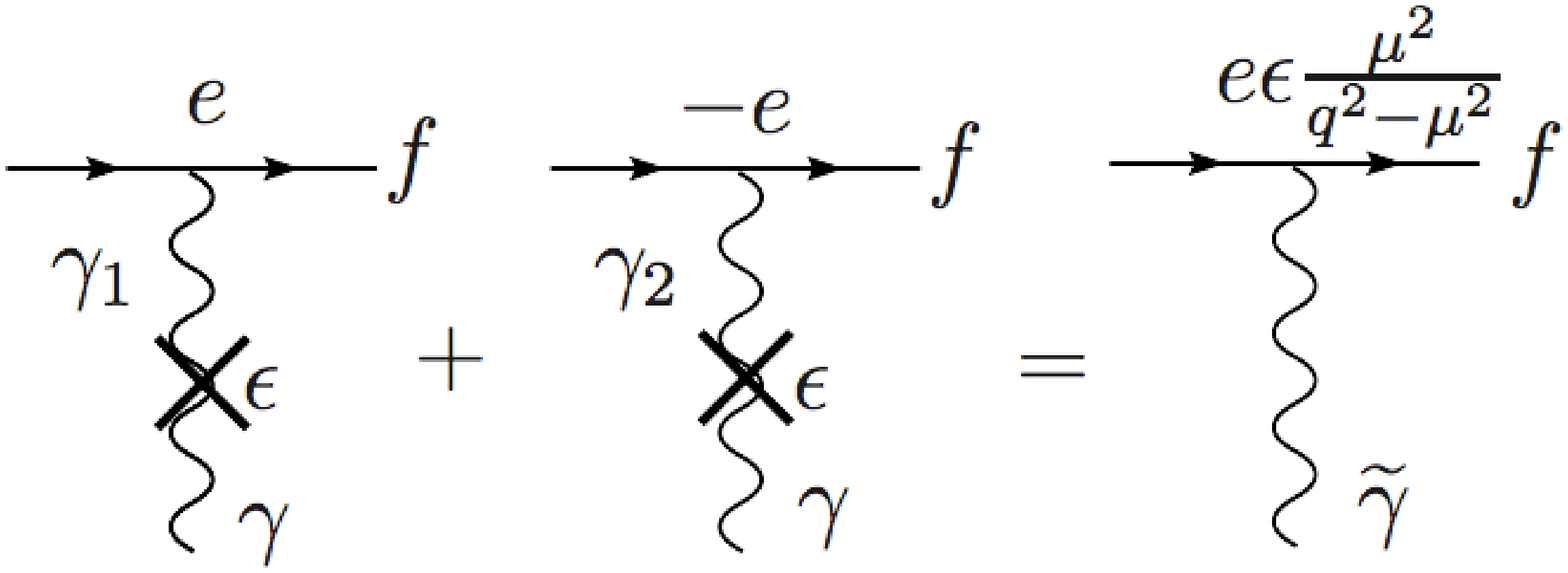}
 \caption{\it             Diagrams of the interaction of $f$ with
 photons.
   \label{fig2} }
\end{figure}

Let us show why we choose these properties. The coupling of $f$ to
photons in the interaction basis is  shown in Fig.(\ref{fig2}).
It proceeds through both paraphotons, with a relative minus sign
among the two diagrams due to the assignment (\ref{Lf}). The induced
electric $f$ charge is thus
\begin{equation}
q_f  =  {\cal U}_{10} - {\cal U}_{20}
\end{equation}

We see from (\ref{U}) that  $m_2=0$ implies  ${\cal U}_{20}=-\epsilon$. However, the value for
${\cal U}_{10}$ has to be discussed separately in the vacuum and the plasma cases. In vacuum, we
have $m_0=0$, so that ${\cal U}_{10}=0$ and thus $q_f = \epsilon$. In this case $A_0$ and $A_2$ are
degenerate and we can make arbitrary rotations in their sector. This corresponds to different
charge assignments that of course leave the physics unchanged. Due to our method of handling the
diagonalizations, eq.\eqref{U} is bad behaved for $m_0=m_2$, except for the case of our interest,
$m_0=m_2=0$, in which the order in which we take the limits $m_0\rightarrow 0$ and $m_2\rightarrow
0$ gives different charge assignments according to the rotational freedom. Here we have made
$m_2\rightarrow 0$ before $m_0\rightarrow 0$ to provide f with a millielectric charge as in
\cite{Holdom:1986eq,Holdom:1985ag}. Changing the order of the limits would end with a paracharge to
electrons.

In the classical and non-degenerated plasmas we consider the dispersion relation can be taken as
$k^2= \omega_P^2 = 4\pi\alpha n_e/m_e $ ($n_e$ and $m_e$ are the density and mass of electrons). If
$m_0= \omega_P$ is much greater that $m_1=\mu$  we get ${\cal U}_{01}\simeq -\epsilon + \epsilon
m_1^2/m_0^2$ and the induced electric charge
\begin{equation}
q_f (k^2 \simeq \omega_P^2)\,  \simeq\,  \frac{\mu^2}{\omega_P^2} \ q_f (k^2 \simeq 0)
 \label{q_f(T)} \ \ \ .
\end{equation}

Provided we have a low energy scale $\mu \ll \omega_P\sim $ keV, we
reach our objective of having a strong decrease of the $f$ charge in
the plasma, i.e., $q_f(\omega_P^2) \ll q_f(0)= \epsilon$.

The cancelation of the two diagrams of Fig.\ref{fig2} requires that the equality $e_1=e_2$ holds up
to terms of order $O(\mu^2/\omega^2_P)$. Note that even if $e_1=e_2$ at some high energy scale
because of a symmetry, a difference in the beta functions could also spoil our mechanism at low
energy. The parafermion $f$ contributes equally to both beta functions so the problem are the
contributions from the sector that gives mass only to $A_1$. However these contributions can be
made arbitrarily small by sending the Higgs mass to infinity in the spirit of the non-linear
realizations of symmetry breaking, by considering Higgsless models like breaking the symmetry
geometrically, or by considering gauge coupling unification $e_1=e_2$ at an energy not far from the
typical solar temperature. A further possibility is to consider $e_1=e_2\ll e$ which would suppress
the loop-induced effects at the prize of making the model less natural.
\section{The role of the low-energy scale}
We now discuss the consequences of our model. The PVLAS experiment is in vacuum, so $f$ has an
effective electric charge $q_f(0)=\epsilon$, which from (\ref{triangle}) has to be
\begin{equation}
\epsilon^2 \simeq   10^{-12}\, \frac{v}{\mathrm eV}
 \label{N1}
\end{equation}

Concerning the astrophysical constraints, we notice that the amplitude for the Primakoff effect
$\gamma Z \rightarrow \phi Z$ is of order $q_f^2=\epsilon^2$ and that there are production
processes with amplitudes of order $\epsilon$ which will be more effective. One is plasmon decay
$\gamma^* \rightarrow \bar f f$. Energy loss arguments in horizontal-branch (HB) stars
\cite{Davidson:1991si,Davidson:2000hf} limits $q_f$ to be below $2\times 10^{-14}$, which
translates in our model into the bound
\begin{equation}
\epsilon\  \frac{\mu^2}{ \mathrm eV^2} <   4\times 10^{-8}
 \label{N2}
\end{equation}
(we have used $\omega_P \simeq 2$ keV in a typical HB core).
Other processes like bremsstrahlung of paraphotons give weaker constraints.

Equations (\ref{N1}) and  (\ref{N2}) do not fully determine the parameters of our model. Together
they imply the constraint
\begin{equation}
v\, \mu^4 <  (\, 0.4 \ {\mathrm eV})^5
\end{equation}
We can now make explicit one of our main results. In the reasonable
case that $v$ and $\mu$ are not too different, we wee that the new
physics scale is in the sub eV range.
\begin{figure}[h]\centering
\includegraphics[width=10cm]{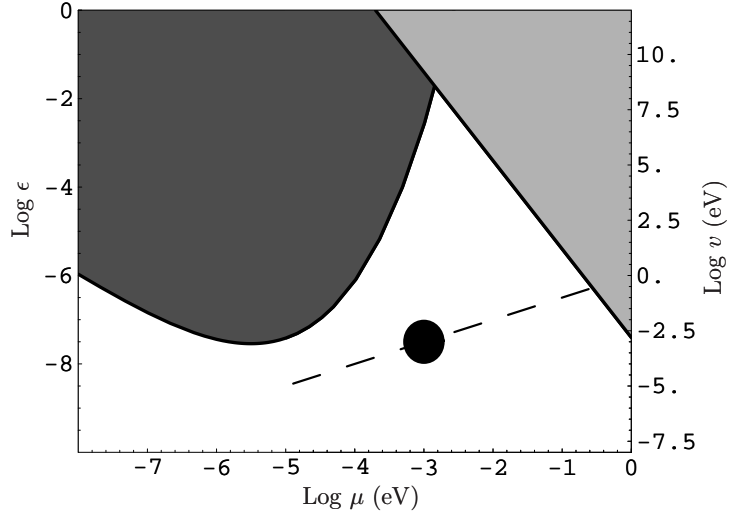}
 \caption{\it         Constraints on the parameters of our model. The black
 area is excluded  by Cavendish-type experiments,
 and the grey area by the astrophysical constraint
 (\ref{N1}). The dashed line corresponds to $v=\mu$, and the dot to
 $v=\mu\simeq 1$ meV.
   \label{fig3} }
\end{figure}

Let us consider now the CAST limit.The CAST helioscope looks for $\phi$'s with energies within a
window of 1-15 keV. In our model, $f$'s and paraphotons are emitted from the Sun, but we should
watch out $\phi$ production. This depends on the specific characteristics of $\phi$. We consider
three possibilities. A) $\phi$ is a fundamental particle. As we said the Primakoff production is
very much suppressed, so production takes place mainly through plasmon decay $\gamma^*\rightarrow
\bar{f}f \phi$. The $\phi$-flux is suppressed, but, most importantly, the average $\phi$ energy is
much less than $\omega_P \simeq.3$ keV, the solar plasmon mass. The spectrum then will be below the
present CAST energy window. B) $\phi$ is a composite $\bar{f} f$ particle confined by new strong
confining forces. The final products of plasmon decay would be a cascade of $\phi$'s and other
resonances which again would not have enough energy to be detected by CAST. C) $\phi$ is a
positronium-like bound state of $\bar{f} f$, with paraphotons providing the necessary binding
force. As the binding energy is necessarily small, ALPs are not produced in the solar plasma.

Let us now turn our attention to other constraints. Laboratory bounds on epsilon-charged particles
are much milder than the astrophysical limits, as shown in \cite{Davidson:1991si,Davidson:2000hf}.
In our model, however, even though paraphotons do no couple to bulk electrically neutral matter, a
massive paraphoton $\widetilde A_1$ couples to electrons with a strength $\epsilon$ and a range
$\mu^{-1}$. This potential effect is limited by Cavendish type experiments
\cite{Williams:1971ms,Bartlett:1988yy}.

In Fig.(\ref{fig3}) we show these limits, as well as the astrophysical bound (\ref{N2}). In the
ordinates we can see both $\epsilon$ and $v$, since we assume they are related by (\ref{N1}). At
the view of the figure, we  find out that there is wide room for the parameters of our model.
However we would like that $v$ and $\mu$ do not differ too much among them. We display the line
$v=\mu$, the region where this kind of naturality condition is fulfilled. The most economical
version of the model would be obtained when the new scales are, on the order of magnitude, about
the scale of the ALP mass of O(1 meV), (\ref{m_phi}). We have also indicated this privileged point
in the parameter space.

Also, we should discuss cosmological constraints, i.e. production of paraphotons and $f$'s in the
early universe. Taking into account that the vertices have suppression factors in the high
temperatures of such environment, we find that there is not a relic density of any of them.

Finally, let us come back to the physics responsible for the $A_1$ mass. If this comes from an
abelian Higgs mechanism then the Higgs boson acquires a millicharge $\varepsilon e_1$ and could be
produced in the Sun and in the early universe, particularly in the period of primordial
nucleosynthesis. However, this is not a problem if the mass of the Higgs is large enough, a
constraint that we required at the end of section \ref{sectiontriangle} when discussing charge
running.

\section{Conclusions}
We have presented a model of new physics containing a paracharged
particle $f$ and two paraphotons, one of which has a mass $\mu$ that
sets the low energy scale of the model. With convenient assignments
of the $f$ paracharges and mixings, we get an induced epsilon-charge
for $f$ that moreover decreases sharply in a plasma with $\omega_P
\gg \mu$. Our model accommodates an axion-like particle with the
properties (\ref{m_phi}) and (\ref{M_PVLAS}), able to explain the
PVLAS results, while at the same time consistent with the
astrophysical and the laboratory constraints, including the limit
obtained by CAST.

We have some freedom in the parameter space of our model; however if
we wish that the energy scales appearing in it are not too
different, we are led to scales in the sub eV range. A preferred
scale is O(meV), because then it is on the same order than the
axion-like particle mass.

If the interpretation of the PVLAS experiment is confirmed, which
means the exciting discovery of an axion-like particle,  then to
make it compatible with the CAST results and with the astrophysical
bounds requires further new physics. In our model, the scale of this
new physics is below the eV.

\textit{Note added}: Recently, a paper has appeared \cite{Abel:2006qt} that justifies our model in
the context of string theory.
\section{acknowledgments}
We acknowledge support by the projects FPA2005-05904 (CICYT) and
2005SGR00916 (DURSI).

%% file: PRD-II.tex
\def\baselinestretch{1}
\chapter{Need for purely laboratory-based axion-like particle searches}
\label{The Need for Purely Laboratory-Based Axion-Like Particle Searches}\vspace{-.5cm}
\renewcommand\arraystretch{1.3}
\begin{flushright} (In collaboration with J.~Jaeckel, E.~Mass{\'o}, A.~Ringwald and F.~Takahashi.\\\vspace{-.1cm}
Published in Phys. Rev. D \textbf{75}, 013004 (2007) \cite{Jaeckel:2006xm}) \end{flushright}

\section{Introduction}
Recently the PVLAS collaboration has reported the observation of  a rotation of the polarization plane of a laser
propagating through a transverse magnetic field \cite{Zavattini:2005tm}. This signal could be explained by the
existence of a new light neutral spin zero boson $\phi$, with a coupling to two photons
\cite{Maiani:1986md,Raffelt:1987im}
\be {\cal L}_I^{(-)}=\f{1}{4M}\phi^{(-)} F_{\mu\nu}\widetilde{F}^{\mu\nu} \hspace{.5cm} \textrm{or} \hspace{.5cm} {\cal
L}_I^{(+)}=\f{1}{4M}\phi^{(+)} F_{\mu\nu}F^{\mu\nu} \label{int-}\ee
depending on the parity of $\phi$, related to the sign of the rotation which up to now has not been
reported\footnote{The PVLAS collaboration has also found hints for an ellipticity signal. The sign of the phase shift
suggests an even particle $\phi^{(+)}$ \cite{PVLASICHEP}.}. Such an Axion-Like Particle (ALP) would oscillate into
photons and vice versa in the presence of an electromagnetic field in a similar fashion as the different neutrino
flavors oscillate between themselves while propagating in vacuum.

The PVLAS signal, combined with the previous bounds from the absence of a signal in the BFRT collaboration
experiment~\cite{Cameron:1993mr}, implies~\cite{Zavattini:2005tm}
\begin{equation}
1\ \mathrm{meV} \lesssim m \lesssim 1.5\ \mathrm{meV},
 \hspace{1.5cm}
2 \times 10^5\ \mathrm{GeV} \lesssim M \lesssim 6\times 10^5\ \mathrm{GeV}, \label{masscoupPVLAS}
\end{equation}
with $m$ the mass of the new scalar.

It has been widely noticed that the interaction \eqref{int-} with the strength \eqref{masscoupPVLAS} is in
\textit{serious} conflict with astrophysical constraints \cite{Raffelt:2005mt,Ringwald:2005gf}, while it is allowed by
current laboratory and accelerator data \cite{Masso:1995tw,Kleban:2005rj}. This has motivated recent work on building
models that evade the astrophysical constraints
\cite{Masso:2005ym,Jain:2005nh,Jaeckel:2006id,Masso:2006gc,Mohapatra:2006pv}, as well as alternative explanations to
the ALP hypothesis \cite{Antoniadis:2006wp,Gies:2006ca,Abel:2006qt}.

At the same time, many purely laboratory-based experiments have been proposed or are already on the way to check the
particle interpretation of the PVLAS signal~\cite{Ringwald:2003ns,Rabadan:2005dm,Pugnat:2005nk,%
Gastaldi:2006fh,Afanasev:2006cv,Kotz:2006bw,Cantatore:Patras,BMV:Patras,Chen:2003tp,Gabrielli:2006im}.
It is important to notice, for the purpose of our paper, that these experiments are optical, and
not high-energy, accelerator experiments.

Quite generally, these experiments will have enough sensitivity to check values of $M$ equal or greater than $10^6$
GeV, but, apart from Ref.~\cite{Ringwald:2003ns}, they do not have the impressive reach of the astrophysical
considerations, implying $M\gtrsim 10^{10}$ GeV. Thus, if the PVLAS signal is due to effects other than $\phi-\gamma$
oscillations and the astrophysical bounds are applicable, these experiments can not detect any interesting signal.

However, the astrophysical bounds rely on the assumption that the vertex \eqref{int-} applies under typical laboratory
conditions as well as in the stellar plasmas that concern the astrophysical bounds. It is clear that, if one of the
future dedicated laboratory experiments eventually sees a positive signal, \textit{this can not be the case}.

In this work we investigate the simplest modification to the standard picture able to accommodate a positive signal in
any of the forthcoming laboratory experiments looking for ALPs, namely that the structure of the interaction
\eqref{int-} remains the same in both environments, while the values of $M$ and $m$ can be different. Interestingly
enough, the environmental conditions of stellar plasmas and of typical laboratory experiments are very different and
thus one could expect a very big impact on $M$ and $m$.

We consider qualitatively the situation in which the dependence of $M$ and $m$ on the environmental parameters produces
a \textit{suppression} of ALP production in stellar plasmas. The main work of the paper is devoted to compute this
suppression using a realistic solar model and to investigate how it relaxes the astrophysical bounds on the coupling
\eqref{int-}. This leaves
 room for the proposed  laboratory experiments to potentially discover such an
axion-like particle.

In section \ref{sectionastrobounds} we revisit the astrophysical bounds and discuss general mechanisms to evade them.
In the following section \ref{numerical}, we present our scenario of environmental suppression and calculate the
modified bounds. We present our conclusions and comment on the reach of proposed future laboratory experiments in
section \ref{summary}.

\section{Astrophysical bounds and general mechanisms to evade them\label{sectionastrobounds}}

\begin{figure}\centering
  \includegraphics[width=6cm]{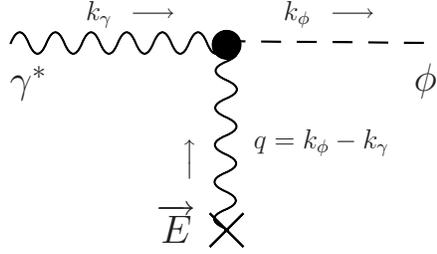}
  \caption{\label{fig0} Primakoff processes in which a photon turns into an
  ALP in the electric field of a charged particle like a proton or electron.}
\end{figure}

Presuming the $\phi\gamma\gamma$ vertex \eqref{int-}, photons of stellar plasmas can convert into ALPs in the
electromagnetic field of electrons, protons and heavy ions by the Primakoff effect, depicted schematically in
Fig.~\ref{fig0}. If $M$ is large enough, these particles escape from the star without further interactions constituting
a non-standard energy-loss channel. This energy-loss channel accelerates the consumption of nuclear fuel and thus
shortens the duration of the different stages of stellar evolution with respect to the standard evolution in which ALPs
do not exist.

In general, the astrophysical observations do agree with the theoretical predictions without additional energy-loss
channels so one is able to put bounds on the interaction scale $M$ \cite{Raffelt:1996wa}. The most important for our
work are those coming from the lifetime of the Sun \cite{Frieman:1987ui}, the duration of the red giant phase, and the
population of Helium Burning (HB) stars in globular clusters \cite{Raffelt:1985nk,Raffelt:1987yu}. The last of them
turns out to be the most stringent, implying
\be M > 1.7\times  10^{10} \ \textrm{GeV} \equiv M_{\mathrm HB}, \label{HBconstraint}\ee
for $m<{\mathcal O}(1\ {\mathrm keV})$. Moreover, if ALPs are emitted from the Sun one may try to reconvert them to
photons at Earth by the inverse Primakoff effect exploiting a strong magnetic field. This is the helioscope idea
\cite{Sikivie:1983ip} that it is already in its third generation of experiments. Recently, the CERN Axion Solar
Telescope (CAST) collaboration has published their exclusion limits \cite{Zioutas:2004hi} from the absence of a
positive signal,
\be M > 8.6\times 10^{9}\ \textrm{GeV}\equiv M_{\mathrm CAST}, \label{CASTconstraint}\ee
for $m<0.02$~eV.

One should be aware that these astrophysical bounds rely on many assumptions to calculate the flux
of ALPs produced in the plasma. In particular, it has been assumed widely in the literature that
the same value of the coupling constant that describes $\phi-\gamma$ oscillations in a magnetic
field in vacuum describes the Primakoff production in stellar plasmas, and the mass has been also
assumed to be the same. We want to remark that this has been mainly an argument of pure simplicity.
In fact, there are models in which $M$ depends on the momentum transfer $q$ at which the vertex is
probed \cite{Masso:2005ym} or on the  effective mass $\omega_P$ of the plasma photons involved
\cite{Masso:2006gc}. These models have been built with the motivation of evading the astrophysical
bounds on  ALPs, by decreasing the effective  value of the coupling $1/M$ in stellar plasmas in
order to solve the inconsistency between the ALP interpretation of PVLAS and the astrophysical
bounds. This has proven to be a very difficult task because of the extreme difference between the
PVLAS value (\ref{masscoupPVLAS}) and the HB (\ref{HBconstraint}) or CAST (\ref{CASTconstraint})
exclusion limits. These models require very specific and somehow unattractive features like the
presence of new confining forces or tuned cancelations (note, however, \cite{Abel:2006qt}). Anyway,
they serve as examples of how $M$ (and eventually $m$) can depend  on ``environmental'' parameters
$\eta = q,\omega_P$, etc... (for other suitable parameters, see Table~\ref{PVLAS_SUN_HB}),
\begin{equation}
M\rightarrow M(\eta), \quad m\rightarrow m(\eta),
\end{equation}
such that the production of ALPs is suppressed in the stellar environment.

\begin{table}\centering
\begin{tabular}{c|c|c|c}
Env. param. &  Solar Core & HB Core & PVLAS \\ \hline $T$ [keV] &  $1.3$ & $8.6$ & $\sim 0$
\\ \hline
$q^2$ [keV$^{2}$] & $\sim 1$ & $\sim 1$ & $ \sim 10^{-18}$ \\
\hline $\omega_P$ [keV] & $0.3$ & $2$ & $0$ \\
\hline $\rho$ [g~cm$^{-3}$] & $1.5\times 10^2$& $10^4$ & $< 10^{-5}$ \\
\end{tabular}
\caption{\label{PVLAS_SUN_HB} Comparison between the values of environmental parameters, such as the temperature $T$,
typical momentum transfer $q$, plasma frequency $\omega_P$, and matter energy density $\rho$, in the stellar plasma and
in the PVLAS experiment. Other parameters to consider could be the Debye screening scale $k_s$, or, to name something
more exotic, the neutrino flux, or  the average electromagnetic field.}
\end{table}

\begin{figure}
  \begin{center}
\includegraphics[width=10.5cm]{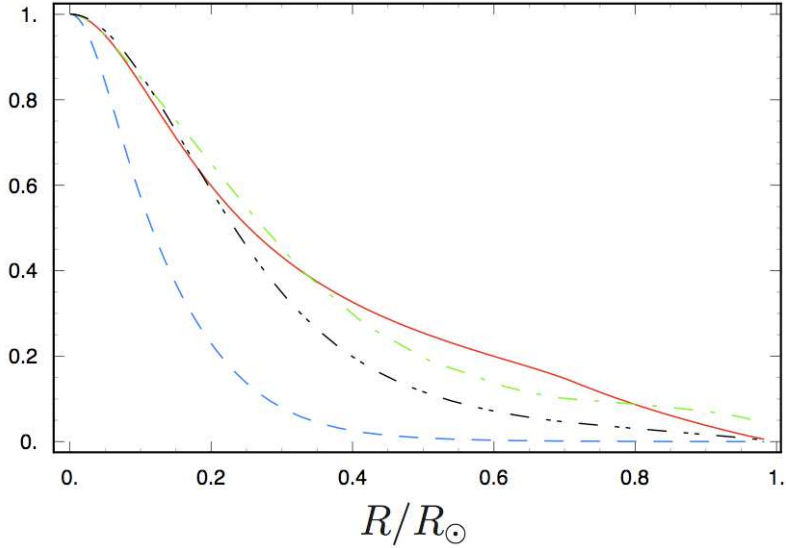}
\end{center}
\vspace{-1cm}
  \caption{\label{fig2} Environmental parameters as a function of the distance to the solar center.
Temperature (solid, red), matter density (dashed, blue),
  Debye screening scale (double dashed, green) and plasma frequency
  (triple dashed, black), normalized to their values in the solar center,
  $T_0=1.35$ keV, $\rho_0=1.5\times 10^{2}$~g\,cm$^{-3}$, $k_{s0}=9$~keV, $\omega_{P0}=0.3$~keV
  for the solar model BS05(OP) of Bahcall {\it{et al.}} \cite{Bahcall:2004pz}.
  }
\end{figure}

In the following, we will not try to construct micro-physical explanations for this dependence but rather write down
simple effective models and fix their parameters in order to be consistent with the solar bounds and PVLAS or any of
the proposed laboratory experiments.

A suppression of the production in a stellar plasma could be realized in two simple ways:
\begin{itemize}
\item[(i)] either the coupling $1/M$ decreases (dynamical suppression) or
\item[(ii)] $m$ increases
to a value higher than the temperature such that the production is Boltzmann suppressed (kinematical suppression).
\end{itemize}

All the environmental parameters considered in this paper are  \textit{much} higher in the Sun than in laboratory
conditions (see Table~\ref{PVLAS_SUN_HB} and Fig.~\ref{fig2}), so we shall consider $M(\eta)$ and $m(\eta)$ as
monotonic increasing functions of $\eta$ with the values of $M(\sim 0)$ and $m(\sim 0)$ fixed by the laboratory
experiments.

Clearly, both mechanisms are efficient at suppressing the production of ALPs in the Sun, but there is a crucial
difference that results in some prejudice against  mechanism (ii). Mechanism (i) works by making the already weak
interaction between ALPs and the photons even weaker. The second mechanism, however, is in fact a strong interaction
between the ALPs and ordinary matter, thereby making it difficult to implement without producing unwanted side effects.
We will nevertheless include mechanism (ii) in our study, but one should always keep this caveat in mind.

As we said, $\eta$ in the stellar plasma is generally much higher than in laboratory-based experiments. It is then
possible that new ALP physics produces  also a big difference between the values of the ALP parameters, $m$ and $M$, in
such different environments.

\begin{figure}[t]
  \centering
   \includegraphics[height=3.3cm]{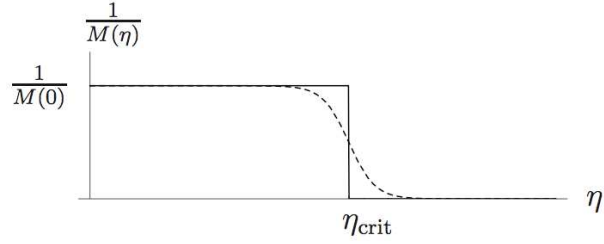}\vspace{-.5cm}
  \caption{\label{fig1} Coupling as a function of an environmental parameter $\eta$:
The simple form used in our calculations (solid line) and a generic, more realistic, dependence (dashed).}
\end{figure}

Let us remark on the  a priori unknown shape of $M(\eta)$ and $m(\eta)$. In our calculations we use a simple step
function (cf. Fig.~\ref{fig1}), which has only one free parameter: the value for the environmental parameter where the
production is switched off, $\eta_{\mathrm{crit}}$. In most situations this will give the strongest possible
suppression. The scale $\eta_{\mathrm{crit}}$ can be associated with the scale of new physics responsible for the
suppression. In what follows, we will consider only the effects of one environmental parameter at once although it is
trivial to implement this framework for a set of parameters.

For simplicity, we restrict the study of the environmental suppression of ALPs to our Sun because we know it
quantitatively much better than any other stellar environment. The group of Bahcall has specialized in the computation
of detailed solar models which provide all the necessary ingredients to compute accurately the Primakoff emission. We
have used the newest model, BS05(OP) \cite{Bahcall:2004pz}, for all the calculations of this work (our accuracy goal is
roughly $10\%$). The variation of some environmental parameters is displayed in Fig.~\ref{fig2} as a function of the
distance from the solar center.

\section{\label{numerical}Numerical results}

Let us first state how a suppression $S$ of the flux of ALPs affects the bounds arising from energy loss considerations
and helioscope experiments. If the flux of ALPs from a stellar plasma is suppressed by a factor $S$, the energy loss
bounds on $M$  are relaxed by a factor of $\sqrt{S}$ while the CAST bound relaxes with $\sqrt[4]{S}$,
\begin{eqnarray}
\label{modloss} M_{\mathrm{loss}} \rightarrow \sqrt{S}\,M_{\mathrm{loss}},&&\hspace{6ex} {\mathrm energy\ loss\ bound},
\\
\label{modcast} M_{\mathrm{CAST}} \rightarrow \sqrt[4]{S}\,M_{\mathrm{CAST}},&&\hspace{6ex} {\mathrm{CAST\ bound}},
\end{eqnarray}
since the former depends only on the Primakoff production, $\sim 1/M^2$, and the latter gets an additional factor $\sim
1/M^2$ for the reconversion at Earth  resulting in a total counting rate $\sim 1/M^4$.

According to \eqref{masscoupPVLAS} and \eqref{CASTconstraint} to reconcile the CAST and PVLAS results we would need
\be S_{\mathrm CAST} < \left(\f{M_{\mathrm{PVLAS}}}{M_{\mathrm{CAST}}}\right)^4 \sim 10^{-20} \ \ \ ,
\label{castfactor} \ee
while to reconcile the PVLAS ALP with the Sun energy loss bound we need a much more moderate \footnote{We have
recalculated the Solar energy loss bound using the latest Solar model BS05(OP). See next section for details.}
\be S_{\mathrm{loss}} < \left(\f{M_{\mathrm{PVLAS}}}{M_{\mathrm{CAST}}}\right)^2 \sim \f{}{}10^{-10} \ \ \ .
\label{lossfactor}\ee

\subsection{Dynamical suppression}

We consider first a possible variation of the coupling that we have enumerated as mechanism (i). Treating the emission
of ALPs as a small perturbation of the standard solar model, we can compute the emission of these particles from the
unperturbed solar data. The $\gamma-\phi$ Primakoff transition rate for both interactions in eq. \eqref{int-} can be
written as (neglecting the plasma mass $\omega_{P}$ for the moment)\footnote{We are using natural units $\hbar = c =1$
with the Boltzmann constant, $k_B=1$.}
\be \label{Gammanoplasma} \Gamma(\omega)_{\gamma-\phi}= \frac{T k^{2}_{s}}{64\pi}
\int^{1}_{-1}\hspace{-7pt}d\cos\theta \f{1+\cos\theta}{\kappa^2+1-\cos\theta}\f{1}{M(\eta)^2},\ee
where $\omega$ is the energy of the incoming photon, and
\begin{equation}
k^{2}_{s}=\frac{4\pi\alpha}{T} (n_{e}+\sum_{i}Z^{2}_{i}n_{i}),
\end{equation}
is the Debye screening scale. $n_i,Z_i$ are the number densities and charges of the different charged species of the
plasma, $\alpha\simeq 1/137$, $n_{e}$ is the electron number density, $\cos\theta$ is the relative angle between the
incoming photon and the outgoing ALP in the target frame (we take the mass of the target to be infinite since the
masses of protons and electrons far exceed the typical momentum transfer of order $\lesssim \mathrm{keV}$; this implies
that the incoming photon and the outgoing axion have the same energy) and $\kappa^2= k_s^2/2\omega^2$. Integration over
the whole Sun with the appropriate Bose-Einstein factors for the number density of photons gives the spectrum of ALPs
(number of emitted ALPs per unit time per energy interval),
\be \label{spectrum}\f{d^2N(\omega)}{d\omega dt}=4\pi\int_0^{R_\odot}\hspace{-9pt}R^2dR \f{\omega^2}{\pi^2}
\f{\Gamma(\omega)_{\gamma-\phi}}{{\mathrm e}^{\omega/T}-1}  \ \ \ .\ee
(Remember that $T$, $k_s^2$, etc. depend implicitly on the distance $R$ from the solar center.)

\begin{figure}
\begin{center}
  \includegraphics[width=10.5cm]{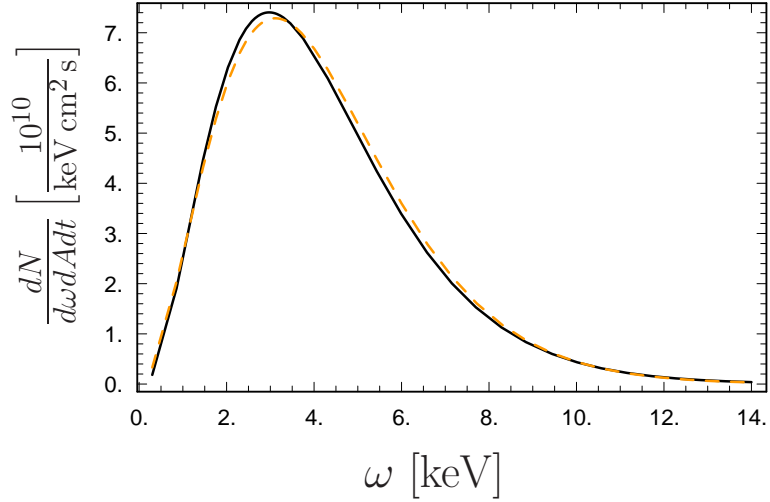}
\end{center}
\vspace{-1cm}
  \caption{\label{figextra} Our spectrum of ALPs at Earth (black solid) agrees reasonably well with that of
  the CAST collaboration \cite{Zioutas:2004hi}
  (dashed orange) for $M=10^{10}\,\mathrm{GeV}$.}
\end{figure}

As a check of our numerical computation we have computed the flux of \textit{standard} ALPs at Earth which is shown in
Fig.~\ref{figextra} and does agree with the CAST calculations \cite{Zioutas:2004hi}.

It is very important to differentiate two possibilities:
\begin{itemize}
\item[\textit{1.}] $\eta$ is a macroscopic (averaged) environmental parameter given by the solar
model and depending only on the distance $R$ from the solar center. Then the suppression acts as a step function in the
$R$ integration \eqref{spectrum} for the flux.
\item[\textit{2.}] $\eta$ depends on the microscopic aspects of the production like the momentum transfer $q^2$.
Then the step function acts inside the integral in eq. \eqref{Gammanoplasma}.
\end{itemize}

We now start with the first possibility and let the second, which requires a different treatment, for subsubsection
\ref{sectionq2}.

\subsubsection{Dynamical suppression from macroscopic environmental parameters}\label{macroscopic}

If $1/M(\eta)$ is a step function, ALP production is switched off wherever $\eta>\eta_{\mathrm crit}$. Let us call
$R_{\mathrm crit}$ the radius at which the coupling turns off, i.e. $\eta(R_{\mathrm crit})=\eta_{\mathrm crit}$. Since
the functions $\eta(r)$ shown in Fig.~\ref{fig2} are monotonous, we can calculate the suppression as a function of
$R_{\mathrm crit}$ and then determine $\eta_{\mathrm crit}=\eta(R_{\mathrm crit})$.

We define the suppression efficiency, $S(\omega,R_{\mathrm crit})$,  as the ratio of the flux of ALPs with energy
$\omega$ with suppression, divided by the one without suppression,
\begin{equation}
S(\omega;R_{\mathrm crit})=\frac{d^2N(\omega;R_{\mathrm crit})}{d\omega\,dt}
\left(\frac{d^2N(\omega)}{d\omega\,dt}\right)^{-1}.
\end{equation}
%

\begin{figure}\centering
    \includegraphics[width=10.5cm]{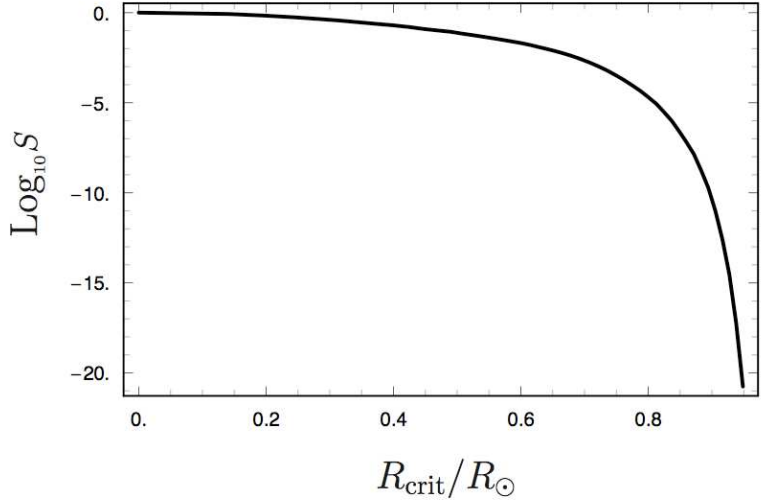} 
\vspace{-.5cm} \caption{\label{fig4}Suppression of the flux of ALPs $S(\omega_0=1\,\mathrm{keV},R_{\mathrm{crit}})$ as
a function of $R_{\mathrm crit}$.}
\end{figure}

The CAST experiment is only sensitive to ALPs in the range of $(1-14)\,\mathrm{keV}$. Hence, we must suppress the
production of ALPs only in this energy range. In order to provide a simple yet conservative bound we use the factor
$S(\omega_{0},R_{\mathrm{crit}})$ evaluated at the energy $1\,\mathrm{keV}\leq \omega_{0}\leq 14\,\mathrm{keV}$ which
maximizes $S$. We have checked that, in all cases of practical interest, $\omega_0$ is the CAST lower threshold, 1~keV.
In Fig.~\ref{fig4}, we plot $S(1\,{\mathrm keV},R_{\mathrm crit})$. In Tab. \ref{tab2} we give some values for $S$
together with the corresponding values of $\eta_{\mathrm{crit}}$.

Looking at Table~\ref{tab2}, we find that it is possible to achieve the suppression required in Eq.~\eqref{castfactor}
and reconcile PVLAS and CAST, but the critical environmental parameters are quite small; for example, the critical
plasma frequency is in the eV range. Moreover, the results  are sensitive to the region close to the surface of the Sun
where $\log(S)$ changes very fast and our calculation becomes somewhat less reliable.

\begin{table}\centering
\begin{tabular}{c|c|c|c|c}
$R_{\mathrm crit}/R_\odot$ & $T_{\mathrm crit}\,[{\mathrm keV}]$ &
$\rho_{\mathrm crit}\,[{\mathrm g}\, {\mathrm cm}^{-3}]$ & $\omega_{P,\mathrm{crit}}\,[{\mathrm keV}]$ & $S$\\
\hline $ 0$ & $1.35$ & $150$ & $0.3$ & $1$ \\
\hline $0.2$ & $0.81$ & $35$ & $0.16$ & $0.67$ \\
\hline $0.5$ & $0.34$ & $1.3$ & $0.03$ & $0.08$ \\
\hline $0.7$ & $0.2$ & $0.2$ & $0.01$ & $2\times 10^{-3}$ \\
\hline $0.8$ & $0.12$ & $0.09$ & $0.008$ & $ 2\times10^{-5}$ \\
\hline $0.85$ & $0.08$ & $0.05$ & $0.006$ & $2\times10^{-7}$ \\
\hline $0.9$ & $0.05$ & $0.03$ & $0.004$ & $4\times10^{-11}$ \\
\hline $0.95$ & $0.025$ & 0.009 & 0.0025 & $\sim 10^{-20}$\\
\end{tabular}
\caption{\label{tab2}Several values of $S(\omega_0=1\,\mathrm{keV},R_{\mathrm{crit}})$ with their respective values of
the suppression scales $\eta_{\mathrm crit}$.}
\end{table}

We now take a look at the solar energy loss bound (\ref{modloss}). The age of the Sun is known to be around $5.6$
billion years from radiological studies of radioactive crystals in the solar system (see the dedicated Appendix in
\cite{Bahcall:1995bt}). Solar models are indeed built to reproduce this quantity (among others, like today's solar
luminosity, solar radius, etc...), so one might think that a model with ALP emission can be constructed as well to
reproduce this lifetime. However, this seems not to be the case for large ALP luminosity \cite{Raffelt:1987yu} and it
is concluded that the exotic contribution cannot exceed the standard solar luminosity in photons. For our purposes this
means
\be L_{\mathrm ALP}<L_\odot=3.846\times10^{26}\ {\mathrm W} \sim 1.60\times 10^{30}\ {\mathrm eV}^2
\label{LifetimeBound},\ee
with
\be L_{\mathrm ALP} \equiv \int_0^\infty d\omega\, \omega\,\frac{d^2N}{d\omega dt}. \ee

We have computed the ALP emission in BS05(OP),
\be L_{\mathrm ALP}= 1.8\times 10^{-3} \left(\f{10^{10}\ \textrm{GeV}}{M}\right)^2 L_\odot . \label{SunConstraint}\ee
This value is slightly bigger than that of Ref.~\cite{vanBibber:1988ge}, which relies on an older solar model
\cite{Bahcall:1981zh}, probably as a consequence of the different data.

\begin{figure}\centering
    \includegraphics[width=10.5cm]{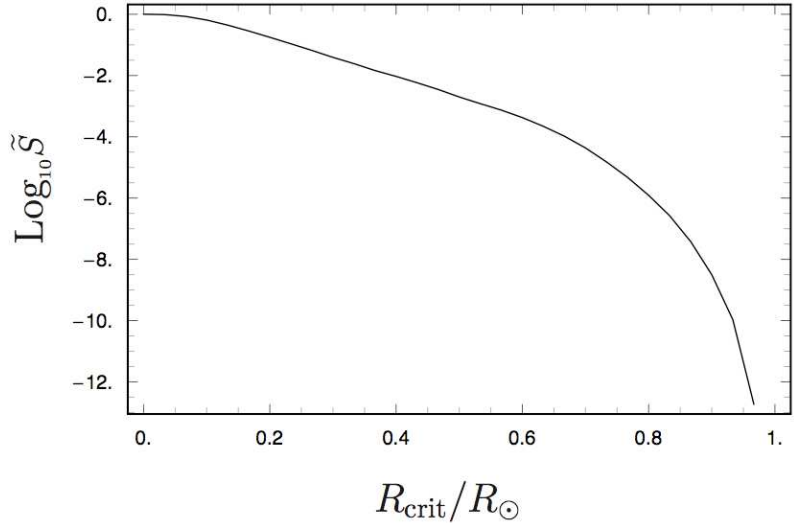}
\caption{\label{totalflux1}Suppression $\tilde{S}$ of the total flux of ALPs as a function of the critical radius
$R_{\mathrm{crit}}$.}
\end{figure}

For the total flux, we find a suppression
\begin{equation}
\tilde{S}(R_{\mathrm{crit}})=\frac{L_{\mathrm{ALP}}(R_{\mathrm{crit}})}{L_{\mathrm{ALP}}} ,
\end{equation}
which we plot in Fig. \ref{totalflux1}.

Remember that in order to avoid a conflict between the PVLAS result and the energy loss argument we required Eq.
\eqref{lossfactor}, $\widetilde S_{\mathrm{loss}} < 10^{-10}$. Looking at Fig. \ref{totalflux1} we find that this bound
alone requires values for the critical environmental parameters that are larger (and therefore less restrictive) than
those from the CAST bound.

\subsubsection{Dynamical suppression from microscopic parameters: $q^2$ \label{sectionq2}}

In the previous subsection, we have considered macroscopic environmental parameters like, e.g., the temperature $T$.
However, suppression could also result from a dependence on microscopic parameters like, e.g., the momentum transfer
$q^2$ in a scattering event (not averaged).

In this section we discuss the well motivated (cf. \cite{Masso:2005ym}) example of a possible dependence $M=M(q^2)$ on
the momentum transfer involved in the Primakoff production (Fig.~\ref{fig0}). Again, we use a step function to model
the dependence on $q^2$,
\be \frac{1}{M(q^2)}=\frac{1}{M(0)}\ \Theta(q^{2}_{\mathrm{crit}}-|q^2|) =\frac{1}{M(0)}\
\Theta(q^2_{\mathrm{crit}}-q_m^2-2k_\phi k_\gamma(1-\cos\theta)) \label{Mq2stepfunction}, \ee
where $k_\phi$, $k_\gamma$ are the moduli of the momenta of the ALP and the photon. $q_m=|k_\phi-k_\gamma|$ is the
smallest possible momentum transfer. Here, we will use the approximation $m=0$, but it will be crucial to take into
account that photons have an effective mass
\begin{equation}
m^{2}_\gamma=\omega^{2}_P=\frac{4\pi\alpha n_{e}}{m_{e}},
\end{equation}
so $q_m(\omega)=\omega-\sqrt{\omega^2-\omega_P^2}$. Note that the plasma mass is crucial because it ensures that
$q_m>0$, i.e. it removes ALP production processes with very small momentum transfer which would be unsuppressed.

With this modification, Eq.~\eqref{Gammanoplasma} reads
\be \Gamma_{\gamma-\phi}(\omega) = \frac{T k^{2}_{s}}{64\pi}
\int_{-1}^{+1}d\cos\theta \f{\sin^2\theta}{(x-\cos\theta)(y-\cos\theta)}\f{1}{M^2(q^2)} \label{gammaunsuppressed} \ee
with  $x=(k_\phi^2+k_\gamma^2)/2k_\phi k_\gamma$ and $y=x+k_s^2/2k_\phi k_\gamma$.
The step function implies that only values of $\cos\theta$ satisfying
\be \cos\theta > 1 - \f{q^2_{\mathrm{crit}}-q^2_m}{2\omega\sqrt{\omega^2-\omega^2_P}} \ee
contribute to the integral. Hence, we find that the effect of the step function \eqref{Mq2stepfunction} is to restrict
the integration limits of Eq.~(\ref{gammaunsuppressed}),
\be \Gamma_{\gamma-\phi}(\omega)=\f{T k_s^2}{64\pi M^2(0)}\int_{\delta(\omega)}^{+1}d\cos\theta
\f{\sin^2\theta}{(x-\cos\theta)(y-\cos\theta)}, \label{gammasuppressed} \ee
with
\be \delta(\omega)= \left(
\begin{array}{cccc} 1 & \textrm{for} & q_{\mathrm{crit}} < q_m(\omega)& \\
 1 - \f{q_{\mathrm{crit}}^2-q^2_m(\omega)}{2\omega\sqrt{\omega^2-\omega^2_P}} & \textrm{for}
& q_{\mathrm{crit}} > q_m(\omega),& 1 - \f{q_{\mathrm{crit}}^2-q^2_m(\omega)}{2\omega\sqrt{\omega^2-\omega^2_P}}>-1
\\
                 -1 \hspace{1cm} & \textrm{for} & q_{\mathrm{crit}} > q_m(\omega),&
1 - \f{q_{\mathrm{crit}}^2-q^2_m(\omega)}{2\omega\sqrt{\omega^2-\omega^2_P}}\leqslant-1
\end{array} \right). \ee
When $\delta(\omega)=1$, the integral is zero and Primakoff conversion is completely suppressed. This happens for
values of the plasma frequency $\omega_P$ and the energy $\omega$ for which the minimum momentum transfer is already
larger than the cut-off scale $q_{\mathrm{crit}}$. We point out that this is an energy dependent statement. For
$\omega\gg\omega_P$ large enough, $q_m$ is small enough to satisfy $q_{\mathrm{crit}}\gg q_m$. When this is the case we
have only partial suppression. The integral goes only over the small interval $[\delta(\omega),1]$ where
$\delta(\omega)\approx 1-q_{\mathrm{crit}}^2/2\omega^2$,  $x\approx 1$ and $y\approx 1+k_s^2/2\omega^2$. Then the
integral can be easily estimated by the value of the integrand at $\cos\theta=1$,
\be \Gamma_{\gamma-\phi}(\omega)\sim \f{T k_s^2}{64\pi M^2} \f{4\omega^2}{k_s^2}\f{q_{\mathrm{crit}}^2}{2\omega^2},
\quad\quad\quad {\mathrm{for}}\,\, \omega\gg \omega_P,\, k_{s},\,q_{\mathrm{crit}}. \ee
Notice that although we have used the strongest possible suppression, a step function, at the end of the day, at high
energies, the transition rate is only suppressed by a factor $q_{\mathrm{crit}}^2/k_s^2$. This means that the $\gamma^*
-\phi$ transition is suppressed at most quadratically.

This holds even for a generic suppressing factor $F(q^2) = M(q^2)/M(0)$. The limitation comes from the part of the
integral which is close to $\cos\theta=1$. There the integrand is a constant, $1+\cos\theta/(y-1)\sim 4\omega^2/k_s^2$.
By continuity, the suppression factor $F(q^2)$, whatever it is, must be close to unity because $q^2$ is very close to
zero and normalization requires $F(q^2=0)=1$. This holds for values of $q^2$ up to a certain range, limited by the
shape of $F(q^2)$. Defining $q_{\mathrm{crit}}^2$ as the size of the interval where $F(|q^2|\lesssim
q_{\mathrm{crit}}^2)\sim 1$, then $q_{\mathrm{crit}}^2=|q^2|$ gives a minimum value for $\cos\theta$ for which the
integrand is nearly constant ($\cos\theta_m \sim 1-q_{\mathrm{crit}}^2/2\omega^2$), leading to
\be \Gamma(\omega)\propto \int^1_{-1}d\cos\theta\,
\f{1+\cos\theta}{y-\cos\theta}\,F(q^2)\gtrsim\int^1_{\cos\theta_m}d\cos\theta\, \f{4\omega^2}{k_s^2}\sim
\f{2q_{\mathrm{crit}}^2}{k_s^2} .\ee
Note that here we have ALP emission from every place of the Sun which is in contrast to the macroscopic environmental
suppression scenario. Indeed, the ALP production rate typically increases towards the solar center.

\begin{figure}
    \centering
    \includegraphics[width=7.5cm]{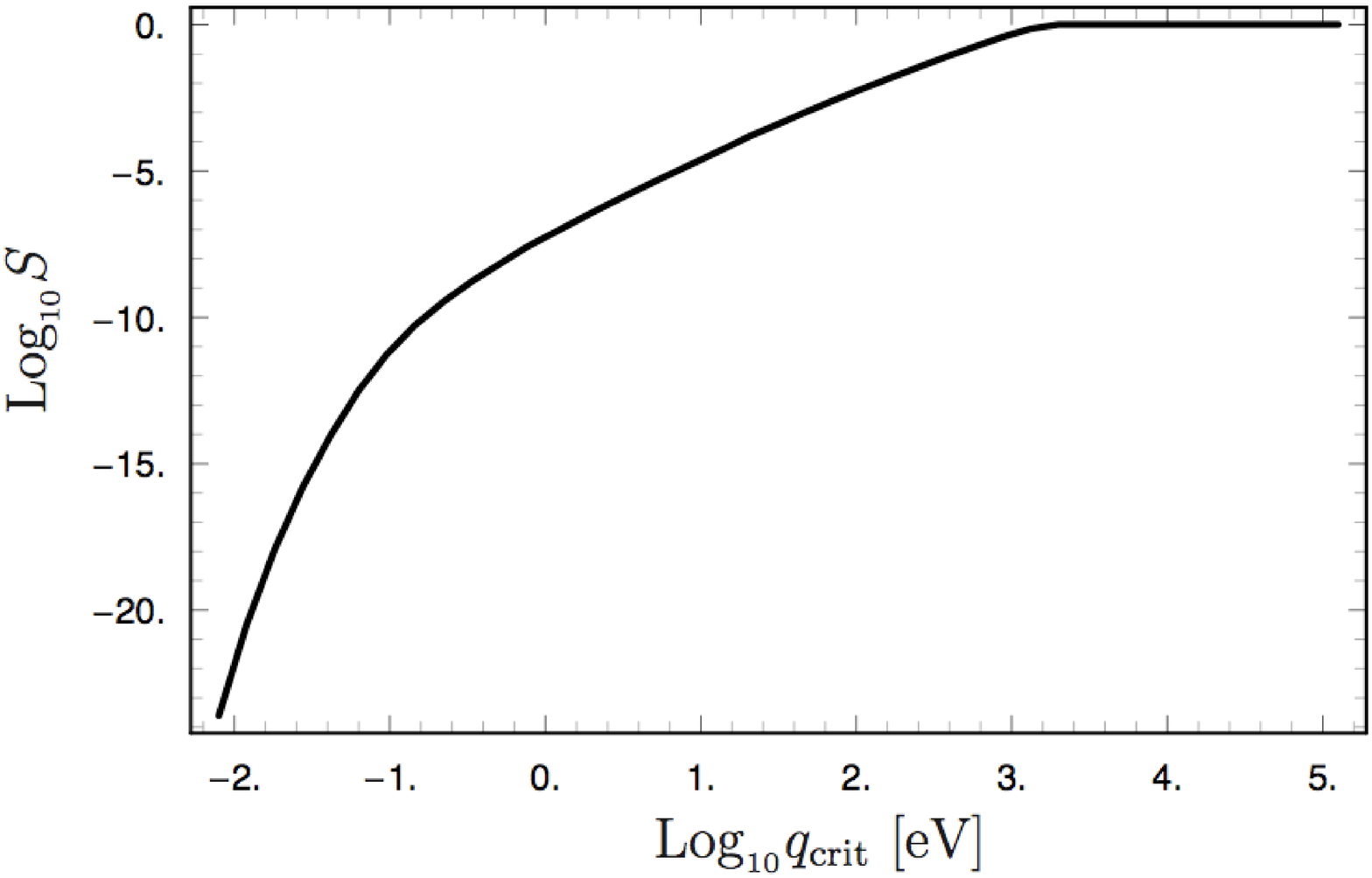}   
    \includegraphics[width=7.5cm]{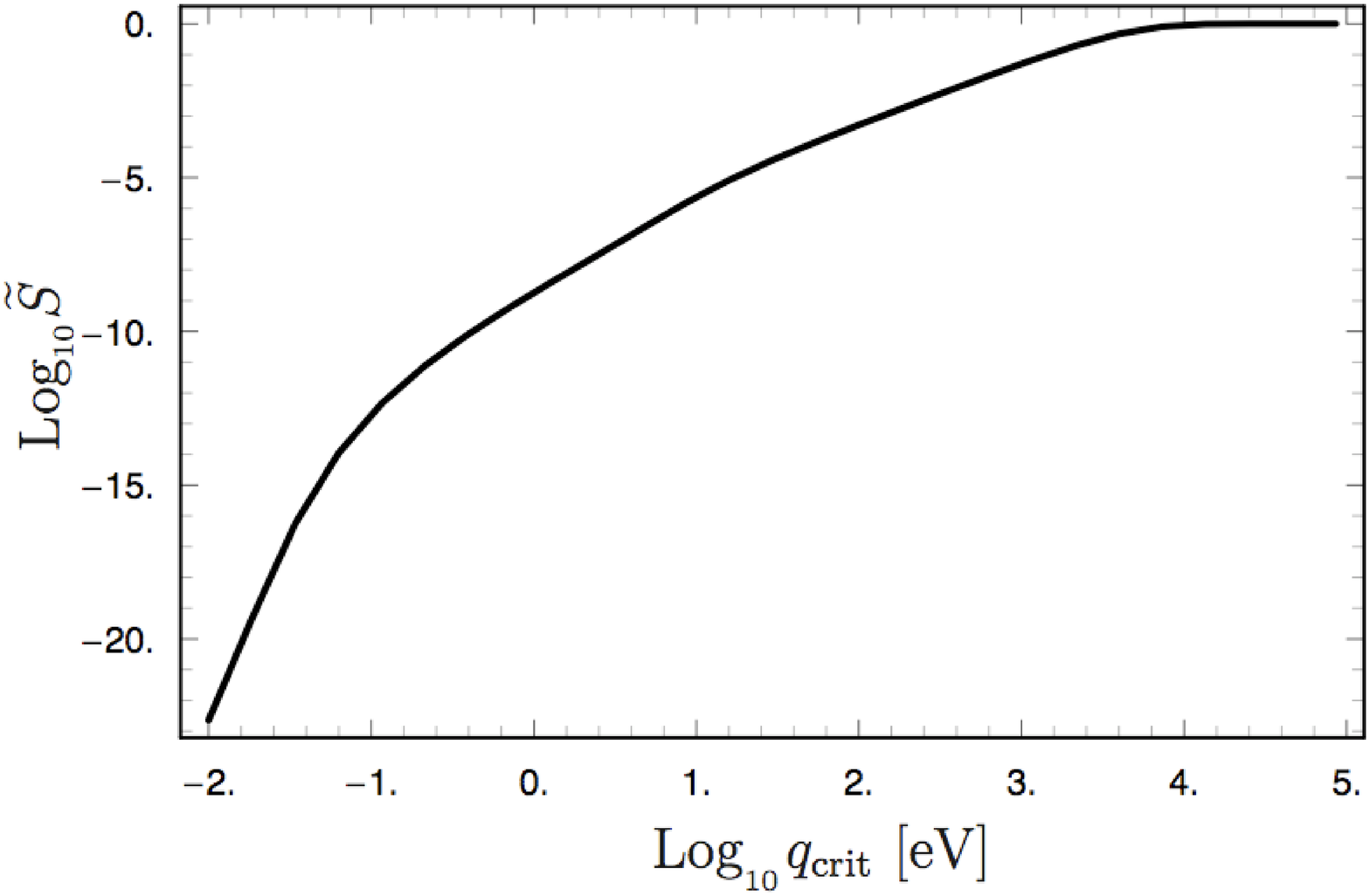}   
    \caption{\label{SUPGAINq2}Suppression factor $S$ for CAST, and $\tilde{S}$ for the energy loss
arguments as a function of $q_{\mathrm{crit}}$.}
\end{figure}

Proceeding along the lines of the previous section we can calculate the suppression factors for the CAST experiment $S$
and the corresponding $\tilde{S}$ that appears in the energy loss considerations. The results are plotted in
Fig.~\ref{SUPGAINq2}.

Using the required suppression \eqref{castfactor}, $S\sim 10^{-20}$, for CAST and \eqref{lossfactor}, $\tilde{S}\sim
10^{-10}$, for the energy loss arguments, we infer that sufficient suppression requires
\begin{equation}
q_{\mathrm{crit}}\lesssim 10^{-2}\, \mathrm{eV}.
\end{equation}
Although this seems rather small it is nevertheless quite big compared to the typical momentum transfer in the PVLAS
experiment,
\begin{equation}
q_{\mathrm{PVLAS}}\approx \frac{m^{2}_{\phi}}{2\omega}\sim 6\times10^{-7}\,\mathrm{eV}.
\end{equation}

\subsection{Kinematical suppression}

So far, we have suppressed the production of ALPs by reducing their coupling to photons. Now, we consider the
possibility that the suppression originates from an increase of the ALP's effective mass. Clearly, if the latter is
larger than the temperature, only the Boltzmann tail of photons with energies higher than the mass can contribute to
ALP production.

If we consider macroscopic environmental parameters $\eta(R)$ and, again, assume the simplest dependence on these
parameters,
\be \label{massdependence} m(\eta < \eta_{\mathrm{crit}})= m\ \ (\sim \textrm{meV}), \hspace{1cm}
m(\eta>\eta_{\mathrm{crit}})=\infty, \ee
the suppression is identical to the one computed in Sect. \ref{macroscopic}, since the Boltzmann tail vanishes for
infinite mass. Accordingly, Figs.~\ref{fig4} and  \ref{totalflux1} give the correct suppression also for the case of an
environment dependent mass.

Before we continue let us point out that a strong dependence of the mass on environmental parameters such as in Eq.
\eqref{massdependence} is problematic because it requires a strong coupling between the ALP and its environment. This
still holds even if we require only $m(\eta>\eta_{crit})\gtrsim 10\,\mathrm{keV}$. The strong coupling is likely to
lead to
unwanted side effects
, as we commented in Sec. II, but let us however discuss some phenomenological aspects which could distinguish
kinematical suppression from a dynamical suppression via the coupling.
As an explicit example, we discuss a dependence on the density $\rho$. The wave equation for the ALP will be
\be \label{waveequation} \Box\phi + m^{2}(\rho(x))\phi=0 . \ee
The effective mass,  $m(\rho(x))$, acts as a potential for $\phi$. This can actually lead to a new way to avoid the
CAST bound. For example consider a situation where ALPs are emitted with energy $\omega$. When they encounter a
macroscopic ``wall'' with $m(\rho_{\mathrm{wall}})>\omega$ on their way to the CAST detector, they will be reflected
due to energy conservation (tunneling through a macroscopic barrier is negligible). In other words, they will not be
able to reach the CAST detector and can not be observed. In this case only the energy loss arguments require a
suppression of the production \eqref{lossfactor} whereas the stronger constraint \eqref{castfactor} from CAST is
circumvented by the reflection.

\begin{figure}
\begin{center}
\includegraphics[width=12cm]{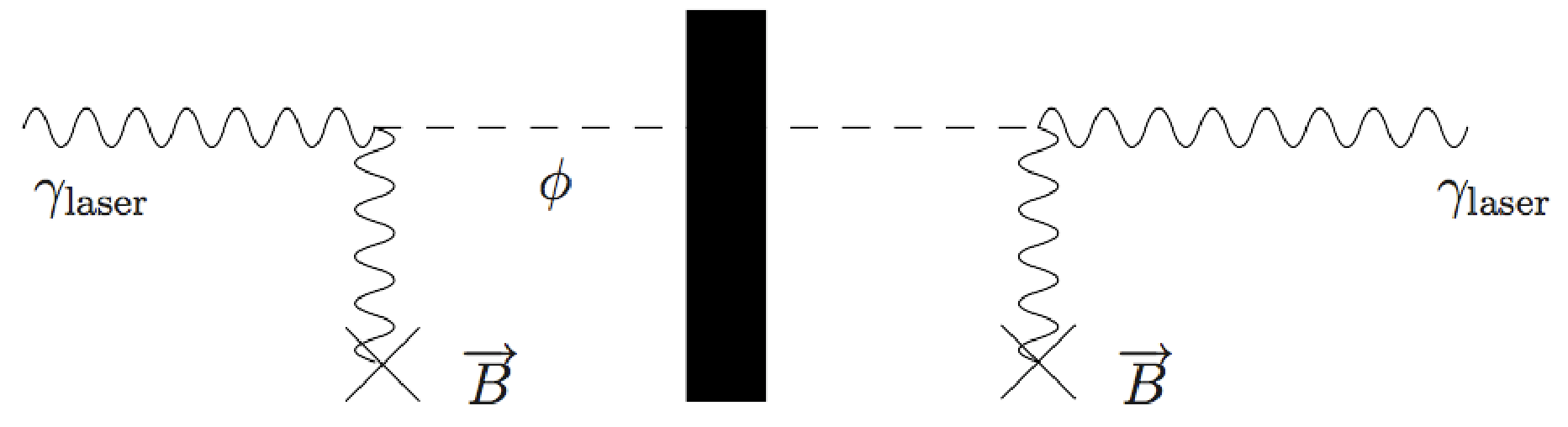}
\end{center}
\vspace{-1cm} \caption{Schematic view of a ``light shining through a wall'' experiment. (Pseudo-)scalar production
through photon conversion in a magnetic field (left), subsequent travel through an (opaque)  wall, and final detection
through photon regeneration (right). \hfill \label{fig:ph_reg}}
\end{figure}

This effect will also play a central role in the interpretation of the PVLAS result in terms of an
ALP. Note that the interaction region (length $L$) of the PVLAS set up is located inside a
Fabry-Perot cavity which enlarges the optical path of the light inside the magnetic field by a
factor $N_r\sim 10^5$ accounting for the number of reflections inside the cavity.. In the standard
ALP scenario, the ALPs created along one path cross the mirror and escape from the cavity. Coherent
production takes place only over the length $L$. The net result produces a rotation non-linear in
$L$ but only linear in $N_r$ \cite{Raffelt:1987im},
\begin{equation}
|\Delta\theta|=N_{r}\left(\frac{B\omega}{M m^2}\right)^{2}\sin^{2}\left(\frac{L m^2}{4\omega}\right).
\end{equation}
However, if $m=m(\rho)$ the ALPs have a potential barrier in this mirror and they will be reflected in the same way as
the photons. In fact, the whole setup now acts like one pass through an interaction region of length $N_{r}L$. The ALP
field in the cavity will increase now non-linearly in $N_{r}L$  modifying the predicted rotation in the following way
\begin{equation}
\label{modified} |\Delta\theta_{\mathrm modified}|=\left(\frac{B\omega}{M m^2}\right)^{2} \sin^{2}\left(\frac{N_{r} L
m^2}{4\omega}\right),
\end{equation}
where $\omega$ is the frequency of the laser. For small enough $m\lesssim {\mathrm few}\times10^{-6}\,\mathrm{eV}$ this
grows as
\begin{equation}
|\Delta\theta_{\mathrm modified}|\approx \frac{N^{2}_{r}L^{2}B^{2}}{16 M^{2}}.
\end{equation}
Under these conditions the PVLAS experiment cannot fix $m$ using the exclusion bounds from BFRT. Using Eq.
\eqref{modified} the rotation measurement \textit{suggests}, however, a much more interesting value
\begin{equation}
M_{\mathrm modified} \sim 10^8\ \mathrm{GeV},\quad{ \mathrm{for} }\quad m\lesssim {\mathrm
few}\times 10^{-6} {\mathrm{eV}},
\end{equation}
where we have used$L\sim 1$~m, $N_r\sim 10^5$ and $\omega\sim 1$ eV for the PVLAS setup. That could be reconciled more
easily with astrophysical bounds within our framework.

Such an effective mass will also play a role in  ``light shining through a wall'' experiments (cf.
Fig.~\ref{fig:ph_reg}). Typically, the wall in such an experiment will be denser than the critical density
$\rho_{\mathrm{crit}}$ required from the energy loss argument. Consequently, an ALP produced on the production side of
such an experiment will be reflected on the wall and cannot be reconverted in the detection region. Hence, such an
experiment would observe nothing if a density dependent kinematical suppression is realized in nature.

\section{\label{summary}Summary and conclusions}

The PVLAS collaboration has reported a non-vanishing rotation of the polarization of a laser beam propagating through a
magnetic field. The most common explanation for such a signal would be the existence of a light (pseudo-)scalar
axion-like particle (ALP) coupled to two photons. However, the coupling strength required by PVLAS exceeds
astrophysical constraints by many orders of magnitude. In this paper, we have quantitatively discussed ways to evade
the astrophysical bounds by suppressing the production of ALPs in astrophysical environments, in particular in the Sun.

The simplest way to suppress ALP production is to make the coupling $1/M$ of ALPs to photons small
in the stellar environment. Motivated by microphysical models~\cite{Mohapatra:2006pv}, we
considered a dependence of $M$ on macroscopic environmental parameters, such as temperature, plasma
mass $\omega_P$, or density $\rho$ whose values typically depend only on the distance from the
center of the Sun. One of our main results is that it is not sufficient to suppress production in
the center of the Sun only. One has to achieve efficient suppression also over a significant part
of the more outer layers of the Sun. As apparent from Tables~\ref{PVLAS_SUN_HB}, \ref{tab2} and
Eq.~\eqref{castfactor}, it is possible to reconcile the PVLAS result with the bound from the CERN
Axion Solar Telescope (CAST) if strong suppression sets in at sufficiently low critical values of
the environmental parameters, e.g. $\rho\sim 10^{-3}$~g/cm$^3$, or $\omega_P\sim $~eV. The bounds
arising from solar energy loss considerations are less restrictive (cf. Eq.~\eqref{lossfactor} and
Figs.~\ref{fig2}, \ref{totalflux1}). As an alternative suppression mechanism, we have also
exploited an effective mass that grows large in the solar environment. This case, too, requires
that the effect sets in already for low critical values of the environmental parameters (cf.
Figs.~\ref{fig4}, \ref{totalflux1}).

A somewhat different possibility is that the coupling $1/M$ depends on a microscopic parameter of
each production event like, e.g., the momentum transfer $q^2$ \cite{Masso:2005ym,Masso:2006gc}. The
typical values of such a microscopic parameter in the Sun may be different from those in laboratory
conditions. For example, the typical $q^2$ in the Sun is $\sim \mathrm{keV}^2$ whereas it is $\sim
10^{-12}\,\mathrm{eV}^2$ in a laser experiment like PVLAS. Suppression is then achieved by making
the coupling small for the typical values in the solar environment. However, untypical events occur
from time to time and the required value, $q_{\mathrm{crit}}\lesssim 10^{-2}\,\mathrm{eV}$ is
smaller than the $q_{\mathrm{crit}}\lesssim 0.4\,\mathrm{eV}$ estimated in \cite{Masso:2005ym} (it
is a bit more difficult to provide a similar estimate for \cite{Masso:2006gc} since there are
energy loss channels other than Primakoff production). This holds even though we have used a
step-function suppression factor which gives a stronger suppression than the form factors obtained
in the microphysical models of \cite{Masso:2005ym,Masso:2006gc}.

Most proposed near-future experiments to test the PVLAS ALP interpretation are of the ``light shining through a wall''
type (cf. Fig.~\ref{fig:ph_reg}). In these experiments, the environment, i.e. the conditions in the production and
regeneration regions, may be modified. The above mentioned critical values are small enough that they may be probed in
such modifications. For example, a density dependence may be tested by filling in buffer gas.

In conclusion, the PVLAS signal has renewed the interest in light bosons coupled to photons. The astrophysical bounds,
although robust, are model-dependent and may be relaxed by many orders of magnitude. Therefore, the upcoming laboratory
experiments are very welcome and may well lead to exciting discoveries in a range which was thought to be excluded.

\clearemptydoublepage

%% file: PRL2.tex
\chapter{Light scalars coupled to photons and non-newtonian forces}
\label{Light scalars coupled to photons and non-newtonian forces}\vspace{-0.5cm}
\begin{flushright} (In collaboration with A.~Dupays, E.~Mass{\'o} and C.~Rizzo.\\\vspace{-0.07cm}
Published in Phys. Rev. Lett. \textbf{98}, 131802 (2007) \cite{Dupays:2006dp})\end{flushright}

Spinless light particles are a common prediction of many theories that go beyond the standard $SU(3)_c\times SU(2)_L
\times U(1)_Y$ gauge theory. Probably the most famous of them is the axion \cite{Weinberg:1977ma,Wilczek:1977pj}, an
unavoidable consequence of the introduction of a new global $U(1)$ symmetry designed to solve the QCD CP-problem
\cite{Peccei:1977hh,Peccei:1977ur}. There are other examples of light particles; some are pseudoscalar like the axion
itself and some are scalar particles. We call them axionlike particles (ALPs).

In general, a spinless particle couples to two photons, and of course in principle it has also
couplings to matter. But the $\gamma\gamma$ coupling is particularly interesting because many
experimental searches for ALPs are based on it. One of these searches is based on the so-called
haloscope \cite{Sikivie:1983ip}, where axions or some other similar particles forming part of the
dark matter in the galactic halo can convert into photons in a cavity with a strong magnetic field.
Another search is to look for ALPs produced in the Sun converting into photons in a detector having
a strong magnetic field; this is called helioscope \cite{Sikivie:1983ip}. Still another experiment
searching for ALPs and that uses the $\gamma\gamma$ coupling looks for optical dichroism and
birefringence in a laser that propagates in a magnetic field \cite{Maiani:1986md}.

While the most recent haloscope \cite{Asztalos:2004} and helioscope \cite{Zioutas:2004hi} experiments have not found
any signal and thus have put limits on model parameters, the third type of experiment described above has reported a
positive signal. Indeed, the PVLAS collaboration finds a rotation of the polarization plane of the laser as well as an
induced ellipticity \cite{Zavattini:2005tm,pvlas-Bir-5thpaper:2006}. There is an exciting interpretation of these
results in terms of a new scalar particle $\phi$, that should have a mass $m_\phi \sim 10^{-3}$ eV and a
$\phi\gamma\gamma$ coupling scale of about $M \sim 10^5$ GeV -see equation (\ref{Lone}) below for the definition of
$M$.

The purpose of this paper is to show that in the case that the particle $\phi$ is indeed a light {\bf scalar}, the
$\phi\gamma\gamma$ coupling leads to the existence of long-range spin-independent non-newtonian forces. Our calculation
will allow us to find a very stringent limit on the coupling, using the experimental null results coming from searches
for new forces.

A scalar particle couples to two photons with the lagrangian
\begin{equation}
{\cal L}_1= {1 \over 4 M}\, \phi F^{\mu\nu} F_{\mu\nu} \label{Lone}
\end{equation}
The key point is that (\ref{Lone}) induces radiatively a coupling to charged particles, for example to protons (see Fig
1). The induced coupling will have the form
\begin{equation}
{\cal L}_2= y\, \phi \bar \Psi \Psi \label{Ltwo}
\end{equation}
where $\Psi$ is the proton field and $y$ the Yukawa coupling. The loop diagram  is logarithmically divergent.

In order to treat physically this divergence, we notice that (\ref{Lone}) corresponds to a non-renormalizable term in
the lagrangian and as such is expected to be valid only up to a high-energy scale $\Lambda$, where new physics has to
appear. We integrate momenta in the loop of Fig.1 only until $\Lambda$, so that we cut the divergence with the scale
$\Lambda$.

The logarithmically divergent part of the diagram of Fig.1 is well-defined and is the leading radiative contribution to
$y$. Approximating $y$ by this term we obtain
\begin{equation}
y= {3 \over 2} \, {\alpha \over \pi}\,  {m_p \over M}\,  \log {\Lambda \over m_p} \label{y}
\end{equation}

In principle there is also a tree level Yukawa term (\ref{Ltwo}) in the theory, and there could be some cancelation
between the tree level and the radiatively induced coupling (\ref{y}). We regard this possibility as very unnatural,
and will use (\ref{y}) for our estimates.

\begin{figure}\centering
  \includegraphics[width=5cm]{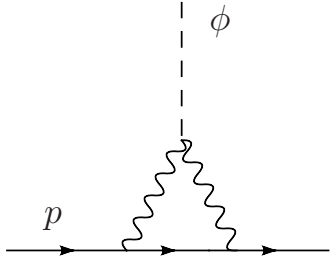}
  \caption{Loop diagram.}
\end{figure}

The Yukawa coupling (\ref{Ltwo}) leads to non-newtonian forces between two test masses $m_1$ and $m_2$, due to $\phi$
exchange. The total potential between 1 and 2 is
\begin{equation}
V(r) = G {m_1 m_2 \over r}\, + \, \frac{y^2}{4\pi} {n_1 n_2 \over r} e^{- m_\phi r} \label{V}
\end{equation}
Here $n_i$ is the total number of protons in the test body $i$. In (\ref{V}) we have neglected the new electron-proton
and electron-electron force since the corresponding value for $y$ in the case of the electron is smaller than (\ref{y})
by a factor $m_e/m_p$. When the two test bodies are constituted each by only one element, with atomic numbers $Z_1$ and
$Z_2$, and mass numbers $A_1$ and $A_2$, we can approximate (\ref{V}) by {\small
\begin{equation}
V(r) \simeq G {m_1 m_2 \over r}\, \left[ 1 + {1 \over G m_p^2} \, \frac{y^2}{4\pi} \, \left( {Z \over A} \right)_1
\left( {Z \over A} \right)_2 e^{- m_\phi r} \right] \label{Vapprox}
\end{equation}
}where now in the second term inside the square brackets i.e., the term containing the correction to the newtonian
potential, we have approximated $m_i\simeq A_i m_p$.

The non-newtonian part of the potential we have obtained has two clear properties: it has a finite range $m_\phi^{-1}$
and depends on the composition of the bodies, i.e. on their $Z/A$ values. We can use the abundant experimental bounds
on fifth-type forces to limit our parameter $M$ as a function of $m_\phi$.

To see how we proceed we find now the limit in the interesting ranges $m_\phi^{-1} \sim
(\mathrm{meV})^{-1} \sim 0.2$ mm, with the PVLAS results in mind. Bounds on new forces have been
obtained by experiments designed to measure very small forces. In the submillimeter range, in 1997
authors of ref. \cite{Carugno:1996uc} using a micromechanical resonator designed to measure Casimir
force between parallel plates gave limits on new forces, and they also estimated the corresponding
limits on the mass and inverse coupling constant of scalar particles. Also, strong experimental
limits on new forces have been published in \cite{Hoyle:2004cw},  where they
 use a torsion pendulum and a rotating attractor in
the framework of tests of the gravitational inverse-square law. The most strict bounds have been obtained very recently
by using torsion-balance experiments \cite{Kapner:2006si,Adelberger:2006dh}.

The limit from the experiment presented in ref. \cite{Kapner:2006si,Adelberger:2006dh} for $m_\phi = 10^{-3}$ eV is
\begin{equation}
\frac{y^2}{4\pi}  {1 \over G m_p^2} \left( {Z \over A} \right)_1 \left( {Z \over A} \right)_2 < 1.3 \times 10^{-2}
\end{equation}
which leads, introducing the conservative values for $(Z/A)$ of 0.4,
\begin{equation}
y < 7.8 \times 10^{-20} \label{y_bound}
\end{equation}

To obtain now a limit on $M$, we will simply put the value of the log in (\ref{y}) equal to 1; this will lead to a
conservative limit since $\Lambda \gg m_p$. With this, we obtain
\begin{equation}
M >  4.2 \times 10^{16}\ \mathrm{GeV} \label{M_bound}
\end{equation}

Such a high lower bound implies that no signal of a $0^+$ particle of mass $m_\phi \sim $ meV should be seen in
experiments like PVLAS \cite{Zavattini:2005tm,pvlas-Bir-5thpaper:2006} or CAST \cite{Zioutas:2004hi}. In order to reach
our conclusion we have to assume that the lagrangian (\ref{Lone})  is valid up to high energy scales $\Lambda \gg m_p$.
\begin{figure}\centering
   \includegraphics[width=11cm]{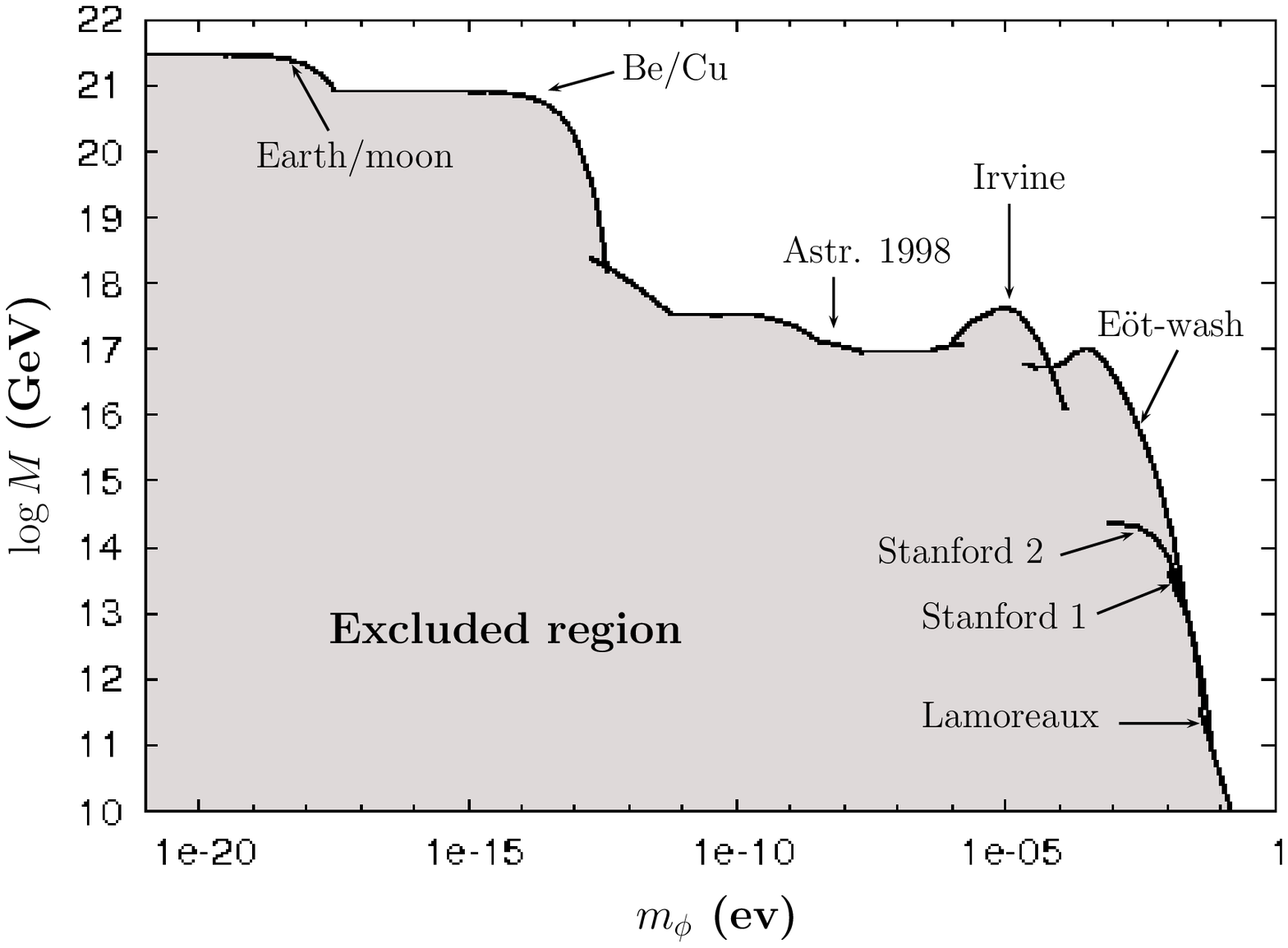}
  \caption{Constraints in the $\log{M}-m_\phi$ plane. Lines labeled Earth/moon
  and Astr. 1998 show constraints from astrophysical observations,
  Refs. \cite{Dickey:1994,Williams:1996} and \cite{Onofrio:2006} respectively. Lines labeled Be/Cu,
  Irvine, E\"ot-wash, Stanford 2, Stanford 2, and Lamoreaux show experimental
  constraints, Refs. \cite{Su:1994}, \cite{Hoskins:1985,Spero:1985}, \cite{Kapner:2006si,Adelberger:2006dh},
  \cite{Smullin:2005},
  \cite{Chiaverini:2003}, and \cite{Lamoreaux:1997,Lamoreaux:1998} respectively. The shaded region is excluded.}
\end{figure}
We can proceed in analogous way and find bounds for other values of $m_\phi^{-1}$. They  are shown in Fig. 2, and as we
see, the limits on $M$ are very tight.

Our final discussion is about modified   $\phi\gamma\gamma$ vertices. Examples of such models have been developed in
\cite{Masso:2005ym,Masso:2006gc} with the motivation of making compatible the PVLAS particle interpretation with the
bounds coming from stellar energy loss and a fortiori with the CAST results. In these models the ALP does couple to new
paraphotons and has not a direct coupling to photons. The   $\phi\gamma\gamma$ vertex arises because there is kinetic
photon-paraphoton mixing. Also, a paraphoton mass $\mu$ induces an effective photon form factor such that the coupling
is reduced for $|q|\gg \mu$.

When there is a form factor with a low scale $\mu$, the analysis shown in this paper should be modified accordingly. In
such a case the lagrangian in (\ref{Lone}) is valid only for photons with momentum $q$ such that  $|q| \ll \mu$. The
induced Yukawa will be suppressed with respect to the value (\ref{y}), and thus the lower bound (\ref{M_bound}) can be
very much relaxed if indeed $\mu$ is a low energy scale.

In order to quantify our last assertion, we have calculated the induced Yukawa coupling when the photons have a
modified propagator
\begin{equation}
{1\over q^2} \rightarrow {1 \over q^2} \,  {\mu^2 \over \mu^2 - q^2} \label{ff}
\end{equation}
Now the diagram of Fig.1 is finite.  The calculation of the coupling to protons  $y'$,
 at leading order in $\mu/m_p$ gives
\begin{equation}
y'=  {\alpha \over 4}\,  {\mu \over M} \label{yp}
\end{equation}

To find the potential between bodies we have to take into account the coupling to protons as well as to electrons,
because (\ref{yp}) is independent of the mass of the fermion. The potential when having a form factor (\ref{ff}) is
given by {\small
\begin{equation}
V(r) \simeq G {m_1 m_2 \over r}\, \left[ 1 + {4 \over G m_p^2} \, \frac{y'^2}{4\pi} \, \left( {Z \over A} \right)_1
\left( {Z \over A} \right)_2 e^{- m_\phi r} \right] \label{Vff}
\end{equation}
} We see that the new non-standard potential (\ref{Vff}) has a Yukawa coupling $y'$ that compared  to (\ref{y})  is
suppressed  by a factor of order $\mu/m_p$. The parameter $\mu$ introduced in  \cite{Masso:2005ym,Masso:2006gc} is not
fully specified by the theory. In order to solve the PVLAS-CAST puzzle and if we do not wish too different scales in
the model, $\mu$ should be in the subeV range with a preferred value $\mu \sim $ meV. For this value, the bound
(\ref{M_bound}) would relax by 12 orders of magnitude, bringing it close to the PVLAS value $M \sim 10^5$ GeV.
Remarkably enough, experiments searching for new forces and testing Casimir forces at the submm lengths may be
sensitive to the potential (\ref{Vff}) that corresponds to a vertex with the form factor (\ref{ff}).

{\bf Note added :} The fact that a scalar-photon-photon coupling gives rise to new forces and leads to a bound on $M$
has been independently realized by Shmuel Nussinov \cite{Zioutas:2006}.


%% file: chapter5.tex
\def\baselinestretch{1}

\chapter{Final Discussion and Conclusions\label{Final Remarks}}

The PVLAS signal has attracted a lot of theoretical work since the publication of their signal. As
a consequence, many arguments have arose that are relevant for my work.  On the other hand it is
mandatory to include some comments on the actual status of the field together with future
prospects. In next section I address both tasks. Finally I present a short summary and my
conclusions.

\section{Final discussion}

The original work of this thesis starts in Chapter \ref{Evading Astrophysical Constraints on
Axion-Like Particles}, corresponding to reference \cite{Masso:2005ym}. In this paper we quantified
the inconsistency between the PVLAS ALP interpretation and the astrophysical bounds (HB lifetime
and Helioscope experiments) and studied two mechanisms for avoiding these bounds. Both of them
require the existence of new particles together with the PVLAS ALP. The energy scale related to
this new physics is closer to the suggested ALP mass $m \sim$ meV than to the inverse coupling
$g^{-1}\sim 10^6$ GeV. This can be achieved considering that the ALP coupling to two photons
proceeds through photon-paraphoton mixing and a \textit{mediator} MCP which couples the paraphoton
field with the ALP.

The first mechanism consists in providing a strong interaction of ALPs with the stellar medium in
such a way that ALPs produced in stellar media can be reabsorbed inside the stellar plasma. The
second proposes that the Primakoff production of ALPs at the relatively high temperatures of
stellar interiors is suppressed with respect to the PVLAS experiment, performed in vacuum. Although
these general mechanisms remain to be valid, (and have probably inspired other works trying to
solve the PVLAS-astrophysics puzzle) the concrete models we presented are not reliable anymore. The
trapping mechanism has to pay the price of introducing a new \textit{relatively strong interaction}
between low mass novel particles and photons. In the same article \cite{Masso:2005ym} we see that a
background of paraphotons and ALPs will be formed in the early universe and this would contradict
the actual observations of the cosmic microwave background, pointing to a transparent universe at
temperatures around $T\sim $ eV.

Nevertheless, we decided to include this model for two reasons.First, because it allowed us to
introduce paraphoton models as our theoretical tool to interpret the PVLAS signal as a smoking gun
of \textit{low energy physics} and second because it illustrates one of the known mechanisms to
avoid the astrophysical bounds, so it might be inspiring for further investigations along that
line. Indeed, the trapping mechanism has originated at least a couple of works trying to evade the
stellar bounds on ALPs \cite{Jain:2005nh,Foot:2007cq} and paraphoton models have been further used
for model building around the PVLAS signal.

The second mechanism has proven to be more fruitful because it does not require a new strong
interaction, but contrarily, it suggests that the new physics can come into the game to suppress
the Primakoff production in stars, i.e. to make the ALP-photon interactions feebler. The concrete
model presented uses a new confining force, in the spirit of QCD, that makes the ALP a composite
particle of millicharged preons, providing a production form factor that suppresses the stellar
emission. This models has nowadays become obsolete. The reason is that, in this case, the main
production channel of ALPs in stars would not be the Primakoff production but the plasmon decay
into a pair of \textit{preons} that would further \textit{hadronize} into many ALPs and other
preonic bound states\footnote{I acknowledge interesting conversations at this respect with Joerg
Jaeckel, Andreas Ringwald, Alex Pomarol, Georg Raffelt.}. Although the final products of the decay
will be very different, such a process occurs at the same rate than the plasmon decay into a pair
of MCPs I used in Chapter \ref{constraints} so that the stellar energy loss bounds apply also in
this case.

In our second paper \cite{Masso:2006gc}, presented in Chapter \ref{Compatibility of CAST search
with axion-like interpretation of PVLAS results}, we recovered the idea of the suppression of
stellar ALP production within the framework of paraphoton models to provide a new model. There, the
coupling between the MCP and photons proceeds through two possible intermediate states, two
paraphotons, in such a way that they interfere negatively canceling the amplitude of any physical
process when the square of the momentum carried by the paraphotons tends to infinity. The
suppression of the stellar production achieved with this model can be very strong but has the
disadvantage that the cancelation between the contributions from the two paraphotons has to be
carefully tuned. Beyond this prejudice, the model is able not only to reconcile the PVLAS ALP with
astrophysics (HB lifetime and Helioscopes) but also with 5$^\mathrm{th}$ force searches if the
PVLAS ALP turns out to be finally a $0^+$ particle \cite{Dupays:2006dp}.

At the moment of writing this paper we were already aware of the above criticism to the suppression
model presented in \cite{Masso:2005ym} so this model focused on suppressing the plasmon decay into
MCP pairs and not the Primakoff production.

An alternative interpretation of the PVLAS signal can be made in terms of the pair production of
millicharged particles (MCPs) with mass $\lesssim 0.1$ eV and charge $\sim 10^{-6}$
\cite{Gies:2006ca}.
Basically this means that our model \cite{Masso:2006gc} is a candidate for explaining the PVLAS
signal \textit{even in the absence of an ALP}. Of course, the MCP-PVLAS interpretation based on
\cite{Masso:2006gc} is more economical than the one originally presented in \cite{Masso:2006gc}
because it requires one particle less.

In order to know if the ALP is superfluous or not we will have to wait until more experimental data
is available. The PVLAS experiment is sensitive to the parity of the ALP or the bosonic/fermionic
character of the MCP. Also, a positive signal in a light-shining-through-walls (LSW) experiment
could be the smoking gun of the existence of the ALP, since MCPs can not contribute to it. The
program of discriminating between these two possibilities has been started in \cite{Ahlers:2006iz},
and I say \textit{has been started} because an exact study of the PVLAS signal in our model
\cite{Masso:2006gc} has not yet been performed after we know that just MCP production can do the
job. A combined analysis of coherent ALP production and MCP production is in current development.
Yet a new source of rotation has been recently discovered, the coherent production of paraphotons
in magnetic fields in models with light paracharged particles \cite{Jaeckel:2007}. This adds up a
new difficulty to this combined analysis because now the model \cite{Masso:2006gc} without an ALP
will give a positive signal in future LSW experiments.

Remarkably, the advent of the MCP interpretation has motivated some works proposing new bounds for
the existence of MCPs \cite{Gies:2006hv,Gluck:2007ia,Melchiorri:2007sq}. In the last reference, the
impact of MCPs in the CMB radiation is studied and it is nowadays the most firm constraint for the
model \cite{Masso:2006gc}. Some work is in progress to ensure if the bounds presented are really
model independent or if they can be circumvented in any region of the parameter space or by means
of soft modifications of the model \cite{Masso:2006gc}. Notably, the CMB constraints do not rule
out the bare ALP interpretation.

Going back to the features of the model \cite{Masso:2006gc} one might feel uncomfortable not only
with so many parameters and fields introduced but also with the apparently arbitrary choice of
charge assignments. At this respect it is interesting to note that our model has been justified
from string theory \cite{Abel:2006qt}, were the existence of additional $U(1)$ symmetries is
generally predicted as well as particles in the required representations.

In \cite{Jaeckel:2006xm} I present a general study of the suppression scenario suggested in
\cite{Masso:2005ym}. We parameterize a general suppression mechanism allowing the ALP mass and
coupling constant to depend on any environmental physical quantity that can make a difference
between the clean PVLAS setup and the hot and dense stellar plasmas. Because in the suppression
scenario the Helioscope bounds are more robust than the energy loss bounds we focus exclusively on
the case of the Sun.

In the models worked out in \cite{Masso:2005ym,Masso:2006gc} we performed a rough estimate of the
stellar lifetime bounds in the presence of suppression of the ALP production: we computed the whole
ALP emission from the values of the relevant parameters at the stellar center. This is indeed a
quite good approximation in the absence of suppression but it is dangerous within our suppression
models for the following reason: our models are designed to be more effective in suppressing ALP
production wherever the conditions of the production are more \textit{extreme}. For instance, in
\cite{Masso:2005ym} the suppression of the Primakoff production depends on the screening scale,
$k_s$, which is maximum in the stellar center because the density of screening charges is much
higher there than in outer shells. Therefore, the suppression computed is overestimated in the
external layers of the stars. In order to provide better estimates we computed more accurate bounds
by integrating the ALP emission over the most recent solar model available.

In \cite{Jaeckel:2006xm} I differentiate two possibilities for a general suppression scenario: (a)
The sensitive parameter is an average environmental parameter like temperature or matter density
that, given a solar model, depends only on the position of the Sun and (b) the parameter is the
momentum transfer of the Primakoff conversion. Our conclusions are very interesting, in the first
case the conciliation of PVLAS and CAST demands that the suppression of the Primakoff ALP
production is complete in almost the whole volume of the Sun (up to a $95\%$ of the solar radius).
In the second, the scale at which the photon propagator must be cut-off is $\sim 10^{-2}$ eV. It is
interesting to note that this is the scale suggested in our model \cite{Masso:2005ym} by means of a
much simpler estimation.

While at the time of publication of \cite{Masso:2005ym} our model was the unique solution for the
PVLAS-astrophysics puzzle some alternatives begun to show up soon after. Interestingly enough, at
least two of them fit exactly our scheme of suppression driven by an environmental parameter in
\cite{Jaeckel:2006xm}. Thus our paper can be used to further constrain these new proposals.

In \cite{Mohapatra:2006pv} the authors propose an intricate mechanism for getting a ALP coupling
$g(T)$ that vanishes abruptly at a temperature between PVLAS and the stars due to the phase
transition of another scalar field. This model fits exactly our model independent scheme
\cite{Jaeckel:2006xm}. Our study clearly predicts that the value of the critical temperature
\textit{must be below} $25$ eV (See table \ref{tab2}) in order to satisfy the CAST bounds because
the ALP emission must be suppressed not only at $T \gg$ keV but essentially at every place of the
whole Sun.

Other model has recently appear in \cite{Brax:2007ak}. There the authors propose that the PVLAS ALP
can be of the chameleon type \cite{Khoury:2003aq}, an exotic $0^+$ field whose mass depends on the
environmental matter density, $m=m(\rho)$. This possibility had been indeed briefly examined in
\cite{Jaeckel:2006xm}. If the mass of the chameleon increases so much under the solar conditions
such that its production is kinetically impossible then we have the suppression required to evade
the CAST bounds. In this case the conclusions of our paper are much more dramatic than in the
previous case. The authors use a typical matter density in the Sun $\rho_\mathrm{Sun}\sim 10$
gcm$^{-3}$ while we find that the suppression should be fully active above $\rho_c\sim 9\ 10^{-3}$
gcm$^{-3}$.

All the models presented in this thesis show that reconciling the PVLAS ALP with astrophysics (in
particular with the Helioscope measurements) does require new physics at accessible scales. As I
pointed out in Chapter \ref{Evading} the ALP physics must dramatically change in between the PVLAS
and stellar environments. Notably, this suggests that the PVLAS particle (ALP or MCP) is not likely
to exist alone and thus the new generation of low energy experiments can eventually discover
several particles.

A program to discriminate between the different possibilities is currently under way. As the
conditions in the solar plasma are not so extreme, they can be testable in laboratories and
therefore new experimental efforts are crucial to test the PVLAS particle interpretation.

Interestingly enough, many experiments have been recently proposed to test the PVLAS signal. At
least three of them have currently began to work. The BMV and Q$\&A$ collaborations are building
PVLAS-type experiments looking for vacuum dichroism and birefringence while the ALPS collaboration
is currently building a LSW experiment. The same PVLAS collaboration will soon perform a LSW run
after a soft modification of their setup. Other LSW experiments have been proposed at Jefferson Lab
and CERN, the LIPPS and OSQAR collaborations. A detailed review of the status of the experimental
tests of the PVLAS signal can be found in \cite{Ringwald:2007pe}.

Finally it is worth mentioning that the PVLAS collaboration has been releasing new data in several
conferences during the last years. Together with the published measurements of the rotation of the
polarization plane of a $\omega\sim 1.2$ eV laser (red) they have presented rotation data for a
smaller wavelength light ($\omega\sim 1.4$ eV, green) and ellipticity measurements with both
lasers. The absolute values of rotation and ellipticity data agree with an ALP interpretation but
it is not the case when one takes into account the measured signs. Ellipticity data favors an $0^+$
ALP while rotation data with the green laser favors a $0^-$ ALP. If these results are confirmed
then the \textit{bare} ALP interpretation would be ruled out. It remains to be investigated if the
model in \cite{Masso:2006gc} could accommodate this results when the effects of MCP and paraphoton
production are taken into account.

The fact that PVLAS ellipticy measurements point to a $0^+$ ALP motivated us to consider the
$5^\mathrm{th}$ bounds presented in \cite{Dupays:2006dp}. As already noted, these bounds are the
strongest constraints for a $0^+$ ALP interpretation. However, they were unknown at the time of
publishing \cite{Masso:2006gc} and \cite{Jaeckel:2006xm} and therefore the requirements we derive
there for a $0^+$ ALP to satisfy the HB lifetime and Helioscope bounds are not sufficient to ensure
a PVLAS ALP interpretation. As a remarkable fact, the model \cite{Masso:2006gc} turned out to evade
the $5^\mathrm{th}$ force bounds and chameleon models as \cite{Brax:2007ak} are in principle able
to satisfy them as well.

\newpage\section{Conclusions}

In this thesis I have been concerned with some of the problems that arise when trying to interpret
the measurements of the PVLAS collaboration \cite{Zavattini:2005tm} in terms of a novel, low mass,
spin-zero particle coupled to light \cite{Maiani:1986md} we call ALP.

My main concern has been to investigate whether the strong astrophysical bounds, excluding such a
particle, could be circumvented in any particle physics model. I have discussed two mechanisms,
trapping the new particles in the stars and suppressing their production in stellar environments
with respect to the PVLAS experiment. The second has given more reliable results.

In order to build models that either trap or suppress the number of stellar ALPs I have invoked new
particles, namely paraphotons and millicharged particles.

Notably the scale of the masses required for these new particles is very low $\lesssim$  eV, a very
suggestive value. I have also made a model-independent study of general suppression mechanisms.
This work also points to new phenomena accessible in earth-ground laboratory experiments.
Therefore, new laboratory experiments have been encouraged.

In the near future, many experiments testing the PVLAS signal will be performed and this thesis
suggests that the PVLAS particle could not be alone. Therefore, having different kinds of
experiments will be crucial for confronting different models.

Recently, new strong constraints on the PVLAS ALP as well as on paraphotons and millicharged
particles have been derived. I have shown that $5^\mathrm{th}$ force searches set strong limits on
the PVLAS particle if it is a $0^+$. On the other hand, the CMB has been proven to be sensitive to
the existence of millicharged particles. Notably, my best model is able to avoid both constraints.

If the PVLAS signal is confirmed and firmly established, I believe that few particle physics models
could solve the strong conflict with astrophysics. In these thesis I have developed solutions that
involve physics at very low energies.


%% file: squizons.tex
\chapter{Schizons} \label{schizons}
If we let our ALP ${\phi_s}$ to couple to light through both interactions in \eqref{ALP-couplings}
parity is not conserved by the ALP interactions. Therefore, the ALP has not definite parity and
because of that it can be called a ``schizon" \cite{Hill:1988bu}. In this case, we find a
complicated $3\times 3$ mixing problem
\be \left[\omega^2+\partial^2_z + 2\omega^2\left( \begin{array}{ccc}
                               (n_\perp-1)  & n_R       &  \f{g^+B_T}{2\omega} \varsigma  \\
                                  n_R       & (n_{||}-1)&  \f{g^-B_T}{2\omega} \\
                       \f{g^+B_T}{2\omega}\varsigma  &  \f{g^-B_T}{2\omega} &-\f{m_{\phi_s}^2}{2\omega^2} \end{array}
 \right)\right] \left(  \begin{array}{c} A_\perp \\ A_{||} \\ {\phi_s} \end{array}\right)(z)=0
\ee
Where in the relativistic limit $\varsigma=\pm1$ for photons traveling along the $\pm z$-direction.
In absence of Cotton-Mouton and faraday effects (vacuum) the problem simplifies notably. By means
of a rotation $R^\varsigma_s$ of angle $\theta^\varsigma_s$ in the $A_{||}-A_\perp$ space it can be
decoupled into
\be \left[\omega^2+\partial^2_z + \left( \begin{array}{ccc}
                              0 & 0 & gB_T\omega \\
                              0 & 0 & 0            \\
                    gB_T\omega & 0 & -m_{\phi_s}^2 \end{array}
 \right)\right] \left(  \begin{array}{c} A_s \\ A_n \\ {\phi_s} \end{array}\right)(z)=0
\ee
where $A_s,A_n$ are the photon states coupling to the schizon  and neutral respectively, and
$g=\sqrt{g_-^{2}+g_-^{2}}$ is the effective coupling constant. We have found exactly equation
\eqref{mixing-equation for agamma_par} with the following replacements
$\{s,n\}\leftrightarrow\{||,\perp\}$, so equations \eqref{A_par_solution},\eqref{phi_solution} are
also solutions for the schizon case. In order to give a result for the induced rotation and
ellipticity in PVLAS we have to rotate back to $A_{||,\perp}$. In the weak mixing regime, I get
\bea \delta\zeta = &\f{\sin{\pa{2\zeta_0-2\theta^\varsigma_s}}}{2}
&\f{2g^2B^2\omega^2}{m_{\phi_s}^4}\sin^2\f{m_{\phi_s}^2z}{4\omega}  \label{schizon_rotation} \\
     \delta\psi  = &\f{\sin{\pa{2\zeta_0-2\theta^\varsigma_s}}}{2}
     &\f{2g^2B^2\omega^2}{m_{\phi_s}^4}\pa{\f{m_{\phi_s}^2z}{2\omega}-\sin\f{m_{\phi_s}^2z}{2\omega}}
\label{schizon_ellipticity} \eea
The schizon couplings add at least two new ingredients to the ALP interpretation. First,
\textit{they provide a phase shift between the photon initial polarization and the final rotation}.
By sending $g_{||,\perp}$ to zero we recover the two pure parity cases, because in this cases
$\sin\theta_s=0,1$, namely $\theta_s=0,\pi /2$ and $\sin({2\zeta_0-0,\pi})=\pm \sin2\zeta_0$. The
rotations shown again to be of opposite sign. Secondly, the photon state coupling to ${\phi_s}$
depends on the direction of propagation ($\theta^\varsigma_s$ depends on the sign $\varsigma$,
namely $\theta^-_s=-\theta^+_s$)
\bea {\phi_s}(\rightarrow)\ \ \  &\mathrm{couples\ \ with}&\ \ \ g^+ A_\perp + g^- A_{||}  \\
     {\phi_s}(\leftarrow)\ \ \   &\mathrm{couples\ \ with}&\ \ \ -g^+ A_\perp + g^- A_{||} \ .\eea
As a consequence the effect of multiple reflections inside a Fabry-Perot cavity can be considerably
diminished \cite{Liao:2007nu}, a fact that worsens the PVLAS discrepancy with astrophysics. This
can be easily seen from \eqref{schizon_rotation} and \eqref{schizon_ellipticity} since the sum of
the two rotations(ellipticities) acquired after \textit{two} consecutive passes in opposite
directions is
\be \delta\zeta(\leftrightarrow) = 2 \cos2\theta_s \times
\f{2g^2B^2\omega^2}{m_{\phi_s}^4}\sin^2\f{m_{\phi_s}^2z}{4\omega}\ee
which is maximum for $\theta_s=0,\pi/2...$, namely, the pure parity cases. A similar expression
holds for the ellipticity. A remarkably fact in this case is that the information of $g$ and
$\theta_s$ cannot be extracted separately and thus PVLAS is not able to discriminate is the ALP is
an schizon or not.
